\documentclass[12pt]{article}

\usepackage{amsmath,amsfonts,amsbsy,amssymb,array,accents,dsfont}
\usepackage{enumerate,array,latexsym,graphicx,mathrsfs,verbatim,psfrag}
\usepackage{bm} 
\usepackage[normalem]{ulem}
\usepackage{multirow}
\usepackage{cancel}

\usepackage{booktabs} 
\usepackage[usenames]{color}
\usepackage[utf8]{inputenc}
\usepackage[english]{babel}
\usepackage{fancybox}

\usepackage{datetime}
\usepackage{pdflscape}

\usepackage[nosort]{cite}
\usepackage{chngpage} 
\usepackage{setspace}
\usepackage{tensor}

\usepackage{enumitem}
\usepackage{subcaption}

\usepackage{enumitem}
\setlist{itemsep=0pt}

\usepackage[colorlinks=true,      linkcolor=darkblue,      urlcolor=darkblue,      
            filecolor=darkblue,      citecolor=darkblue,       pdfstartview=FitH,     
						pdfpagemode=UseNone,      bookmarksopen=true]{hyperref}  
\usepackage[all]{hypcap}     

\newcommand{\captionfonts}{\small}
\makeatletter  
\long\def\@makecaption#1#2{%
  \vskip\abovecaptionskip
  \sbox\@tempboxa{{\captionfonts #1: #2}}%
 \ifdim \wd\@tempboxa >\hsize
    {\captionfonts #1: #2\par}
  \else
    \hbox to\hsize{\hfil\box\@tempboxa\hfil}%
  \fi
  \vskip\belowcaptionskip}
\makeatother   

\DeclareMathSymbol{\medhatsym}{\mathord}{largesymbols}{"62} 

\DeclareMathSymbol{\medtildesym}{\mathord}{largesymbols}{"65}

\newcommand{\comm}[1]{} 

 \def\indirect{\emph{indirect} }
 
 \def\direct{\emph{direct} }

\def\IR{\mathbb{R}}

\def\({\left(}
\def\){\right)}
\def\[{\left[}
\def\]{\right]}

\def\One{{\hbox{ 1\kern-.8mm l}}}

\def\barray{\begin{array}}
\def\earray{\end{array}}
\def\be{\begin{equation}}
\def\ee{\end{equation}}
\def\bea{\begin{eqnarray}}
\def\eea{\end{eqnarray}}
\def\bal{\begin{align}}
\def\eal{\end{align}}

\def\-{\,-\,}
\def\={\,=\,}
\def\+{\,+\,}
\def\equi{\,\equiv\,}

\numberwithin{equation}{section} 

\makeatletter
\g@addto@macro\bfseries{\boldmath}
\makeatother

\definecolor{cardinal}{rgb}{0.6,0,0}
\definecolor{darkgreen}{rgb}{0,0.4,0}
\definecolor{purple}{rgb}{0.5, 0, 0.5}
\definecolor{golden}{rgb}{0.92, 0.7, 0}
\definecolor{midnight}{rgb}{0, 0, 0.5}
\definecolor{darkblue}{rgb}{0, 0, 0.8}

\def\IR{\mathds{R}}

\def\cA{{\cal A}}
\def\cB{{\cal B}}

\def\cK{{\cal K}}

\def\cM{{\cal M}}
\def\cN{{\cal N}}
\def\cO{{\cal O}}

\def\cQ{{\cal Q}}

\def\cS{{\cal S}}

\def\cO{{\cal O}}

\begin{document}

\phantom{AAA}
\vspace{-10mm}

\begin{flushright}
%
%
\end{flushright}

\vspace{0cm}

\begin{center}

\baselineskip=34pt
\parskip=3pt

{\huge {\bf Gravitational Footprints of Black Holes and Their Microstate Geometries}}

\baselineskip=23pt
\parskip=3pt

\vspace{1cm}

\vskip 1.2cm

 {\large Ibrahima Bah$^{a}$,~~ Iosif Bena$^{b}$,~~ Pierre Heidmann$^{a}$, ~~ Yixuan Li$^{b}$ ~~and~~\\  Daniel R. Mayerson$^{b}$ }
 
 \baselineskip=15pt
\parskip=3pt

\vskip 0.8cm
{\it  $^{a}$ Department of Physics and Astronomy,\\
Johns Hopkins University,\\
3400 North Charles Street, Baltimore, MD 21218, USA} \\
\vskip 0.1cm
{\it  $^{b}$ Institut de Physique Th\'eorique,\\
Universit\'e Paris Saclay, CEA, CNRS\\
Orme des Merisiers,  
91191 Gif-sur-Yvette Cedex, France} 
\vskip 0.1cm
\vskip 0.8cm

\vspace{5mm} 
{\footnotesize\upshape\ttfamily iboubah, pheidma1 @ jhu.edu;~~iosif.bena, yixuan.li, daniel.mayerson @ ipht.fr} \\

\end{center}
\vspace{1.2cm}
 

\begin{adjustwidth}{3mm}{3mm} 
 
\vspace{-1.2mm}
\noindent

We construct a family of non-supersymmetric extremal black holes and their horizonless microstate geometries in four dimensions. The black holes can have finite angular momentum and an arbitrary charge-to-mass ratio, unlike their supersymmetric cousins. These features make them and their microstate geometries astrophysically relevant. Thus, they provide interesting prototypes to study deviations from Kerr solutions caused by new horizon-scale physics. In this paper, we compute the gravitational multipole structure of these solutions and compare them to Kerr black holes. The multipoles of the black hole differ significantly from Kerr as they depend non-trivially on the charge-to-mass ratio. The horizonless microstate geometries have the same multipoles as their corresponding black hole, with small deviations set by the scale of their microstructure.

\end{adjustwidth}

\thispagestyle{empty}
\newpage


\baselineskip=14pt
\parskip=2.5pt

\setcounter{tocdepth}{2}
\tableofcontents

\baselineskip=15pt
\parskip=3pt


\newpage
\section{Introduction}
\label{sec:Intro}
%

The observation of gravitational waves by the LIGO collaboration \cite{Abbott:2016blz} from colliding black holes has lead to a paradigm shift in how we think about black holes: they are physical objects that can be observed and studied in nature.  Furthermore, gravitational-wave interferometers \cite{Abbott:2007kv} and telescope arrays \cite{EHT2019a} have began to measure effects that are sensitive to possible new physics at the scale of the horizon, and this sensitivity is only bound to increase with the advent of space-based  gravitational-wave interferometers \cite{Audley:2017drz} and third-generation ground-based ones \cite{Punturo:2010zz}.

On the other hand, black holes have served as the primary theoretical lab for exploring quantum gravity.  Their theoretical studies have led to many interesting puzzles and paradoxes which have offered important windows into quantum gravity; chief among them are the origin of the microstates that make up the Bekenstein-Hawking entropy, and the unitary problem of black hole evaporation.  The study of microstates is an inherently top-down question as it requires an understanding of the microscopic degrees of freedom of quantum gravity. In contrast, the unitary problem can be studied from a bottom-up perspective by exploring consistency of quantum mechanics near black hole environments.  An important lesson from the latter is that black-hole evaporation is compatible with the standard rules of quantum mechanics only when there is new structure at the scale of the black horizon \cite{Mathur:2009hf}.  This is also articulated more recently in the firewall paradox \cite{Almheiri:2012rt}.

One of the crowning successes of string theory is reproducing the Bekenstein-Hawking entropy for a wide classes of supersymmetric black holes from stringy microscopic states \cite{Strominger:1996sh}.  A fundamental question that follows is: What are the gravitational properties of such microstates? While generically it is hard to characterize their quantum mechanical properties, there can exist wide classes of microstates that are sufficiently coherent to admit classical descriptions as smooth horizonless geometries \cite{Bena:2007kg,Bena:2016ypk}. These states are called microstate geometries.  These geometries are indistinguishable from their corresponding black hole up to the region of the would-be horizon, where the spacetime ends in a smooth horizonless cap that contains non-trivial topological cycles wrapped by fluxes.  

The key lesson from the top-down studies of microstate geometries and as well as the bottom-up analysis is: \emph{black holes as observed in nature may correspond to ultra compact objects made up of new phases of matter from quantum gravity}.  In the fast emerging field of gravitational-wave astronomy and astrophysics, an important goal will be to characterize the possible deviations from astrophysical Kerr black holes in general relativity (GR) coming from the new black hole microstructure in string theory.

An important class of observables that can distinguish microstate geometries with horizon-size structure from classical GR black holes is the tower of gravitational multipole moments. These observables are already astrophysically interesting and will become more so in the era of gravitational-wave astronomy. In this paper, we aim to characterize gravitational multipole moments of microstate geometries that correspond to a class of four-dimensional non-supersymmetric spinning extremal black holes, dubbed almost-BPS \cite{Goldstein:2008fq,Bena:2009en,Bena:2009ev,DallAgata:2010srl,Vasilakis:2011ki}.  There are several benefits in studying the almost-BPS black holes and their microstates:

 First, almost-BPS black holes and their microstate geometries are constructed from supergravity by subtly breaking supersymmetry while maintaining the general linear structure that allows for solvability \cite{Goldstein:2008fq,Bena:2009en}.  This has several benefits as compared to their widely studied supersymmetric\footnote{For recent study of gravitational multipoles for supersymmetric microstate geometries see \cite{Bena:2020see,Bianchi:2020bxa,Bena:2020uup,Bianchi:2020miz}.  For a more general discussion of observables of supersymmeric microstate geometries see \cite{Mayerson:2020tpn}.} cousins. Indeed, the supersymmetric four-dimensional black holes in string theory are non-rotating, extremal and carry large charges;\footnote{These charges emerge from string theory and can be associated to dark Maxwell fields.} these features undermine their phenomenological interest.  In contrast, almost-BPS black holes can have angular momentum, small charge-to-mass ratios,\footnote{Even if almost-BPS black holes are extremal, their four-dimensional charges can be made arbitrarily small compared to their mass by turning on non-trivial scalar fields that modify the effective electromagnetic coupling in four dimensions. We discuss this in detail in section \ref{h-physics}.} and can therefore have the same conserved charges as astrophysical black holes. In this regard, it is phenomenologically relevant to study the multipole structure of these black holes and compare it to the multipole structure of Kerr black holes. This initial analysis will provide the baseline to study deviations of multipoles caused by  horizon-scale microstructure in the almost-BPS microstate geometries when compared to the almost-BPS black holes and to the Kerr black holes.

Another benefit of studying almost-BPS solutions is to contrast them with the phenomenological modeling of Exotic Compact Objects (ECOs) \cite{Cardoso:2017njb,Cardoso:2019rvt,Mayerson:2020tpn}.  These are bottom-up objects that force structure at horizon scale and thereby violate black hole no-hair theorems and the Buchdahl bound \cite{Buchdahl:1959zz}. Their description require exotic matter with no UV physics.  Almost-BPS solutions on the other hand can be understood directly in string theory.  

The multipole structure of almost-BPS black holes and their microstate geometries is very rich. This is owed to the non-trivial four-dimensional angular momentum that is absent in the non-spinning supersymmetric black holes. We show that the multipoles of almost-BPS black holes have a similar functional dependence on the mass-to-spin ratio as the multipoles of Kerr(-Newman) black holes. Moreover, unlike the latter, almost-BPS multipoles  also have an interesting dependence on the charge-to-mass ratio. Furthermore, the generic almost-BPS black holes have all non-zero multipoles, including the odd-parity multipoles which vanish in Kerr; the presence of these multipoles indicates a breaking of equatorial symmetry ($\theta \leftrightarrow \pi-\theta$) which could have interesting observable consequences.  This is a significant deviation from Kerr black holes.  

Multipoles of almost-BPS microstate geometries have a highly non-trivial dependence on the internal degrees of freedom of the geometry.  These deviate from the multipoles of the almost-BPS black hole at the same scale as the size of the microstructure in the near horizon region.\footnote{Extremal black holes have a throat of infinite length, that an infalling observer can traverse in finite proper time, so the``size'' of the microstructure above the horizon is subtle to define \cite{Dimitrov:2020txx}.} 
We show that the deviations are rather ``random'' as they depend on the geometry of the topologically non-trivial bubbles that give the horizon-scale structure, and they can be either positive or negative; this contradicts a recent conjecture that multipoles of microstate geometries are larger than the multipoles of their corresponding black holes \cite{Bianchi:2020bxa}. In a broader scope, our constructions and studies in this paper allow us to understand the physics of potential microstructure of the Kerr black hole much better than what one could hope from naively extrapolating the supersymmetric microstate results as was previously done in \cite{Bena:2020see,Bianchi:2020bxa,Bena:2020uup,Bianchi:2020miz}.

The structure of the paper is as follows. In the next subsection we summarize our main results. In section \ref{sec:Generalities}, we review non-supersymmetric four-dimensional solutions obtained by the almost-BPS ansatz and some generalities about the derivation of multipole moments. In section \ref{sec:ABPSBlackHoles}, we construct non-supersymmetric rotating almost-BPS black holes that look almost neutral and we compare their physics to that of Kerr-like GR black holes. In section \ref{sec:ABPSmicrostates}, we construct a family of multicenter microstate geometries and discuss their physics and their multipole structure with respect to the almost-BPS black holes they correspond to. In particular, we compare and contrast properties of the almost-BPS microstate multipoles with previous work and conjectures on (supersymmetric) microstate geometry multipoles. Finally, we conclude in section \ref{sec:conclusion} with a brief overview of possible future directions.

\subsection{Summary of our results}

Using the almost-BPS ansatz in $\cN=2$ four-dimensional supergravity, we construct four-charge non-supersymmetric rotating black holes and smooth horizonless microstate geometries thereof. We describe the new (astro)physics brought about by the horizon-scale structure compared to the classical black holes of general relativity. The expected modifications can be separated in two categories:
\begin{itemize}
\item Deviations between the physics of almost-BPS black holes and the physics of Kerr black holes.
\item Deviations that arise from the presence of a smooth horizonless microstructure replacing the horizon of almost-BPS black holes.  
\end{itemize}

\subsubsection*{The physics of almost-BPS black holes}

Our discussion starts with non-supersymmetric four-charge extremal black holes constructed from the almost-BPS ansatz.  We highlight the role of a dimensionless parameter, $h$, that allows to dial the ratio between the four-dimensional mass and charges. Therefore, our string-theory black holes can have the same mass, charges and spin as an almost-neutral Kerr-Newman black hole. 

We first discuss the geometrical differences between the near-horizon regions of these two solutions. Being extremal, 
rotating almost-BPS black holes have no ergosphere and the area of their horizon and their cosmic censorship bounds scale differently with the charges compared to Kerr black holes. Second, the multipoles of almost-BPS black holes have a similar dependence  on the mass/spin ratio compared to Kerr black holes. However, they also depend non-trivially on the mass/charge ratio (determined by the parameter $h$) unlike Kerr(-Newman) solutions. Moreover, all multipoles of the rotating almost-BPS black hole are non-zero, irrespective of their parity. 

We highlight the important similarities and differences that exist between the multipole moments of generic almost-BPS black holes in string theory and those of Kerr black holes.

For $h=1$, the almost-BPS black hole is the \emph{purest spinning black hole}: all its multipoles except the mass and angular momentum vanish. This is unique among rotating gravitational solutions (for example Kerr has infinite towers of non-zero multipole moments).

\subsubsection*{The physics of almost-BPS microstate geometries}

In the second part of the paper, we discuss the physics of almost-BPS microstate geometries and compare it to the physics of their corresponding almost-BPS black holes.
\begin{itemize}
\item The multipole moments of microstate geometries are equal to the multipoles of the black hole they correspond to with deviations proportional to the ``size'' of the microstructure above the would-be black-hole horizon. More concretely,
\begin{equation}
\text{Mult}\left(\text{Microstate}\right) \= \text{Mult}\left(\text{BH}\right) \,\left(1 \+ \cO\left(\delta_\text{micro} \right) \right)\,,
\end{equation}
where $\delta_\text{micro} \ll 1$ is the scale of the solution which characterizes how ``close'' in moduli space the microstate geometry is to the black hole \cite{deBoer:2008zn,Li:2021gbg}. When $\delta_\text{micro} \rightarrow 0$, the microstate geometry becomes identical to the black hole. 
\item When $\delta_\text{micro}$ is finite, the microstate-dependent contribution proportional to $\cO\left(\delta_\text{micro}\right)$ can modify the classical-black-hole result. In particular, in \cite{Bianchi:2020miz,Bianchi:2020bxa} it was conjectured that all microstate geometries have bigger multipoles than the black hole with the same charges.\footnote{This conjecture came from comparing supersymmetric non-scaling microstate geometries to Kerr black holes. As mentioned at the beginning of the summary section, a careful analysis should be done in two steps: (i) comparing scaling microstate geometries to their corresponding black hole in string theory, and then (ii) comparing this black hole to the astrophysical Kerr black hole with the same conserved charges (or at least the same mass and angular momentum).
}
We have constructed a number of almost-BPS microstate geometries that explicitly violates this conjecture. The multipoles of these geometries deviate from their expected average black hole values by $\cO\left(\delta_\text{micro}\right)$ terms that behave as  small  microstate-dependent ``noise-type'' contributions and that can be either positive or negative.
\end{itemize}

\section{The class of solutions and multipole moments}\label{sec:Generalities}

In this section, we review the construction of non-supersymmetric almost-BPS four-dimensional solutions and give a few generalities about the computation of gravitational multipole moments.

\subsection{Almost-BPS solutions}

We work with non-supersymmetric solutions of string theory that are asymptotic to four-dimensional Minkowski space times a six-dimensional CY manifold. More specifically, our solutions fit within the so-called \emph{almost-BPS} ansatz, in which one can construct multi-center non-supersymmetric black holes as well as horizonless microstate geometries \cite{Goldstein:2008fq,Bena:2009en,Bena:2009ev,DallAgata:2010srl,Vasilakis:2011ki}. Despite breaking supersymmetry, the equations governing the solutions in this ansatz can be solved using a linear algorithm. This linear structure comes because of an underlying nilpotent algebra \cite{Bossard:2012ge}, and is present in several other ansatze governing non-supersymmetric solutions \cite{Bossard:2011kz, Bossard:2012xsa}.

In general the almost-BPS ansatz can be used obtain solutions to any $U(1)^n$ five-dimensional supergravity, but in this paper we will focus on the STU solutions that arise from a compactification of M-theory on $T^6\times S^1$ \cite{Cremmer:1984hj,Duff:1995sm,Chow:2014cca}. The solutions have a Taub-NUT base space, and have an additional isometry, which allows us to compactify them to four dimensions. 
We describe in detail the four-dimensional STU Lagrangian in appendix \ref{sec:Gen}, starting from the string theory realization of these solutions. This Lagrangian contains, besides the \emph{four-dimensional metric}, \emph{four vector gauge fields} (hence generic solutions have four electric and four magnetic charges) and \emph{three complex scalar fields} that are all non-trivially coupled. 

The metric of an almost-BPS solution is described by eight scalar functions $(V,K^1,K^2,K^3,Z_1,Z_2,Z_3,\mu)$, together with an angular momentum one-form $\varpi$:
\be
\begin{split}
d s_{4}^{2} &=- {\mathcal{I}_4}^{-\frac{1}{2}} \,\left( dt+\varpi \right)^2 \+ {\mathcal{I}_4}^{\frac{1}{2}} \,ds_3^2 \,, \qquad \mathcal{I}_4 \equi Z_{1} Z_{2} Z_{3} V -\mu^{2} V^2\,,
\label{eq:metricSTU}
\end{split}
\ee
where $ds_3^2$ is the metric of a flat three-dimensional base that we parameterize by the spherical coordinates
\begin{equation}
ds_3^2 \= d\rho^2 \+ \rho^2 \left( d\theta^2 + \sin^2 \theta \,d\phi^2\right)\,.
\end{equation}
The gauge fields and scalars also have a specific form (which depends on all of the functions $(V,K^1,K^2,K^3,Z_1,Z_2,Z_3,\mu)$), which we give in appendix \ref{sec:4dframe}.

These almost-BPS solutions have \emph{ten conserved quantities}: four electric charges $q_\Lambda$ ($\Lambda=0,\cdots,3$) and four magnetic charges $p^\Lambda$ in addition to a four-dimensional mass $M$ and an angular momentum $J$. We work in units where Newton's constant in four dimensions is $G_4 \= 1$. Thus, asymptotically we always have:
\begin{equation}
\mathcal{I}_4\, \sim\, 1 \+ \frac{4M}{\rho}\,,\qquad \mathcal{I}_4^{-\frac{1}{2}} \,\varpi \bigl|_{d\phi} \,\sim\, \frac{2J}{\rho}\,\sin^2 \theta\,,
\label{eq:Mass&AMfromI4omega}
\end{equation}

The almost-BPS ansatz contains a non-BPS extremal four-charge black hole (the ``almost-BPS black hole''), as well as horizonless solutions that are smooth in string theory\footnote{In four dimensions, the centers of these solutions correspond to D6 branes and D4 branes with Abelian worldvolume flux, hence are singular. However, the metric near each of these centers becomes completely smooth and non-singular when uplifted to a six-dimensional duality frame.} and allow \emph{scaling limits} where they can approach the black hole geometry arbitrarily well \cite{Bena:2009ev}. Thus, we can think about these solutions as ``almost-BPS microstate geometries''. 
Crucially, the equations of motion for these microstate geometries have an ``almost-linear'' structure (see \eqref{eq:EOM}), which allows one to generate multi-center configurations.  
In appendix \ref{sec:ABPSmulticenter}, we review the method to solve these equations for solutions where all the centers are collinear \cite{Bena:2009en}.

\subsection{Gravitational multipole moments}\label{sec:intromultipoles}

One important physical property that we will study in detail for almost-BPS solutions is the structure of their gravitational multipoles. For asymptotically-flat four-dimensional solutions, these can most easily be read off by using the  ACMC-coordinate\footnote{Asymptotically-Cartesian and Mass-Centered} formalism developed by Thorne \cite{Thorne:1980ru}, which we briefly review here. All of the almost-BPS solutions constructed thus far are stationary and axisymmetric, so we will only discuss ACMC spacetimes satisfying these symmetries.

We will follow \cite{Bena:2020uup} by first writing the metric in AC coordinates: these are asymptotically Cartesian coordinates that are not necessarily mass-centered. This means that the mass dipole moment, $\tilde M_1$, does not necessarily vanish (as it must in ACMC coordinates). One can then obtain ACMC coordinates from any AC coordinate system by a simple shift of origin.

In such AC coordinates\footnote{Note that here we are only discussing ACMC-$\infty$ coordinates, from which all of the multipoles $M_\ell,S_\ell$ can be read off. More generically, one can also have ACMC-$N$ (or AC-$N$) coordinate systems from which we can only read off the multipoles to order $N+1$. Fortunately, for almost-BPS solutions the coordinates we find are AC-$\infty$.}, the asymptotic expansion of the metric is:
\begin{align}
\label{eq:gttmultipoles}  g_{tt}  &= -1 + \frac{2M}{\rho}+ \sum_{\ell\geq 1}^{\infty}  \frac{2}{\rho^{\ell+1}} \left( \tilde M_\ell P_\ell + \sum_{\ell'<\ell} c^{(tt)}_{\ell \ell'} P_{\ell'} \right), \\
\label{eq:gtphimultipoles}  g_{t\phi} &= -2\rho \sin^2\theta\left[ \sum_{\ell\geq 1}^{\infty} \frac{1}{\rho^{\ell+1}} \left( \frac{\tilde S_\ell}{\ell} P'_\ell +\sum_{\ell'<\ell} c_{\ell \ell'}^{(t\phi)}  P'_{\ell'}\right) \right],\\
 \label{eq:gspacespacemultipoles}  g_{\rho\rho} &= 1 + \sum_{\ell\geq 0}^\infty\frac{1}{\rho^{\ell+1}} \sum_{\ell'\leq \ell} c_{\ell \ell'}^{(\rho\rho)} P_{\ell'}, & 
   g_{\theta\theta} &= \rho^2\left[ 1 + \sum_{\ell\geq 0}^\infty\frac{1}{\rho^{\ell+1}} \sum_{\ell'\leq \ell} c_{\ell \ell'}^{(\theta\theta)} P_{\ell'}\right],\\ 
g_{\phi\phi} &= \rho^2\sin^2\theta\left[ 1 + \sum_{\ell\geq 0}^\infty\frac{1}{\rho^{\ell+1}} \sum_{\ell'\leq \ell} c_{\ell \ell'}^{(\phi\phi)} P_{\ell'} \right], &
\nonumber g_{\rho\theta} &= (-\rho\sin\theta) \left[ \sum_{\ell\geq 0}^\infty\frac{1}{\rho^{\ell+1}} \sum_{\ell'\leq \ell} c_{\ell \ell'}^{(\rho\theta)} P'_{\ell'}\right],
\end{align}
The argument of the Legendre polynomials $P_\ell$ (and their derivatives) appearing above is always  $\cos\theta$. The terms that contain $c^{(ij)}_{\ell \ell'}$ correspond to non-physical ``harmonics'', and depend on the particular AC(MC)-$N$ coordinates used. Even though these coefficients $c^{(ij)}_{\ell \ell'}$ are unphysical, it is a condition of ACMC coordinates that only $c^{(ij)}_{\ell \ell'}$ appear with $\ell'<\ell$.\footnote{For example, the $g_{\rho\rho}$ component of the Kerr metric in Boyer-Lindquist coordinates does not satisfy this condition, since there is a $g_{\rho\rho}$ component already at order $\rho^{-2}$ \cite{Thorne:1980ru,Bena:2020uup}.}

As mentioned above, AC coordinates are also ACMC if and only if the mass dipole vanishes, $\tilde M_1=0$. The gravitational multipoles $M_\ell,S_\ell$ are then simply the $\tilde M_\ell,\tilde S_\ell$ quantities appearing above. However, if our AC coordinate system has $\tilde M_1\neq 0$, it is easy to obtain ACMC coordinates by a simple shift of the origin along the $z$-axis by $z_0=-\tilde M_1/\tilde M_0$. We can then express the true multipoles $M_\ell,S_\ell$ in terms of the $\tilde M_\ell,\tilde S_\ell$ for any AC coordinate system \cite{Bena:2020uup}:
\begin{align}
 \label{eq:truemultipoles} M_\ell &= \sum_{k=0}^\ell \binom{ \ell}{k} \tilde M_k \left(-\frac{\tilde M_1}{\tilde M_0}\right)^{\ell-k}, & S_\ell & = \sum_{k=0}^\ell \binom{ \ell}{k} \tilde S_k \left(-\frac{\tilde M_1}{\tilde M_0}\right)^{\ell-k}
\end{align}
The coordinate-independent multipoles then consist of the mass multipoles $M_\ell$ (of which the mass is $M_0=M$) and the current (or angular momentum) multipoles $S_\ell$ (of which the angular momentum is $S_1=J$).

\section{Almost-BPS extremal black hole}\label{sec:ABPSBlackHoles}

In this section, we review the stationary and axisymmetric almost-BPS black hole constructed in \cite{Bena:2009ev}. We shall add to the solutions of \cite{Bena:2009ev} non-trivial asymptotic values for the eight scalar functions $(V,K^I,Z_I,\mu)$. These asymptotic values, parameterized by $h$, will add interesting physical decorations to the black hole solution as they can be used to dial the ratio between the mass and the charges in four dimensions.  We will further compute the multipole moments of this black hole, and compare them to those of Kerr-Newman black holes.

\subsection{The solution}
\label{sec:BHsol}

We consider a specific family of almost-BPS black holes that has the form \eqref{eq:metricSTU}, with:
\be 
V \= h +\frac{Q_0}{\rho}\,,\qquad Z_I \= \frac{1}{h}+\frac{Q_I}{\rho}\,,\qquad K^I \= 0 \,, \qquad \mu \,V \=  m_\infty  \+ \alpha \frac{\cos\theta}{\rho^2}\,, \qquad \varpi \= -\alpha {\sin^2\theta\over \rho} d\phi\,.
\label{eq:BHHarmonicFunc}
\ee
Note that in order to have a physical solution, one needs $ {\mathcal{I}_4} > 0$ in \eqref{eq:metricSTU}. A necessary condition is that $h$ and $Q_\Lambda$ have the same sign, which we will assume to be positive.

\noindent These solutions are asymptotically flat when $\mathcal{I}_4 \rightarrow 1$ in \eqref{eq:metricSTU}. This requires
\be 
h^{-2}-m_\infty^2\,=\, 1\,\qquad \Longleftrightarrow \qquad m_\infty \= \pm \frac{\sqrt{1-h^2}}{h}\,.
\label{eq:minffix}
\ee
We can define the warp factor
\be 
\begin{split}
\Delta &\equi \rho^2 \, \sqrt{V Z_1 Z_2 Z_3 - \mu^2 V^2} \\
&\=  \sqrt{\left(Q_0+h\,\rho\right) \left(Q_1+\frac{\rho}{h}\right) \left(Q_2+\frac{\rho}{h}\right) \left(Q_3+\frac{\rho}{h}\right) \-(m_\infty \,\rho^2+ \alpha \cos\theta)^2}\,,
\label{eq:DeltaDef}
\end{split}
\ee
and then express the four-dimensional metric as:
\be 
\label{eq:almostBPSBHmetric} ds_4^2 \= - \frac{\rho^2}{\Delta} \,\left( dt-\alpha \frac{\sin^2\theta}{\rho} \,d\phi \right)^2 \+ \Delta \,\biggl[\,\frac{d\rho^2}{\rho^2} + d\theta^2 + \sin^2 \theta \, d\phi^2 \,\biggr]  \,,
 \ee
The spherical coordinates we use are isotropic coordinates for the black hole, where the horizon is at $\rho=0$; the timelike Killing vector $\partial_t$ vanishes at this locus. As mentioned above, this solution is supported by three non-trivial complex scalars and four gauge fields, which are given in detail in  appendix \ref{sec:ABPSBHdetails}.
 
 \subsection{Properties}
 The mass and angular momentum of this solution can be read off from \eqref{eq:Mass&AMfromI4omega} and are:
\be 
 M \= \frac{Q_0 + h^2 (Q_1+Q_2+Q_3)}{4\,h^3}\,,\qquad  J \= -\frac{\alpha}{2}\,.
 \label{eq:BHMass&AM}
\ee
Moreover, the electromagnetic charges $p^\Lambda,q_\Lambda$ sourcing the solutions are given by (see appendix \ref{sec:ABPSBHdetails}):
\be 
(p^0 ,p^1,p^2,p^3; q_0, q_1 ,q_2 ,q_3) \= (Q_0, 0 ,0 ,0; 0, Q_1 ,Q_2, Q_3 )\,.
\label{eq:BHcharges}
\ee
The black hole thus has one magnetic charge and three electric charges. To consider this black hole as an astrophysically relevant one, these charges should not be thought of as standard model charges; rather, they should be viewed as ``hidden'' or dark charges, and their corresponding gauge fields considered as dark photons. Interestingly, the bounds on such dark charges (especially if they only interact gravitationally with standard model fields) from gravitational wave observations \cite{Cardoso:2016olt} or black hole imaging \cite{KNGRMHDpaper, bozzola2020general} are still very weak --- in particular, black holes with large (even near-extremal) dark charges have not necessarily been ruled out yet.

The event horizon is located at $\rho =0$, and has the topology of an $S^2$.
The horizon area is
\be 
A_H \,=\, 4 \pi\, \sqrt{Q_0 Q_1 Q_2 Q_3 \- \alpha^2}\,.
\label{eq:AreaHor}
\ee
which is the same as the area of the horizon of a supersymmetric extremal D6-D2-D2-D2-D0 black hole in four dimensions. This comes from the fact that the near-horizon geometry of the almost-BPS black hole is identical to the near-horizon geometry of its BPS cousin when uplifted to five dimensions.

The metric (\ref{eq:almostBPSBHmetric}) is already given in AC coordinates (as introduced in section \ref{sec:intromultipoles}). We can then easily read off the coefficients $\tilde M_\ell$ and $\tilde S_\ell$:
\begin{align}
 \label{eq:BHACmassmult} \widetilde{M}_0 &= M \,,& \widetilde{M}_1 &= - \frac{1}{2} \,m_\infty\,\alpha\,,&   \widetilde{M}_\ell \bigl|_{\ell\geq 2} &= 0\,,\\
 \label{eq:BHACcurrentmult} \widetilde{S}_0 &= 0 \,,&\widetilde{S}_1 &= J\,,&  \widetilde{S}_\ell \bigl|_{\ell\geq 2} &= 0\,.
\end{align}
Upon using  \eqref{eq:minffix} to express everything in terms of $h$, the multipoles  (\ref{eq:truemultipoles}) become:
\begin{equation}
\label{eq:almostBPSmultipoles} M_\ell \= \left( \mp 1 \right)^\ell \left( 1- \ell\right) \,M\, \left( \frac{1-h^2}{h^2}\right)^{\frac{\ell}{2}} \,\left(\frac{J}{M} \right)^\ell\,,\qquad S_\ell \=\left( \mp 1 \right)^{\ell-1} \ell \,J\, \left(\frac{1-h^2}{h^2}\right)^{\frac{\ell-1}{2}} \,\left(\frac{J}{M} \right)^{\ell-1}\,,
\end{equation}
where the expressions are valid for every $\ell\geq 0$ and where $\mp 1$ corresponds to the choice of branches for $m_\infty$ in \eqref{eq:minffix}.\footnote{More precisely, $\left( \mp 1 \right)^\ell$  is $(-1)^\ell$ if $m_\infty = \frac{\sqrt{1-h^2}}{h}$ and $(+1)^\ell$ if $m_\infty = -\frac{\sqrt{1-h^2}}{h}$.}
Note that for $h=1$, when the mass is determined by the sum of the charges \eqref{eq:BHMass&AM}, all multipoles (except $M_0$ and $S_1$) vanish, despite the presence of a finite angular momentum. This makes this solution unique among all spinning gravitational solutions and, as we explained in the Introduction, we can think about it as  the \emph{purest spinning black hole}.

\subsection{Comparing with Kerr black holes}

One can see that almost-BPS black holes have a common point with the electrically and magnetically charged Kerr-Newman solution when $Q_0=Q_1=Q_2=Q_3$. Taking in addition $\alpha=m_\infty=0$ (implying $h=1$) gives precisely the extremal Reissner-Nordstrom metric in isotropic coordinates. However, more generic almost-BPS black holes have significant differences that might affect their gravitational footprints compared to black holes in general relativity. In this section we compare the almost-BPS black holes to Kerr and Kerr-Newman black holes in order to further assess their phenomenological viability.

\subsubsection{Cosmic censorship and ergosphere}
\label{sec:CompwithKerr}

A Kerr-Newman black hole of mass $M$, angular momentum $J$ and carrying any number of charges $\cQ_I$ has an allowed regime of parameters dictated by the cosmic censorship bound:
\begin{equation}
M^2 - \sum_I \cQ_I - \frac{J^2}{M^2} \geq 0\,.
\end{equation}
This is very different from the cosmic censorship bound of the almost-BPS black hole, which is given by demanding that \eqref{eq:AreaHor} remains real and can be expressed in terms of the angular momentum $J$ and conserved charges $p^0, q_i$ in \eqref{eq:BHcharges} as:
\begin{equation}
p^0 q_1 q_2 q_3 - 4J^2 >0\,.
\end{equation}
Note that the mass $M$ does not appear explicitly in this formula.

Moreover, a Kerr-Newman black hole generically has an ergosphere, where the asymptotically timelike Killing vector $\partial_t$ becomes spacelike. The almost-BPS black hole does \emph{not} have an ergosphere, despite having a non-zero angular momentum. It would be interesting to understand how this could give rise to potential dynamical differences between Kerr(-Newman) and almost-BPS black holes, for example for photon orbits or gravitational-wave emission.

\subsubsection{Multipole moments}\label{sec:BHmult}
We calculated the multipoles of the almost-BPS black hole in (\ref{eq:almostBPSmultipoles}). We can compare these to the multipoles of a Kerr(-Newman) black hole, of which the non-zero multipoles can be written as:
\begin{equation}
\text{Kerr:} \quad M^{\text{Kerr}}_{2 \ell}=(-1)^\ell\,M\,\left(\frac{J}{M}\right)^{2\ell}, \quad S^{\text{Kerr}}_{2 \ell+1}=(-1)^\ell\,J\,\left(\frac{J}{M}\right)^{2\ell}\,, \quad M^{\text{Kerr}}_{2\ell+1} = S^{\text{Kerr}}_{2\ell} = 0\, .
\end{equation}
Note that the gravitational multipoles of Kerr and (charged) Kerr-Newman are the same, and thus (perhaps surprisingly) they are independent of the black hole charges \cite{Sotiriou:2004ud}.
For an almost-BPS and Kerr-Newman black hole of equal mass $M$ and angular momentum $J$, we can compare the even-mass and odd-current multipole moments:
\begin{equation}
\label{eq:multipoleBHoverKN}\frac{M_{2\ell}}{M^{\text{Kerr}}_{2 \ell}} \= (-1)^{\ell+1} \,(2\ell-1) \,\left(\frac{1-h^2}{h^2}\right)^{\ell}\,,\qquad \frac{S_{2\ell+1}}{S^{\text{Kerr}}_{2 \ell+1}} \= (-1)^{\ell}\,(2\ell+1) \,\left(\frac{1-h^2}{h^2}\right)^{\ell}\,.
\end{equation}
We see that the multipoles differ significantly because of the overall sign difference and the presence of $h$. As we will discuss below, $h$ is related to the ratio of mass and charges of the black hole.
 For example, in the small $h$ limit, $0<h\ll 1$, the mass $M$ over charge $Q$ (where $Q$ stands for any of the $p^\Lambda,q_\Lambda$ charges) ratio scales as $M/Q\sim 1/h^3$ (see (\ref{eq:MQscalingh})), so that (\ref{eq:multipoleBHoverKN}) scales as:
\be 0<h\ll 1:\qquad \frac{M_{2\ell}}{M^{\text{Kerr}}_{2 \ell}} \sim (-1)^{\ell+1} \,(2\ell-1) \,\left(  \frac{M}{Q}\right)^{\frac{2\ell}{3}}.\ee
Hence, unlike  Kerr-Newman black holes, the almost-BPS multipoles have a non-trivial dependence on the charges of the solution, that comes through the dependence on $h$.

In the opposite regime, $h= 1$, we see from (\ref{eq:almostBPSmultipoles}) that all of the almost-BPS multipoles vanish with the exception of the mass $M_0=M$ and of angular momentum $S_1=J$. It is worth pointing out that this is an unique physical system --- to our knowledge, no other known (super)gravity solution can have a non-zero angular momentum $S_1\neq 0$ without \emph{any} other multipole turned on. Hence, we can think about the $h=1$ black hole as the \emph{purest} spinning black hole. 

We can tune $h$ in (\ref{eq:multipoleBHoverKN}) to set the mass quadrupole moments equal, $M_{2} = M^{\text{Kerr}}_{2}$; this sets $h^{-1} = \sqrt{2}$. Of course, then the higher-order multipoles will differ by larger and larger factors as we increase $\ell$. We note that while current and near-future observations (will) constrain the quadrupole moment rather well (constraining $M_2/M^3$ within $10^{-4}$), it is unlikely that for example eLISA will be able to constrain many higher multipole moments to a similar degree \cite{Babak:2017tow,Barack:2006pq, Gair:2012nm, Barausse:2020rsu}.\footnote{Although note that in most if not all such modeling, it is assumed that the odd parity multipoles $M_{2n+1},S_{2n}$ vanish, so it is unclear to what extent observations will be able to distinguish spacetimes where these are non-zero.}

Finally, we also note that the odd mass multipoles $M_{2\ell+1}$ and even current multipoles $S_{2\ell}$ of the almost-BPS black hole do not vanish, in contrast to the Kerr(-Newman) black hole. These multipoles are odd-parity: they correspond to terms in the metric that break the equatorial symmetry ($\theta\leftrightarrow \pi-\theta$) and thus can give rise to interesting new equatorially asymmetric phenomena \cite{Cunha:2018uzc,Raposo:2018xkf,Junior:2021atr,Chen:2020aix}. Heuristically, the presence of these odd-parity multipoles are a consequence of the curious fact that the \emph{center of mass of the black hole} is not the same location as the \emph{center} of the black hole. The coordinates used in (\ref{eq:almostBPSBHmetric}) are centered around the location of the black hole horizon at $\rho=0$. However, as discussed above, these coordinates are AC and not ACMC (since the mass dipole $\tilde M_1\neq 0$ in these coordinates). Rather, the ACMC coordinates where the dipole vanishes are related to these coordinates by a shift of the origin by a distance proportional to $\tilde M_1/\tilde M_0$ (see (\ref{eq:truemultipoles})).

\subsubsection{The physics of the $h$ parameter}

\label{h-physics}

From the discussions above, it is clear that the parameter $h$, which can be freely chosen in the almost-BPS solution (\ref{eq:BHHarmonicFunc}) between $0<h\leq 1$, has a great influence on the physical properties of the almost-BPS black hole. First of all, it sets the ratio between the mass \eqref{eq:BHMass&AM} and the conserved charges \eqref{eq:BHcharges}:
\begin{equation}
\label{eq:MQscalingh}\frac{M}{(p^\Lambda,q_\Lambda)} \= \underset{h\ll 1}{\cO}\left( \frac{1}{h^3} \right)\,,\qquad \text{and}\qquad \frac{M}{4} \underset{h=1}{\=} \sum_\Lambda (p_\Lambda +q^\Lambda)\,.
\end{equation}
Taking $h$ small introduces a large discrepancy between the four-dimensional mass and the charges, $M\gg (p^\Lambda,q_\Lambda)$. Consequently, our solutions can describe black holes with very small charges and large mass. However, even it these charges are small, our black holes are still extremal.

Similarly, one can compute the ratio of the mass squared over the area of the event-horizon \eqref{eq:AreaHor}, assuming that the angular momentum parameter $\alpha\sim\mathcal{O}(h^0)$:
\begin{equation}
\frac{M^2}{A_H} \=  \underset{h\sim 0}{\cO}\left( \frac{1}{h^6} \right)\qquad \text{and}\qquad \frac{M^2}{A_H} \=  \underset{h=1}{\cO}(1)\,.
\end{equation}
Therefore, $h$ also changes the relative size of the horizon with respect to the mass; as the mass becomes much larger than the charges, $M\gg (p^\Lambda,q_\Lambda)$, the horizon area also becomes relatively small, $A_H \ll M^2$ (note that the Schwarzschild black hole has $A^\text{S}_H = 16\pi M^2$.) 

From a four-dimensional perspective, $h$ corresponds to non-trivial profiles for the three scalar fields of the solutions \eqref{eq:scalarBHApp}. Because the STU Lagrangian \eqref{eq:LagrangianSTU} has non-trivial couplings between scalar and gauge fields, these profiles change the effective electromagnetic couplings,  increasing the effect of the charges on the deformation of the spacetime. More precisely, taking $h$ small increases the impact of a small charge on the geometry.

This parameter $h$ can have great implications in new black hole astrophysics. In this paper, we have mostly focussed on its effect on the multipole structure of the almost-BPS black holes and how it differs from that of Kerr black holes. It would be interesting to study further processes, such as gravitational wave emission, tidal Love numbers, or scattering, to understand further the implications of this parameter.

Because the almost-BPS and BPS black holes have a similar structure, one can wonder what the effect would be of such a scalar profile on a four-dimensional supersymmetric black hole. It appears that a similar parameter $h$ for the BPS solutions can be used to freely dial the ratio between the four-dimensional mass and charge ratio and construct an effective neutral solution. However, the magnetic charge ($Q_0$) needs to be turned off for the BPS solutions when $h\neq 1$, so the area of the horizon will vanish and the BPS black hole corresponds to a microscopic black hole (which, although microscopic, would have potentially a large mass and small charges).

\section{Smooth microstate geometries}\label{sec:ABPSmicrostates}

The almost-BPS black hole is only one member of the very large family of almost-BPS solutions \cite{Goldstein:2008fq,Bena:2009en,Bena:2009ev,DallAgata:2010srl,Vasilakis:2011ki}. This family of solutions is controlled by a specific ansatz (see \eqref{eq:metricSTU}, \eqref{eq:STUscalars}, and \eqref{eq:STUgaugefields}), and the equations of motion have a nested linear structure (see \eqref{eq:EOM}), similar to that of BPS solutions. This allows the construction of almost-BPS multicenter solutions \cite{Bena:2009en}, which include multicenter black holes and black rings, as well as solutions that are smooth, horizonless microstate geometries. In this section, we will review their construction heuristically (relegating most of the technical details to the appendices) and discuss some of their basic properties.

\subsection{Construction}
In this section, we construct horizonless almost-BPS microstate geometries. First, we review the principles behind this construction and then we give the specific details of the family of microstate geometries that we consider in this paper.

\subsubsection{Heuristics: blowing up topological cycles}
\label{sec:PrincipleSmooth}

The microstate geometries we are interested in are solutions that match the almost-BPS black hole up to the region close to the would-be horizon. In this region, the multicenter configuration has a non-trivial, horizonless structure. In four dimensions, those solutions are singular at all the almost-BPS centers. When embedded in higher dimensions, these singularities have a very specific form, and become smooth regions of spacetime in different duality frames. 

\begin{figure}[ht]\centering
 \includegraphics[width=\textwidth]{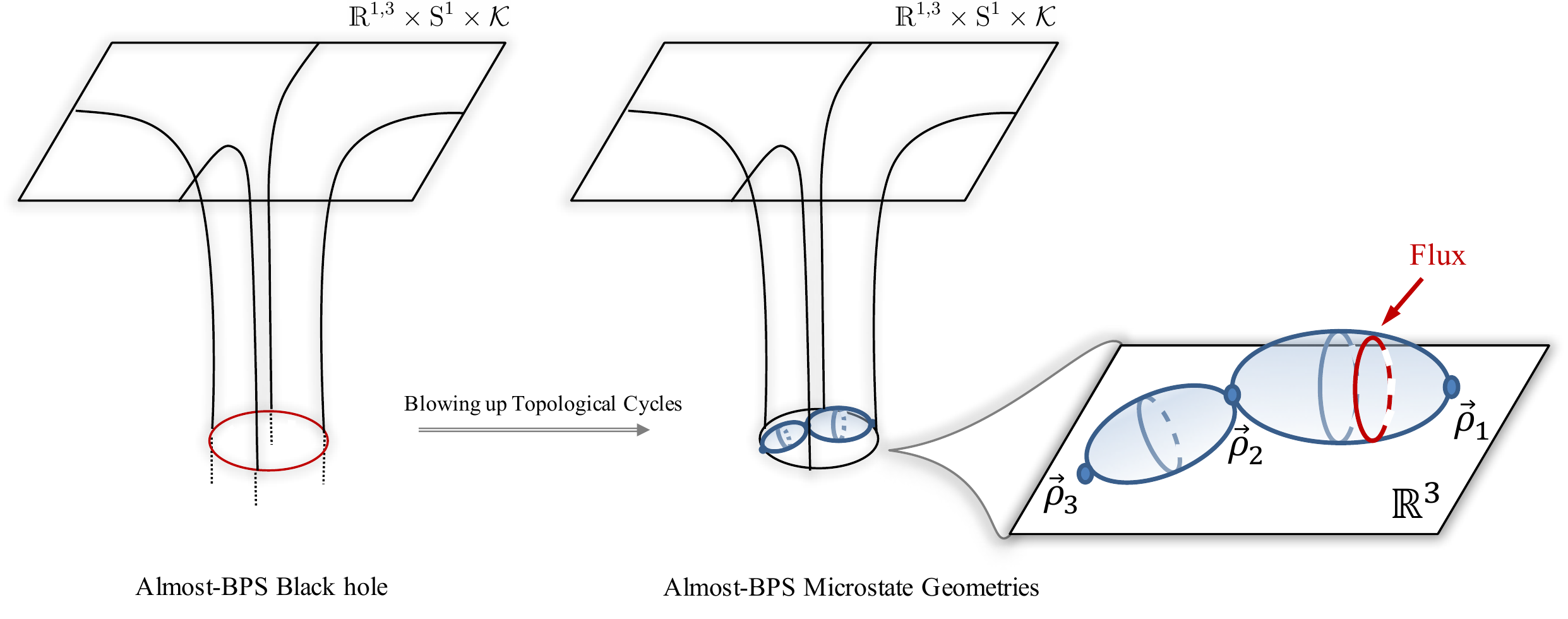}
\caption{Schematic description of the time-slices of an almost-BPS black hole and their corresponding multicenter smooth solutions. The horizon is resolved by blowing up topological cycles wrapped by fluxes in higher dimensions, where $\cK$ denotes these extra compact dimensions.}
 \label{fig:MGPhilo}
\end{figure}

The construction of such solutions can be rather technical. However, the main philosophy is simple to depict (see Fig.\ref{fig:MGPhilo}). By embedding the four-dimensional STU Lagrangian in higher dimensions, some of the four-dimensional scalars and gauge fields become geometric, corresponding to metric components along the extra dimensions. In particular, certain singularities of the four-dimensional solution have scalars and gauge fields that diverge in such a way that the uplift of the solution to higher dimensions is smooth, and the singularity corresponds to an ``end of spacetime''. The best example of this is the D6 brane compactified on a six-torus, which appears singular from a four-dimensional perspective, but is smooth when uplifted to M-theory.

As depicted in Fig.\ref{fig:MGPhilo}, having several end-of-spacetime loci gives a \emph{bubbling topology} induced by the behavior of the extra dimensions. These bubbles are kept from collapsing by being wrapped by electromagnetic fluxes, which generate the same asymptotic charges as the four-dimensional black hole, but without a horizon.

In a sense, the microstate geometry \emph{blows up} or \emph{resolves} the black-hole singularity, dissolving the horizon into smooth topological cycles wrapped by fluxes in higher dimensions. A crucial point with microstate geometries is that they allow a \emph{scaling limit}, where the centers can come arbitrarily close to each other \cite{Bates:2003vx,Bena:2006kb,Bena:2007qc}  from the point of view of the $\IR^3$ base of the solution, $\left| \vec{\rho}_i-\vec{\rho}_j\right| \sim \lambda \ll (M,Q,J)$, which makes the solution resemble the black hole more and more, but still allows it to end in a smooth horizonless cap \cite{Bena:2006kb}.

In the almost-BPS Ansatz, the main ingredients that are used as smooth end-to-spacetime loci are \emph{supertube centers} and \emph{Taub-NUT centers}.  A \emph{Taub-NUT center}, located at the position $\vec{\rho}_0$, sources only $V$ and $\mu V$ \eqref{eq:metricSTU} such as
\be V= \ldots + \frac{Q^{(0)}_0}{|\rho-\vec{\rho}_0|} + \ldots\,,\qquad \mu V= \ldots + \frac{m^{(0)}}{|\rho-\vec{\rho}_0|} + \ldots\,. \ee
This will end up sourcing a magnetic charge $p^0$ (through the equations of motion, see \eqref{eq:EOM}).
The $i^\text{th}$ center is a \emph{supertube center of species ``$I$''}, with $I=1,2,3$, located at a position $\vec{\rho}_i$. It has a source in $K^I$, two sources in $Z_J$ and $Z_K$ with $I \neq J \neq K$, and one source in $\mu V$ \cite{Bena:2008dw}:\footnote{These supergravity sources were shown in \cite{Bena:2008dw} to correspond to the backreaction of the supertubes constructed using the DBI action \cite{Mateos:2001qs,Emparan:2001ux}.}
 \be K^I \=   .. + \frac{k^{(i)}}{|\rho-\vec{\rho}_i|} + ..\,,\quad Z_J \=  .. + \frac{Q^{(i)}_J}{|\rho-\vec{\rho}_i|} + ..\,,\qquad Z_K \=  .. + \frac{Q^{(i)}_K}{|\rho-\vec{\rho}_i|} + ..\,, \quad \mu V \=   .. + \frac{m^{(i)}}{|\rho-\vec{\rho}_i|} + ..\,.\ee
 This will source a magnetic charge $p^I$ and two electric charges $q_J,q_K$. Therefore, by combining Taub-NUT centers and supertube centers of at least two species, we will be able to construct solutions that have the same charges as the almost-BPS black hole.

When embedded in five dimensions, the metric near a Taub-NUT center is that of smooth $\IR^{4,1}$. For each of the species of supertube centers, one can dualize to a six-dimensional supergravity where they are smooth  \cite{Bena:2008dw} providing that their charges satisfy:\footnote{At first sight, it may appear that we cannot have a unique duality frame where a solution with multiple species of supertubes is smooth at each supertube. However, one can perform a ``generalized spectral flow'' duality \cite{Bena:2008wt}, that transforms each type of supertube into a smooth center \cite{Bena:2009fi}. These transformations give a new solution that does not belong to the almost-BPS ansatz \cite{Bena:2009fi,DallAgata:2010srl}, but from a four-dimensional perspective they leave the metric invariant and simply reshuffle the scalars and the vectors \cite{DallAgata:2010srl}. Hence, for simplicity, we will continue working with almost-BPS supertubes and Taub-NUT centers.} 

\be m^{(1)} \=\frac{Q_J^{(1)}Q_K^{(1)}}{2k^{(1)}} .\label{eq:SupRegGen}\ee

\subsubsection{Two supertubes in Taub-NUT}\label{sec:concreteMGs}
We now construct a specific family of horizonless, multicenter geometries that have the same charges, mass and angular momentum as the almost-BPS black hole.  In addition, the family of solutions  we want to construct consists of \emph{scaling solutions}, which means we can make the distance between the centers arbitrarily small in the $\IR^3$ base, so that the multi-center solution matches the single-center black hole arbitrarily well.

As discussed above, the supertube and Taub-NUT centers of the solution look singular in four dimensions, but the origin of these singularities is understood: they are 16-supercharge fluxed D4 or fluxed D6 branes that make perfect sense in string theory, and moreover can be uplifted to smooth solutions in higher dimensions.

The technical details of the construction can be found in  appendix \ref{sec:ABPSmulticenter}. We will consider multicentered configurations with centers on the $z$-axis, so that the resulting solution is axisymmetric. We use the smallest number of ingredients that allow us to construct a horizonless solution with the same charges as the almost-BPS black hole: one Taub-NUT center and two supertube centers:

\begin{itemize}
\item The Taub-NUT center is at the origin of our spherical coordinates in the $\IR^3$ base of the solution. It has  charge $Q_0$ in $V$ and a momentum parameter $m^{(0)}$ in $\mu V$.
\item  The first supertube center is of species 2, and is on the z-axis at position $z=a_2$. It carries a magnetic charge parameter $k^{(2)}$ in $K^2$, two electric charges $Q^{(2)}_1$ and $Q^{(2)}_3$ in {the harmonic functions appearing in} $Z_1$ and $Z_3$, respectively, and a momentum parameter $m^{(2)}$ in the harmonic part of $\mu V$. 
\item  The second supertube center is of species 3, and is located at  $z=a_3$. It has a magnetic charge parameter $k^{(3)}$ in $K^3$, two electric charges $Q^{(3)}_1$ and $Q^{(3)}_2$ in {the harmonic functions appearing in}  $Z_1$ and $Z_2$, respectively, and a momentum parameter $m^{(3)}$ in the harmonic part of $\mu V$. 
\end{itemize}
The regularity of the supertubes requires, from \eqref{eq:SupRegGen}
\be 
m^{(2)} \= \frac{Q_1^{(2)}Q_3^{(2)}}{2 k^{(2)}}\,,\qquad m^{(3)} \= \frac{Q_1^{(3)}Q_2^{(3)}}{2 k^{(3)}}\,.
\label{eq:supreg}
\ee

For simplicity we consider $a_3>a_2>0$. We introduce the local spherical coordinates around the $I^{th}$ center, $(\rho_I,\theta_I,\phi)$, $$\rho_I \equi \sqrt{\rho^2 +a_I^2 - 2 \rho a_I \cos\theta},\qquad \cos \theta_I \equiv \frac{\rho \cos \theta-a_I}{\rho_I}.$$ 

The interested reader can find details on the resolutions of the equations of motion in  appendix \ref{sec:ABPSmulticenter}, and the gauge fields and scalars of the three-center solutions in appendix \ref{sec:3centerdetails}. The metric is still given by \eqref{eq:metricSTU}:
\begin{equation}
d s_{4}^{2} =- {\mathcal{I}_4}^{-\frac{1}{2}} \,\left( dt+\varpi \right)^2 \+ {\mathcal{I}_4}^{\frac{1}{2}} \,\left[d\rho^2 \+ \rho^2 \left( d\theta^2 + \sin^2 \theta \,d\phi^2\right) \right] \,, \qquad \mathcal{I}_4 \equi Z_{1} Z_{2} Z_{3} V -\mu^{2} V^2\,,
\end{equation}
where now the functions $(V,Z_I,\mu)$ and the rotation one-form $\varpi$ are given by: 
\begin{alignat}{2}
V &=&& h ~+~ \frac{Q_0}{\rho} \,, \qquad Z_1 = \, \frac{1}h ~+~ \frac{Q_1^{(2)}}{\rho_2}~+~ \frac{Q_1^{(3)}}{\rho_3} ~+~ \Big(h + \frac{Q_0 \, \rho}{a_2 \, a_3} \Big)   \frac{k^{(2)}\,k^{(3)} }{ \rho_2\, \rho_3} \,,\nonumber \\
Z_2 &=&& \frac{1}h ~+~ \frac{Q_2^{(3)}}{\rho_3} \, , \quad Z_3~=~  \frac{1}h ~+~ \frac{Q_3^{(2)}}{\rho_2} \, ,\nonumber\\
\mu \,V &~=~ && m_\infty + \frac{m^{(0)}}{\rho}+ \frac{m^{(2)}}{\rho_2}+ \frac{m^{(3)}}{\rho_3} \+ \left( \frac{k^{(2)}}{2h\rho_2}+\frac{k^{(3)}}{2h\rho_3}\right) \,V \+ \frac{h}{2 \,\rho_2 \rho_3}\left(k^{(2)} Q^{(3)}_2 +k^{(3)} Q^{(2)}_3 \right) \label{eq3:3supZmuomega}\\
& && \- \frac{Q_0\,  \cos \theta}{(a_3-a_2) \,\rho_2 \rho_3}\left(k^{(2)} Q^{(3)}_2 -k^{(3)} Q^{(2)}_3 \right)  \+ \frac{Q_0\, ( \rho^2+a_2 a_3)}{2(a_3-a_2) \,\rho \rho_2 \rho_3}\left(\frac{k^{(2)}}{a_2} Q^{(3)}_2 -\frac{k^{(3)}}{a_3} Q^{(2)}_3 \right) \,,  \nonumber\\
\varpi &~=~ && -\sum_{I=2,3} \frac{k^{(I)}}{2}\left[\cos\theta_I + Q_0 \frac{\rho-a_I \cos\theta}{h\,a_I \rho_I} \right] d\phi \+ \varpi_0 \, d\phi \+ \left( m^{(0)} \cos\theta +m^{(2)} \cos\theta_2 +m^{(3)} \cos\theta_3 \right)\,d\phi  \nonumber\\
& && \+ \sum_{J\neq I} \frac{Q_I^{(J)}k^{(I)}}{2(a_J-a_I)\rho_I \rho_J} \biggl[h(\rho^2+a_I a_J-(a_I+a_J)\rho\cos\theta) \nonumber \\
& && \hspace{4.3cm} - Q_0 \frac{\rho (a_J + a_I \cos2\theta) -(\rho^2 +a_I a_J)\cos \theta}{a_I} \biggr] d\phi \,.\nonumber 
\end{alignat}

There are regularity conditions (see \eqref{regularity1}) that the geometry must satisfy, as well as the condition to be asymptotic to flat four-dimensional spacetime; together, these give 5 algebraic conditions. Three directly fix $\varpi_0$, $m_\infty$ and $m^{(0)}$; the other are called \emph{bubble equations} and are non-linear relations that constrain the distances between the centers:
\be
\begin{split}
 m_\infty^2&~=~ \pm \,\frac{1-h^2}{h^2}\,, \\
\varpi_0 & ~=~   \frac{k^{(2)}}{2} \,\left(\frac{h \,Q^{(3)}_2}{a_3-a_2}-\frac{Q_0}{h a_2} \right)\- \frac{k^{(3)}}{2} \left(\frac{h \,Q^{(2)}_3}{a_3-a_2}+\frac{Q_0}{h a_3}\right)\,,\\
\frac{m^{(0)}}{Q^{(0)}} &~=~ -\frac{k^{(2)}}{2ha_2} \,\left(\frac{h \,Q^{(3)}_2}{a_3-a_2}+1 \right)\+ \frac{k^{(3)}}{2ha_3} \left(\frac{h \,Q^{(2)}_3}{a_3-a_2}-1\right)\,, \\
m^{(2)} & = \frac{Q_1^{(2)}Q_3^{(2)}}{2 k^{(2)}} ~=~ - \frac{1}{2(a_3-a_2)}\left(k^{(3)} Q^{(2)}_3\left(h+\frac{Q_0}{a_3}\right)-k^{(2)} Q^{(3)}_2 \left(h+\frac{Q_0}{a_2}\right)\right)\+ \frac{k^{(2)}}{2h} \left(h+\frac{Q_0}{a_2} \right)  \,,\\
m^{(3)} & = \frac{Q_1^{(3)}Q_2^{(3)}}{2 k^{(3)}}~=~  \frac{1}{2(a_3-a_2)}\left(k^{(3)} Q^{(2)}_3\left(h+\frac{Q_0}{a_3}\right)-k^{(2)} Q^{(3)}_2 \left(h+\frac{Q_0}{a_2}\right)\right)\+ \frac{k^{(3)}}{2h} \left(h+\frac{Q_0}{a_3} \right)  \,.
\end{split}
\label{eq:RegSupSol}
\ee
We can also calculate the integer magnetic charges of the supertube centers, by integrating the corresponding gauge field around the centers \cite{Bena:2009en,Vasilakis:2011ki}. These magnetic charges, $\kappa^{(I)}$, are: 
\be 
\kappa^{(I)} \equiv \left(h+\frac{Q_0}{a_I}\right) k^{(I)}\,.
\label{eq:effectivedipolecharges}
\ee

The solutions are guaranteed to be physical when the warp factors, $Z_I$, and the quartic invariant, $\mathcal{I}_4$, satisfy everywhere the inequalities \cite{Bena:2005va,Berglund:2005vb}:
\be 
\label{eq:condCTC}
Z_I \, V \geq 0\,,\qquad  \mathcal{I}_4 \equi  Z_1 Z_2 Z_3 \, V \- \mu^2 V^2 \,\geq\, |\varpi |^2\, .
\ee
By expanding this expression near each pole one can obtain inequalities constraining the supertube charges. The easiest way to satisfy these is to take all electric charges, $Q_I^{(J)}$, to be positive,  and have one magnetic charge to be negative.\footnote{By reshuffling the bubble equations above, one can see that at least one charge needs to be negative.}

\subsection{Properties}

The almost-BPS multicenter solutions constructed above have a rather complicated and unintuitive form. It is therefore important to point out their key physical properties:

First of all, note that the system is only constrained by the two bubble equations \eqref{eq:RegSupSol}, so that the phase space of solutions is very large. An important subclass of solutions, called \emph{scaling solutions}, are the ones in which the intercenter distances can be made arbitrarily small; we will mainly be interested in such scaling solutions.

When we approach this \emph{scaling limit} and bring the centers close to each other, we can call $\lambda$ the overall size of the cluster of centers in the $\IR^3$ base space, so that each center position satisfies $a_I \= \cO(\lambda)$. Far away from the centers, $\rho \gg \lambda$, the main functions \eqref{eq3:3supZmuomega} determining the metric match the functions that enter in the almost-BPS black hole solution with the same charges, up to $\cO(\lambda)$ corrections. In other words, the solutions are virtually indistinguishable from the black hole at these scales. In the IR, when $\rho = \cO(\lambda)$, the structure of the multicenter configuration starts to be visible and distinguishable from the black hole horizon.
We will quantify these statements further by comparing the conserved quantities and the multipole moments of the solutions to the black hole results.

\subsubsection{Conserved quantities}

We compute the ADM mass, the angular momentum and the four electric and magnetic charges of the solution. We can again obtain the mass and angular momentum via \eqref{eq:Mass&AMfromI4omega}:
\be 
\begin{split}
 M &\= \frac{1}{4 h^3}\left[Q_0 + h^2 \,\left(Q^{(2)}_1+Q^{(3)}_1+Q^{(3)}_2+Q^{(2)}_3+Q_0\, \frac{k^{(2)} k^{(3)}}{a_2 a_3} \mp 2\sqrt{1-h^2}\,\left(k^{(2)}+k^{(3)} \right)\right) \right]\,,\\
J &\= \frac{Q_0}{4 h} \, \left( \frac{2\, h\,k^{(3)}  \,Q^{(2)}_3}{a_3} + k^{(2)} + k^{(3)} \right) + \frac{h}{4} \left( k^{(3)} Q^{(2)}_3 -k^{(2)} Q^{(3)}_2 \right)  \,,
\end{split}
\label{eq:mass&AMsupertubes}
\ee
where the ``$\mp$'' depends on which choice of branch have been chosen for $m_{\infty}$ in \eqref{eq:RegSupSol}. Note that these expressions are obtained ``on shell'', after enforcing the bubble equations \eqref{eq:RegSupSol}.
The four magnetic and electric charges are given by (see \eqref{eq:4dchargesgeneric}):
\be 
\begin{split}
(p^0 ,p^1,p^2,p^3) \= &  \left( Q_0 \, ,\,0 \, ,\, -h\, k^{(2)} \, ,\, 
-h \,k^{(3)} \right)\\
 (q_0, q_1 ,q_2 ,q_3)\= & \left(  \frac{k^{(2)} +k^{(3)}}{h} \, ,\,Q^{(2)}_1 +Q^{(3)}_1 + Q_0 \frac{k^{(2)} k^{(3)}}{a_2 a_3}\, ,\, Q^{(3)}_2 \, ,\, Q^{(2)}_3\right)
 \end{split}
 \label{eq:elecmagchargesSup}
\ee
Note that there are more charges turned on compared to the almost-BPS black hole solution of \eqref{eq:BHcharges}. This is because of the presence of the magnetic dipole charges, $k^{(2)}$ and $k^{(3)}$, which are crucial elements that allow the spacetime to be smooth around the centers in higher dimensions. However, although they must be non-vanishing, we can freely take them small compared to the main charges: $k_i \ll Q_0,Q^{(j)}_k$. In this limit, the D0 charge can be ignored as well, and we have
\be 
\begin{split}
&(p^0 ,p^1,p^2,p^3; q_0, q_1 ,q_2 ,q_3) \,\sim\,(Q_0, 0 ,0 ,0; 0, Q^{(2)}_1 + Q^{(3)}_1 + Q_0 \frac{k^{(2)} k^{(3)}}{a_2 a_3} ,Q^{(3)}_2, Q^{(2)}_3 )\,, \\
&M \,\sim\, \frac{1}{4h^3}\left[Q_0 + h^2 \,\left(Q^{(2)}_1+Q^{(3)}_1+Q^{(3)}_2+Q^{(2)}_3+Q_0\frac{ k^{(2)} k^{(3)}}{a_2 a_3}\right) \right]\,, \\
& J \sim  \frac{h}{4} \left( k^{(3)} Q^{(2)}_3 -k^{(2)} Q^{(3)}_2 \right) + Q_0\frac{ k^{(3)}\,Q_3^{(2)}}{2 a_3} \,,
\end{split}
\ee
so that the microstate geometry has {the same}  conserved charges as the black hole.

\subsubsection{Multipole moments}\label{sec:MSmult}

In appendix \ref{sec:MultipoleMulticenter}, we derive the multipole moments of generic multicenter solutions in Taub-NUT. In this section, we apply these formulas to our specific three-center solutions. 

Note that the coordinates used in the almost-BPS ansatz \eqref{eq:metricSTU} are automatically AC coordinates as defined in section \ref{sec:intromultipoles}. Therefore, one can read off the coefficients $\tilde M_\ell, \tilde S_\ell$ from simply expanding the metric in powers of $1/\rho$, and obtain the true multipoles using \eqref{eq:truemultipoles}:
\be
M_\ell = \sum_{k=0}^\ell \binom{\ell}{k}  \tilde M_k \left(-\frac{\tilde M_1}{\tilde M_0}\right)^{\ell-k}, \qquad S_\ell = \sum_{k=1}^\ell \binom{\ell}{k}  \tilde S_k \left(-\frac{\tilde M_1}{\tilde M_0}\right)^{\ell-k}  \,.
\ee
For our specific three-center solution, the relevant AC-coordinate frame coefficients, $\tilde M_\ell$, are then given by \eqref{eq:almost-BPS_microstates_mass_multipoles}:
\be \label{eq:almost-BPS_microstates_mass_multipolesEx}
\begin{split}
4 \tilde{M}_\ell \=  
& \left(\frac{Q_0}{h^3} -2 m_\infty m^{(0)} \right) {a_0}^\ell
+  \sum_{\underset{J,K=1}{I=2}}^{3} \frac{|\varepsilon_{IJK}|}{2h} \left(Q^{(I)}_J +Q^{(I)}_K -h\,m_\infty\left( k^{(I)}+  \frac{Q^{(I)}_J Q^{(I)}_K}{k^{(I)}} \right)\right) {a_I}^\ell \\
&+\sum_{I,J=2}^{3} \frac{Q_0\,|\epsilon_{IJ}|   }{h} \left[ \frac{k^{(I)} k^{(J)}}{2a_I a_J} q^{(2)}_\ell(a_I,a_J) -h m_\infty \frac{k^{(J)} Q^{(I)}_J}{a_I-a_J} \(\frac{q^{(2)}_\ell(a_I,a_J)}{a_J} -  \frac{2\ell}{2\ell-1} q^{(2)}_{\ell-1}(a_I,a_J) \) \right]
\end{split}
\ee
where $\epsilon_{IJ}$ and $\epsilon_{IJK}$ are the Levi-Civita tensors of dimension two and three respectively and we have defined $a_0=0$ as the coordinate of the Taub-NUT center, with also ${a_0}^\ell \ \equiv \delta_{\ell 0}$. We define the polynomial 
\be
q^{(2)}_{\ell}(a_I,a_J)\equiv \binom{2\ell}{\ell}^{-1} \sum_{p+q=\ell} \binom{2p}{p} \binom{2q}{q} {a_I}^p {a_J}^q \,.
\ee
The coefficients $\tilde S_\ell$, are given by \eqref{eq:almost-BPS_microstates_current_multipoles}
\be \label{eq:almost-BPS_microstates_current_multipolesEx}
\begin{split}
4\tilde{S}_\ell \= &-2 m^{(0)}\,{a_0}^\ell+  \sum_{\underset{J,K=1}{I=2}}^{3} \frac{|\varepsilon_{IJK}|}{2} \left( k^{(I)} - \frac{Q^{(I)}_J Q^{(I)}_K}{k^{(I)}}\right)\,{a_I}^\ell 
\\
&- \sum_{I,J=2}^{3} Q_0\,|\epsilon_{IJ}|   \frac{k^{(J)} Q^{(I)}_J}{a_I-a_J} \(\frac{q^{(2)}_\ell(a_I,a_J)}{a_J} -  \frac{2\ell}{2\ell-1} q^{(2)}_{\ell-1}(a_I,a_J) \) . 
\end{split}
\ee

It is of particular interest to consider the scaling limit of (\ref{eq:almost-BPS_microstates_mass_multipolesEx})-(\ref{eq:almost-BPS_microstates_current_multipolesEx}), or more generally of (\ref{eq:almost-BPS_microstates_mass_multipoles})-(\ref{eq:almost-BPS_microstates_current_multipoles}), when $a_I \= \cO( \lambda)$ and $\lambda \rightarrow 0$. Using the regularity conditions (\ref{eq:RegSupSol}) (or (\ref{regularity1})), one can see that the AC coefficients behave as (for $h\neq 1$): 
\be \label{eq:Mtildescalinglimithneq1}
\begin{aligned}
 \tilde M_0 &= M^{\text{BH}}\left(1 +  \mu_0 \lambda\right) + \mathcal{O}(\lambda^2),\\
 \tilde M_1 &= \tilde M_1^{\text{BH}}\left(1 + \mu_1 \lambda\right) + \mathcal{O}(\lambda^2), & \tilde S_1 &= J^{\text{BH}}\left(1 + \sigma_1 \lambda\right) + \mathcal{O}(\lambda^2),\\
 \tilde M_\ell &= \tilde M_1^{\text{BH}} \, \frac{(\tilde M_1^{\text{BH}})^{\ell-1}  }{(M^{\text{BH}})^{\ell-1}} \,\mu_\ell \,\lambda^{\ell-1}  + \mathcal{O}(\lambda^\ell), & \tilde S_\ell &= J^{\text{BH}}\frac{(\tilde M_1^{\text{BH}})^{\ell-1}  }{(M^{\text{BH}})^{\ell-1}} \sigma_\ell\lambda^{\ell-1}  + \mathcal{O}(\lambda^\ell),
\end{aligned}
\ee
where $M^{\text{BH}}, \tilde M_1^{\text{BH}}, S_1^{\text{BH}}=J^{\text{BH}}$ are the non-vanishing AC coefficients for the almost-BPS black hole {with the same charges} in \eqref{eq:BHACmassmult} and \eqref{eq:BHACcurrentmult}, and $\mu_\ell,\sigma_\ell$ are microstate-dependent dimensionless numbers. The ACMC multipole moments then behave as:
\begin{align}
 \label{eq:scalingmultACMC-M0} M_0 &=M^{\text{BH}}\left(1 +  \mu_0 \lambda\right) + \mathcal{O}(\lambda^2),\\
  \label{eq:scalingmultACMC-S1}  S_1 &= J^{\text{BH}}\left(1 + \sigma_1 \lambda\right) + \mathcal{O}(\lambda^2),\\
 \label{eq:scalingmultACMC-Ml}  M_\ell &= M_\ell^{\text{BH}}\left(1 + \left[ (1- \ell)\mu_0+\ell \left(\mu_1 - \frac12 \mu_2\right)\right]\lambda \right) + \mathcal{O}(\lambda^2),\\
  \label{eq:scalingmultACMC-Sl}  S_\ell &= S_\ell^{\text{BH}}\left(1 + \left[ (1- \ell)\mu_0+\sigma_1+(\ell-1) \left(\mu_1 - \frac12 \sigma_2\right)\right]\lambda \right) + \mathcal{O}(\lambda^2),
 \end{align}
where the black hole ACMC multipoles $M^\text{BH}_\ell,\,S^\text{BH}_\ell$ were given in \eqref{eq:almostBPSmultipoles}. Note that (\ref{eq:Mtildescalinglimithneq1})-(\ref{eq:scalingmultACMC-Sl}) are only valid for $h\neq 1$; when instead $h=1$ and thus $m_\infty = 0$, it is easy to show that $\tilde M_\ell\sim\mathcal{O}(\lambda^\ell)$ and $\tilde S_\ell\sim\mathcal{O}(\lambda^{\ell-1})$ so that also $M_\ell\sim  \mathcal{O}(\lambda^\ell)$ (for $\ell\neq 1$) and $S_\ell\sim\mathcal{O}(\lambda^{\ell-1})$ (for $\ell\geq 1$) --- interestingly, this is similar but not exactly the same as the scaling with $\lambda$ that one has for multipoles of scaling supersymmetric microstate geometries \cite{Bena:2020see,Bena:2020uup}, which is $(M^\text{SUSY}_\ell,S^\text{SUSY}_\ell)\sim \mathcal{O}(\lambda^\ell)$.

We can summarize the behavior of the multipole moments of scaling almost-BPS microstate geometries in an (intuitive) conjecture:

\emph{All multipoles of scaling microstate geometries match the values of the black hole they correspond to, up to small deviations proportional to the scale for which the microstructure starts to be manifest and resolve the horizon into smooth topologies.}

Clearly, smooth horizonless solutions can mimic classical black hole characteristics with a very high accuracy. As $\lambda$ gets vanishingly small, this implies that the microstructure of the microstate geometry can become virtually indetectable, at least as far as the  multipole moments are concerned.

Moreover, nothing dictates a priori the value of $\mu_\ell$ and $\sigma_\ell$ as they depend on the internal degrees of freedom of the family of solutions. As we illustrate later, they can either be positive or negative. In particular, as we show explicitly in section \ref{sec:multipolesclaims}, this leads to counterexamples to the two claims/conjectures in \cite{Bianchi:2020miz}, where it was suggested that multipole moments of smooth horizonless geometries will be larger than those of the corresponding black hole.

\subsection{Explicit examples}
We give several examples of three-center almost-BPS microstate geometries, whose general form is given in \eqref{eq3:3supZmuomega}. We also give the parameters of the almost-BPS black hole with the same charges.

Finding explicit parameters that give rise to physical three-center solutions in the family constructed above is relatively easy. Our family initially contains 16 parameters. After imposing the supertube regularity \eqref{eq:supreg} and the regularity conditions \eqref{eq:RegSupSol}, we end up with 9 free parameters. The physicality condition \eqref{eq:condCTC} gives a bound on the parameters; it is sufficient to assume that all  charges are positive except one magnetic dipole charge,  that we will assume to be $\kappa^{(2)}$. Moreover, we aim to construct scaling solutions for which the centers can be tuned to come arbitrarily close to each other.

\subsubsection{A simple example}

The first solution we consider is given by the following charges:\footnote{Recall that the effective dipole charges $\kappa_I$ are given by \eqref{eq:effectivedipolecharges}.}
\be \label{eq:simpleexamplecharges}
-4\, \kappa^{(2)} \= 4\, \kappa^{(3)} \= 2 \,Q_0 \= \frac{4}{3}\,Q^{(2)}_1 \= Q^{(3)}_1 \= \frac{4}{3}\,Q^{(2)}_3 \= 20000\,,\qquad Q^{(3)}_2 \= 15001\,,\qquad h \= 0.01 \,,
\ee
The bubble equations \eqref{eq:RegSupSol} fix the distance between the centers
\be 
a_2 \simeq 0.21 \,,\qquad a_3 \simeq 0.28 \,,
\ee
which gives, from \eqref{eq:effectivedipolecharges}
\be 
k^{(2)} \simeq -0.11\,,\qquad k^{(3)} \simeq 0.14\,,
\ee
The mass and angular momentum of the solution \eqref{eq:mass&AMsupertubes} are
\be 
M \, \simeq \, 2.5 \times 10^{9}\,,\qquad J \= -3.7 \times 10^7 \,,
\label{eq:Ex1Mass&Ang}
\ee
and the eight charges \eqref{eq:elecmagchargesSup} are 
\be 
\begin{split}
(p^0 ,p^1,p^2,p^3) \,\simeq & \left( 10000 \, ,\, 0\, ,\,0.0011 \, ,\, -0.0014 \right)\\
 (q_0, q_1 ,q_2 ,q_3)\,\simeq\, &\left(3.5\, ,\,32500\, ,\, 15001\, ,\, 15000\right)
 \end{split}
\ee
The scaling point can be obtained by shifting $Q_3^{(2)} \rightarrow 15000$ which gives $a_2,a_3 \rightarrow 0$.

The solutions match very closely the non-BPS extremal black hole detailed in section \ref{sec:BHsol} with a mass and an angular momentum given by \eqref{eq:Ex1Mass&Ang}, and one magnetic charge $Q_0 = p^0$ and three electric charges $Q_I = q_I$. As detailed in section \ref{sec:CompwithKerr}, the ratio between the mass and charges is of order $h^{-3} = 10^6$. 

\subsubsection{A one-parameter family}

We can easily expand the above example to a one-parameter family of microstate geometries, where we allow $h$ to vary while we keep the other charges in (\ref{eq:simpleexamplecharges}) fixed.
For example, when $h\ll 1$, the intercenter distances are then approximately:
\be 
a_2 \simeq 21 \,h\,,\qquad a_3 \simeq 28\,h\,,
\ee
The mass, angular momentum and eight charges of the solutions are, at leading order in $h$:
\be 
\begin{split}
M &\, \simeq \, \frac{2500}{h^3}\,,\qquad  J \,\simeq\, -3.7 \times 10^7 \,,\\
(p^0 ,p^1,p^2,p^3) &\,\simeq\,  \left( 10000 \, ,\,0 \, ,\, 0\, ,\, 
0\right) \\
 (q_0, q_1 ,q_2 ,q_3)&\,\simeq\, \left(  3.5 \, ,\,32500\, ,\, 15001\, ,\, 15000\right)
\end{split}
\ee
Interestingly, $h$ does not change the topology of the IR geometry since it acts as a scaling factor. As expected and discussed above, by fine-tuning $h$ to be small, one can construct solutions that look almost neutral from a four-dimensional perspective.

\subsection{Aspects of microstate multipoles}\label{sec:multipolesclaims}

In this section, we discuss some aspects of the gravitational multipoles for the almost-BPS solutions. For the black holes, the multipole formulas were derived in section \ref{sec:BHmult}, whereas the microstate geometry multipoles can be found in section \ref{sec:MSmult}.
We will focus in this section on discussing multipoles (and certain ratios) as studied in \cite{Bianchi:2020bxa,Bianchi:2020miz} for families of supersymmetric black holes. 
In appendix \ref{sec:multipoleratios}, we also show that the multipole \emph{ratio} analysis of \cite{Bena:2020see,Bena:2020uup} for supersymmetric black holes and their multicentered microstate geometries can also be extended straightforwardly to the multipole ratios of almost-BPS black holes and their microstate geometries described here. It would be interesting to expand this analysis (in the spirit of \cite{Bena:2020uup} for supersymmetric black holes); we leave this for future work.

It will be convenient to define the following dimensionless, positive quantities:\footnote{These were denoted by $\mathfrak{M}_\ell$ and $\mathfrak{S}_\ell$ in \cite{Bianchi:2020miz}.}
\begin{align} 
\mathcal{M}_\ell &:= \left| \frac{M_\ell M_0^{\ell-1}}{S_1^{\ell}} \right|, & \mathcal{S}_\ell &:= \left|  \frac{S_\ell M_0^{\ell-1}}{S_1^{\ell}} \right|,
\end{align}
In  \cite{Bianchi:2020bxa,Bianchi:2020miz} these multipole ratios were computed for supersymmetric multi-center microstates, and compared to those of the non-supersymmetric Kerr(-Newman) black holes of the same mass and angular momentum. Note that for any Kerr(-Newman) black hole, $\mathcal{M}_{2n}=\mathcal{S}_{2n+1}=1$ and $\mathcal{M}_{2n+1}=\mathcal{S}_{2n} = 0$. For the families of microstates considered in \cite{Bianchi:2020bxa,Bianchi:2020miz}, two claims were made:
\begin{enumerate}[label=\textbullet\, \textbf{C.\roman*}:,ref=C.\roman*]
\item \label{claim1} Generically $\mathcal{M}_2>1$. In other words, the (absolute value of) the quadrupole moment $M_2$ of microstate geometries is generically \emph{larger} than that of Kerr with the same mass and angular momentum.
Similar statements are valid for higher-order multipole moments that are non-zero for Kerr-Newman (in particular, $\mathcal{S}_3>1$ is mentioned explicitly) \cite{Bianchi:2020bxa,Bianchi:2020miz}.
 \item \label{claim2} Both $\mathcal{M}_\ell$ and $\mathcal{S}_\ell$ (for any $\ell$) are \emph{always} monotonically increasing functions of the intercenter distance $\lambda$ for scaling solutions \cite{Bianchi:2020miz}. In particular, a corollary is that $\left(\partial_\lambda \mathcal{M}_\ell(\lambda)\right)_{\lambda=0}>0$ and similar for $\mathcal{S}_\ell$.
\end{enumerate}
While these claims are based on extrapolations of suggestive features of supersymmetric microstate geometries, here we will show that they are contradicted explicitly by the physics of the more realistic non-supersymmetric microstate geometries we built.

To illustrate this, we take a family of microstate geometries (as constructed in section \ref{sec:concreteMGs}) with charge parameters:\footnote{We wish to thank J. F. Morales for pointing out the unphysicality of the microstate geometry family we had in an earlier version of this paper.}
\be \label{eq:fammicrostates}
\begin{aligned}
m_\infty &= (-1)^n\frac{\sqrt{1-h^2}}{h}, & \kappa^{(2)} &= -7000, & \kappa^{(3)} &= 5000,\\
Q_0 &= 7000, & Q^{(2)}_1& = 6000x,& Q^{(2)}_3 &= 6000,\\
   a_0 &= 0, &  a_2 &=  \lambda, &  a_3 &= 2 \lambda. 
 \end{aligned}\ee
The parameters $Q^{(3)}_1, Q^{(3)}_2$ are then determined by the bubble equations \eqref{eq:RegSupSol}. Thus, this family of solutions depends on the choice of branch for $m_\infty$ through $n$ (where $n=1,2$) and has three free parameters, given by $h$, $x$, and the distance $\lambda$ between centers. The mass, angular momentum, and electromagnetic charges of this family is given by:
\be 
\begin{split}
M &\, \simeq \, \frac{250}{7h^3}\left( 49+h^2 \frac{115 +36x (13x-3)}{6x-5}\right) + \mathcal{O}(\lambda)\,, \quad J \, \simeq -15\times 10^6   + \mathcal{O}(\lambda) \\
(p^0 ,p^1,p^2,p^3) &\,\simeq\,  \left( 7000 \, ,\,0 \, ,\, 0\, ,\, 
0\right) + \mathcal{O}(\lambda) \\
 (q_0, q_1 ,q_2 ,q_3)&\,\simeq\, \left(  0 \, ,\,\frac{1000 \left(36 x^2+25\right)}{6 x-5}\, ,\, \frac{36000 x}{7}-\frac{30000}{7}\, ,\, 6000\right)+ \mathcal{O}(\lambda)
\end{split}
\ee 
When the centers merge at $\lambda=0$, the solution corresponds to the almost-BPS black hole with the conserved charges above. We will show that by varying the parameters $h$ and $x$ in this family of microstates, one can easily invalidate  both claim \ref{claim1} and \ref{claim2}.

\begin{figure}[ht]\centering
\begin{subfigure}{0.48\textwidth}\centering
 \includegraphics[width=\textwidth]{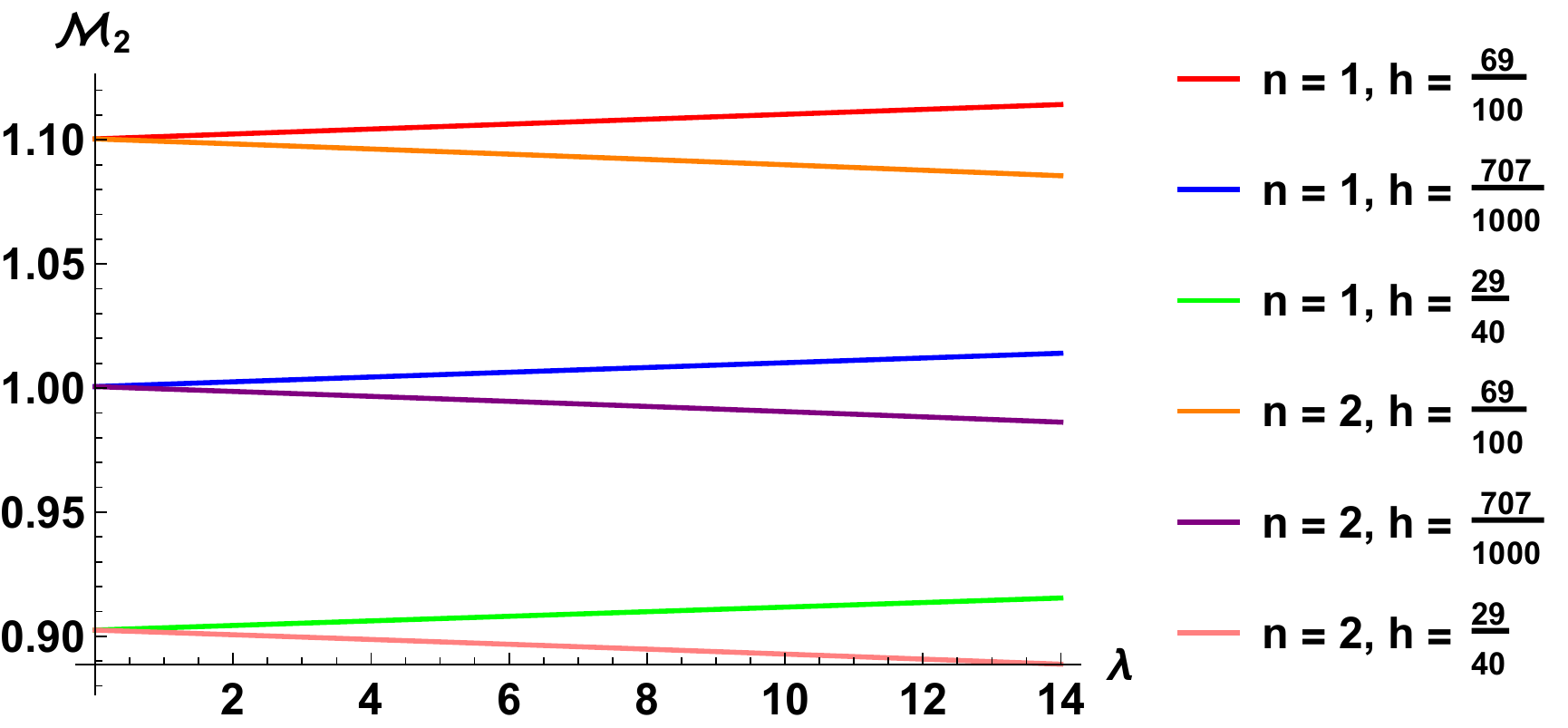}
 \caption{$\dfrac{\mathcal{M}_2}{\mathcal{M}_2^{\text{Kerr}}}$}
 \end{subfigure}
 \begin{subfigure}{0.48\textwidth}\centering
   \includegraphics[width=\textwidth]{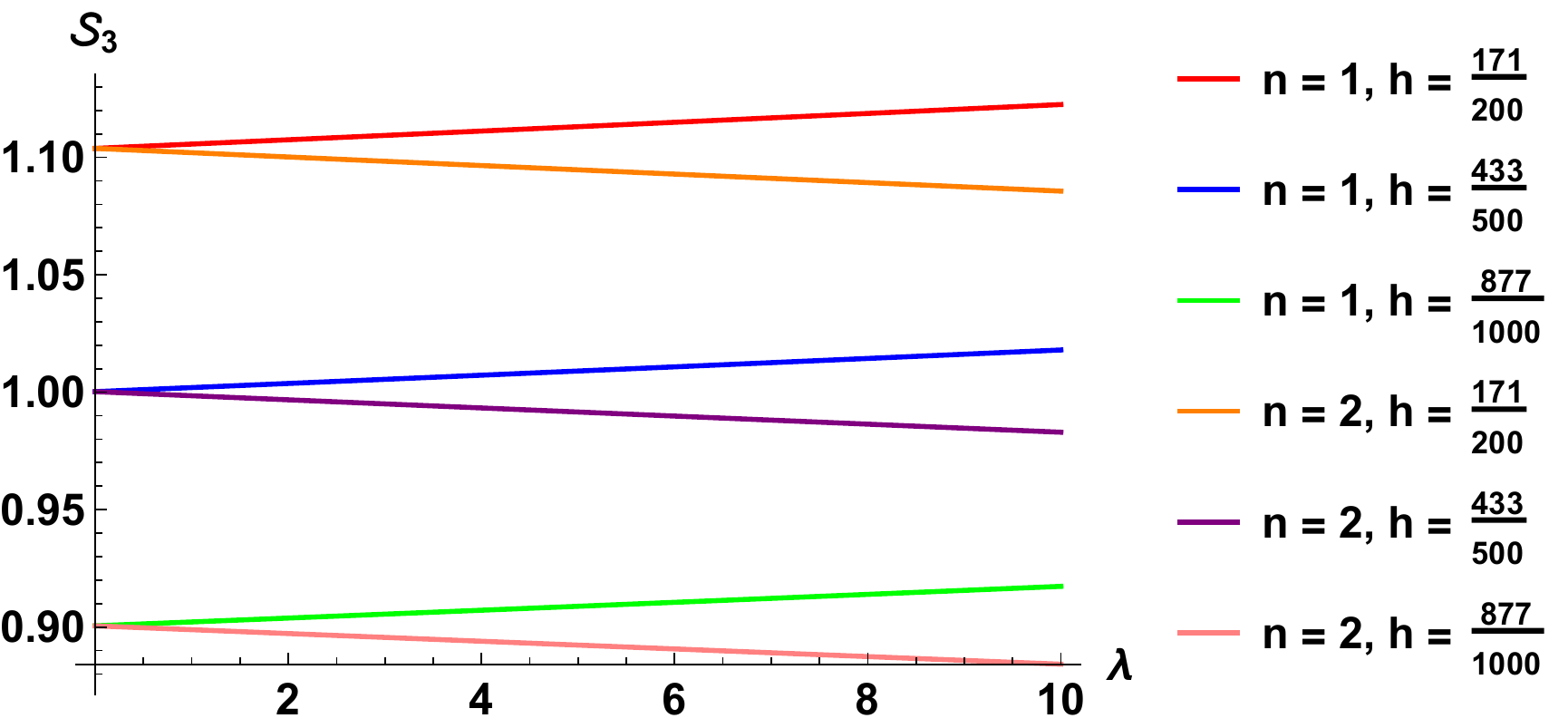}
    \caption{$\dfrac{\mathcal{S}_3}{\mathcal{S}_3^{\text{Kerr}}}$}
 \end{subfigure}
 \caption{Plots of $\mathcal{M}_2/\mathcal{M}_2^{\text{Kerr}}$ and $\mathcal{S}_3/\mathcal{S}_3^{\text{Kerr}}$ (where the corresponding Kerr(-Newman) black hole is chosen to have the same mass and angular momentum as the microstate) as a function of $\lambda$ for $x=1$ and the different values of $h$ and $n$ indicated by the legend. }
 \label{fig:MSfrakcompareKerr}
\end{figure}

For the almost-BPS black hole, $\mathcal{M}_\ell/\mathcal{M}_\ell^{\text{Kerr}}$ was calculated in (\ref{eq:multipoleBHoverKN}). We can already see from this expression that the value of, say, $\mathcal{M}_\ell$ for the almost-BPS black hole can be made smaller or larger than the corresponding Kerr(-Newman) value (for a black hole of equal mass and angular momentum), by adjusting the value of $h$. For the microstates (whose $\lambda\rightarrow 0$ limit corresponds to an almost-BPS black hole), this is then obviously also true; see Fig. \ref{fig:MSfrakcompareKerr}. This shows that claim \ref{claim1} is not generically true for almost-BPS microstates and more generically for non-supersymmetric microstate geometries.

\begin{figure}[ht]\centering
\begin{subfigure}{0.48\textwidth}\centering
 \includegraphics[width=\textwidth]{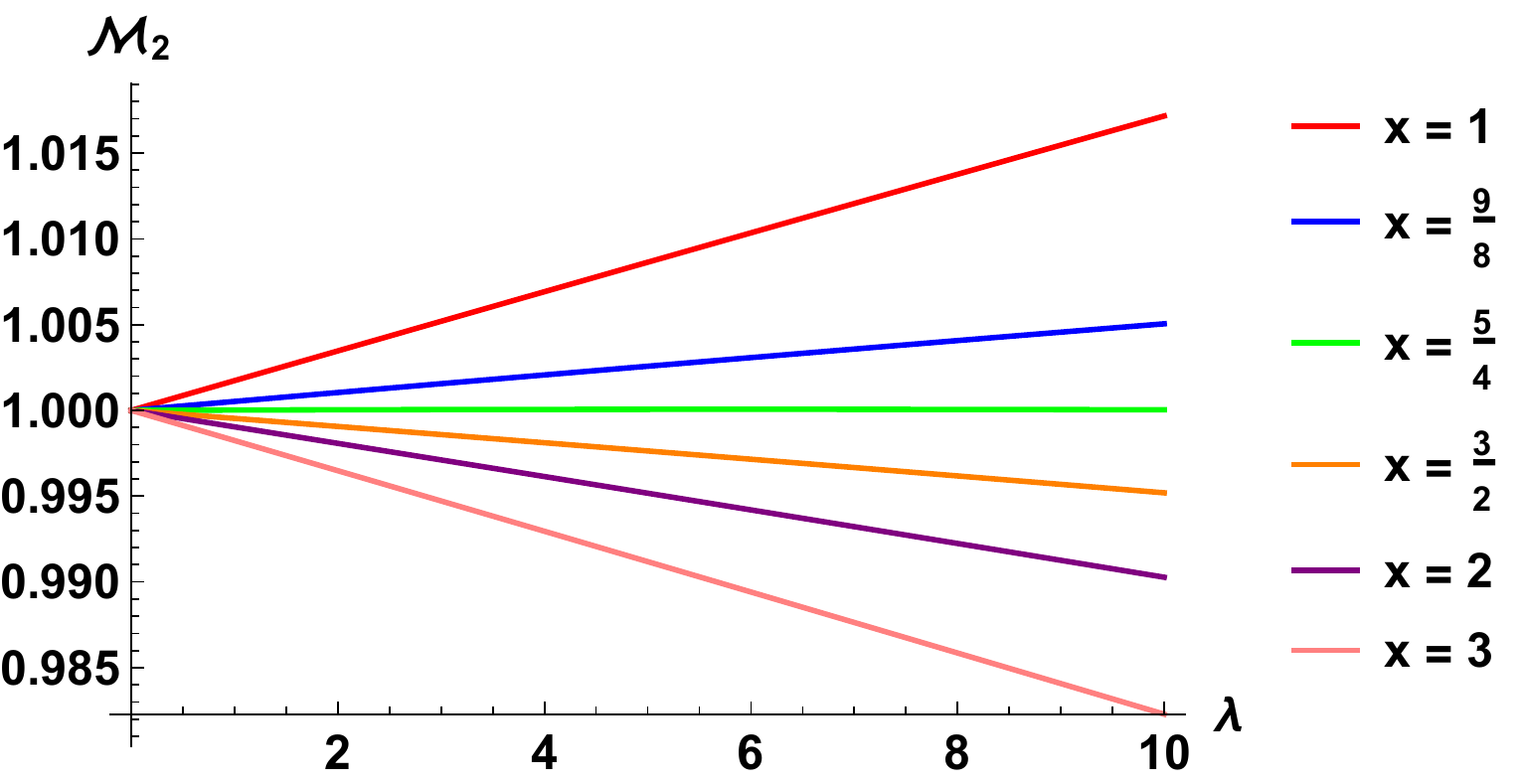}
 \caption{$\dfrac{\mathcal{M}_2(\lambda)}{\mathcal{M}_2(\lambda=0)}$}
 \end{subfigure}
 \begin{subfigure}{0.48\textwidth}\centering
   \includegraphics[width=\textwidth]{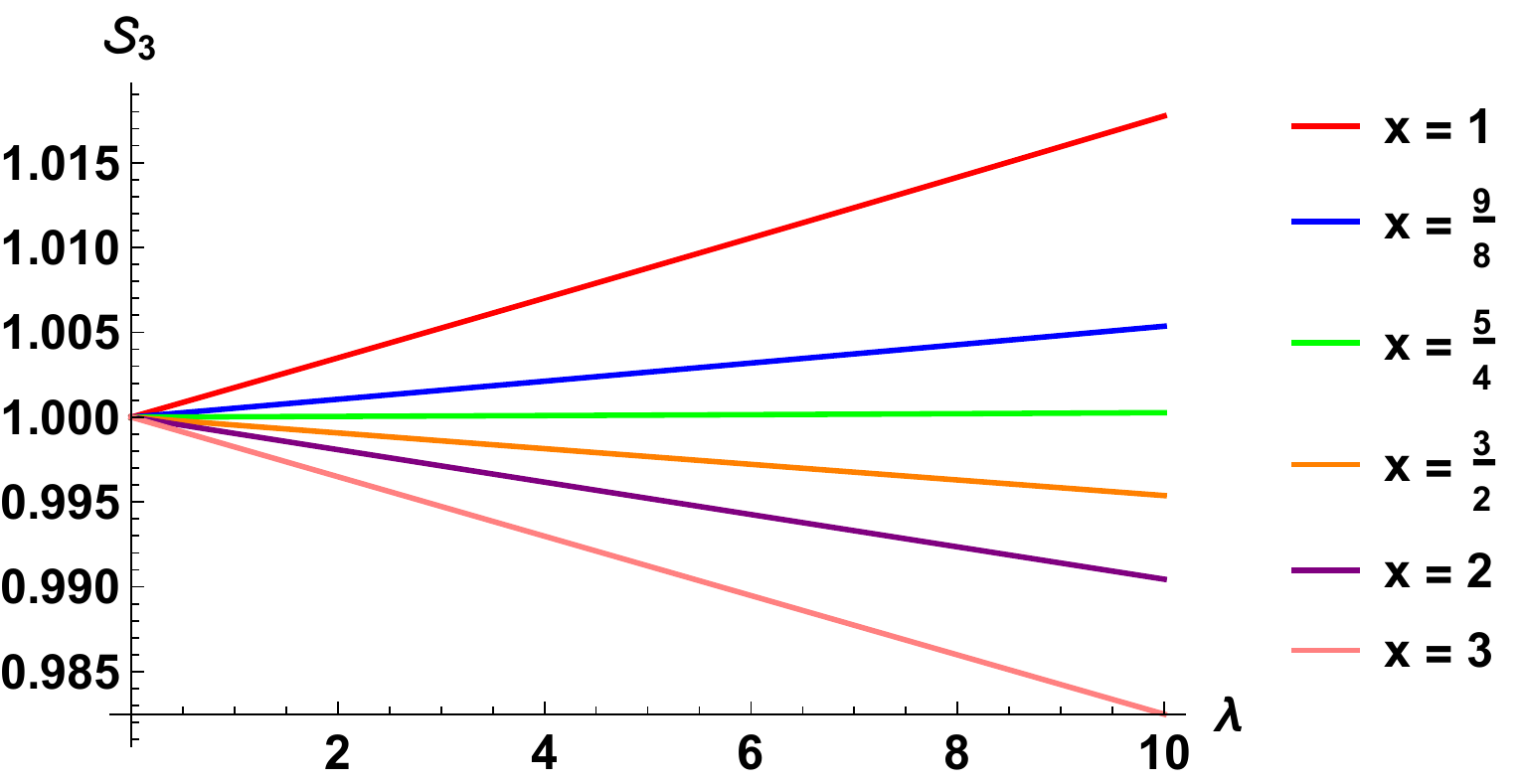}
    \caption{$\dfrac{\mathcal{S}_3(\lambda)}{\mathcal{S}_3(\lambda=0)}$}
 \end{subfigure}
 \caption{Plots of $\mathcal{M}_2$ and $\mathcal{S}_3$ (normalized with respect to their $\lambda=0$ black hole values) as a function of $\lambda$ for $h=433/500$, $n=1$, and $x$ taking on various values indicated by the legend.}
 \label{fig:MSfrakcompare}
\end{figure}

In Fig. \ref{fig:MSfrakcompare}, we plotted $\mathcal{M}_2$ and $\mathcal{S}_3$ (normalized by the $\lambda=0$ black hole value) for $h=433/500$, $n=1$, and various values of $x$; we can clearly see that $\mathcal{M}_2$ and $\mathcal{S}_3$ are \emph{not} always a monotonically increasing function of $\lambda$; in particular, these are counterexamples to claim \ref{claim2}.

One of the main hopes coming from the conjectures of  \cite{Bianchi:2020bxa} was that microstate geometries might be observationally distinguishable from astrophysical (Kerr) black holes by their multipole moments; in particular, (most of) the microstate geometries considered in \cite{Bianchi:2020bxa} had quadrupole moments larger than those of the corresponding Kerr black hole with equal mass and angular momentum, which in turn led to claim \ref{claim1}. The analysis in \cite{Bianchi:2020bxa,Bianchi:2020miz} and the claims \ref{claim1} and \ref{claim2} seemed to imply that the black hole solution is somehow an \emph{extremum} on the space of solutions. However, we show here that there exist microstate geometries for which this is not true.

In particular, we show that the multipole moments of the almost-BPS black hole and its microstate geometries can be larger or smaller than those of Kerr for equal mass and angular momentum. Nevertheless, we want to emphasize that the multipole moments are generically \emph{different} from those of the Kerr black hole, so that they remain a good distinguishing criteria and can be observationally relevant.

\section{Conclusions}\label{sec:conclusion}

In this paper, we have argued for using the almost-BPS class of black holes and microstate geometries as phenomenological models of black holes. These solutions have two advantages over the more commonly-used supersymmetric microstate geometries: First, our solutions are not supersymmetric and in particular can have a large mass over charge ratio. Second, and most importantly, the almost-BPS solutions can have arbitrarily large rotation, in stark contrast to the supersymmetric geometries which have small and limited angular momentum (which moreover must vanish in the scaling limit).

There has already been a considerable body of work studying supersymmetric microstate geometries as phenomenological models \cite{Mayerson:2020tpn}. While these geometries already give rise to interesting observable phenomena, it is important to start considering string theory geometries that are more realistic and can be made to resemble actual astrophysical black holes more closely. The almost-BPS class of black holes and their microstates thus present a prime target for phenomenological study.

We initiated the phenomenological study of almost-BPS geometries here by describing their multipoles (and multipole ratios) in some detail. As we have mentioned in section \ref{sec:BHmult}, the particular family of almost-BPS black holes we have considered can be tuned (by choosing the parameter $h$) such that the quadrupole moment matches that of a Kerr black hole (with equal mass and angular momentum); the deviations of the almost-BPS black hole from Kerr then only show up at higher multipole moments --- such deviations at higher multipole orders will be harder (but not impossible) to detect at future experiments.

In addition, the almost-BPS black holes break equatorial symmetry $\theta\leftrightarrow \pi-\theta$ by having non-zero parity-odd multipoles ($M_{2n+1}$ and $S_{2n}$). Equatorial symmetry breaking of black holes has been largely unexplored phenomenologically; it would be interesting to understand if this can lead to clear observational signatures \cite{danieloddparity}.

Recent studies of supersymmetric microstate geometries include the analysis of their shadows and images \cite{Bacchini:2021fig} (relevant for EHT observations), and scalar field echoes on these geometries \cite{Ikeda:2021uvc} (relevant for the ringdown phase in mergers). These studies showed the intricate and subtle behavior that allows the supersymmetric microstate geometries to behave very similarly to a black hole, but nevertheless allowing for certain observable signatures when far from the scaling point. Expanding these studies to include almost-BPS black holes and their microstates would show how moving away from supersymmetry and adding rotation will alter these mechanisms, and is an important next step to understand possible observable signatures of string theoretic black hole models.

\section*{Acknowledgments}
We thank J. F. Morales for discussions and especially for pointing out an inconsistency in an earlier version of this paper.
The work of the CEA-Saclay members of this collaboration (IB, DM and YL) is supported in part by  the  ANR  grant  Black-dS-String  ANR-16-CE31-0004-01, by the  John Templeton Foundation grant 61149, and by the ERC Grants 787320-QBH Structure and 772408-Stringlandscape and 679278 Emergent-BH.
The work of the Johns Hopkins members of this collaboration (IB and PH) is supported in part by the NSF grant PHY-1820784.


\appendix

\section{The almost-BPS ansatz in different dimensions}
\label{sec:Gen}

Our microstate geometries can be dualized to many duality frames, some of which are better suited for their description than others. In particular, in the duality frame in which the charges of the black hole correspond to D1 branes, D5 branes, momentum and KKM charge, some of the supertube centers that appear singular from a four- or five-dimensional perspective are smooth. These IIB solutions on  T$^4$  can be trivially truncated to six-dimensional $\cN =(1,0)$ supergravity coupled to one extra tensor multiplet, which then can be reduced to four-dimensional $\cN=2$ supergravity with three extra vector multiplets.

\subsection{Six-dimensional frame}
\label{sec:6dframe}

Six-dimensional $\cN=(1,0)$ supergravity coupled to a tensor multiplet has the following bosonic fields coming from the graviton multiplet and the extra tensor multiplet \cite{Nishino:1984gk,Ferrara:1997gh,Gutowski:2003rg,Cariglia:2004kk}:
\begin{itemize}
\setlength\itemsep{0em}
\item A gravitational field $g_{\mu\nu}$.
\item 2 two-form gauge fields $B^{I}_{\mu \nu}$ and their field strengths $G^I= d_6 B^I$.
\item 2 scalars $v^I$ in the coset space $SO(1,1)/SO(1)$. It is convenient to group them into a constrained $SO(1,1)$ matrix:
\be 
\begin{split}
\cS & \= \begin{pmatrix}
v_I  \\
x_I 
\end{pmatrix} \, , \qquad I=0,1 \, \, , \\
v_I v^I & \= 1\, , \qquad v_I v_J -x_I x_J \= \eta_{I J}\, , \qquad v^I x_I \=0\, ,
\end{split}
\ee
where the scalar indices, $I$ or $J$, are raised by the  $SO(1,1)$ Minkowski metric in light-cone coordinates with the mostly-minus signature, 
\be 
\eta\=\begin{pmatrix}
    0 & 1 \\
    1 & 0 
  \end{pmatrix}.
  \label{eq:c2eta6d}
\ee
\end{itemize}
The scalars are involved in the tensor dynamics through the metric $$\cM_{IJ} \=\left(\eta \cS^T \cS \eta \right)_{IJ}  \=  v_I v_J \+x_I x_J \= 2\, v_I v_J \- \eta_{I J}\,, $$ which dictates the twisted self-duality conditions of the tensors
\be 
\cM_{IJ} G^J \= \eta_{IJ}\star_6 G^J\,.
\ee
This implies that the tensor $v_I G^I$ is self-dual and belongs to the gravity multiplet whereas the tensor $x_I G^I$ is anti self-dual and belongs to the tensor multiplet. One can write down a ``pseudo-action" \cite{Ferrara:1997gh,Cariglia:2004kk} 
\be 
(16\pi G_{6})\,S_{6} \= \int d^{6}x\, \sqrt{-g}\,\left( R \- \eta_{I J}\, \partial_\mu v^I \partial_\mu v^J \- \frac{1}{3} \cM_{I J}\, G^I_{\mu \nu \rho} G^{J\,\mu \nu \rho}\right)\,,
\label{eq:c26dAction}
\ee
The dynamics of the solutions of the action given in \eqref{eq:c26dAction} is governed by the following Einstein-Maxwell-scalar equations \cite{Ferrara:1997gh,Cariglia:2004kk}\footnote{We used the self-duality condition to simplify the equations. This also reduces the usual Maxwell equations for the tensor fields to the Bianchi identity.} 
\be 
\begin{split}
R_{\mu \nu} \+ \partial_{\mu} v^I \partial_{\nu} v_I \-\cM_{IJ}\, G^I_{\mu \alpha \beta}  {G^J_{\nu}}^{\alpha \beta}&\= 0\, ,\\
x_I^M \, d_6 \star_6 d_6 v^I \+ 4 \,x_I^M v_J \, G^I \wedge \star_6 G^J &\= 0\,, \\
d_6 G^I &\= 0.
\label{eq:c3EOM6d}
\end{split}
\ee 
We work with the floating-brane ansatz \cite{Bena:2009fi} that encompasses BPS and almost-BPS solutions. The axisymmetric solutions we consider have three spatial isometries and a flat three-dimensional base. The two isometries are parametrized by the coordinate $y$ and the angle $\psi$, whereas the flat space is parameterized by the spherical coordinates $(\rho,\theta,\phi)$:
\be 
\begin{split}
v^1 &\= \sqrt{\frac{Z_2}{2 Z_1}}\,,\qquad G^1 \=  \frac{1}{\sqrt{2}} \left[\,\star_4 d_4 Z_2 \- d_6 \left(\sqrt{\frac{Z_2}{Z_1}}(dt +\omega) \wedge (dy+\beta) \right) \+ (dy+\beta )\wedge \Theta^1 \, \right]\,,\\
v^2 &\= \sqrt{\frac{Z_1}{2 Z_2}}\,,\qquad G^2 \=  \frac{1}{\sqrt{2}} \left[\,\star_4 d_4 Z_1 \- d_6 \left(\sqrt{\frac{Z_1}{Z_2}}(dt +\omega) \wedge (dy+\beta) \right) \+ (dy+\beta )\wedge \Theta^2 \, \right]\,, \\
ds_6^2 &\= - \frac{1}{Z_3 \sqrt{Z_1 Z_2}} \,(dt +\omega)^2\+ \sqrt{Z_1 Z_2}\, ds_4(\cB)^2 \+ \frac{Z_3}{\sqrt{Z_1 Z_2}} (dy + \beta - Z_3^{-1}(dt+\omega))^2\,,
\label{6dSTU}
\end{split}
\ee
where $ds_4(\cB)^2$ is a Gibbons-Hawking metric
\be 
ds_4(\cB)^2 \= V^{-1} (d\psi -w^0 )^2 \+ V \left(d\rho^2 +\rho^2(d\theta^2 +\sin^2\theta \,d\phi^2) \right)\,, \qquad \star_3 d_3 w^0  \= \pm \, d_3 V\,
\label{eq:GHbase}
\ee
and we have defined 
\be 
\omega \= \mu (d\psi -w^0)\+ \varpi\,,\qquad \beta \= K^3 \,(d\psi-w^0)+ w^3\,,\qquad \Theta_a \= d_4\left(K^a \,(d\psi-w^0)+ w^a \right)\,. \label{eq:oneform6dApp}
\ee
The ``$\pm$'' for the connection $w^0$ corresponds to different choice of orientation that leads to different types of solution. In our conventions the minus sign gives supersymmetric solutions while the plus sign gives almost-BPS solutions \cite{Goldstein:2008fq,DallAgata:2010srl}.

\subsection{Five-dimensional frame}
\label{FiveDFrame}

The STU model can be embedded in five-dimensional $\cN =2$ supergravity coupled to two extra vector multiplets. It can be obtained from a KK reduction along $y$ of the six-dimensional frame studied above \cite{deLange:2015gca} or more generically from the low-energy limit of M theory on T$^6$ \cite{Antoniadis:1995vz}.

Five-dimensional $\cN=2$ supergravity coupled to 2 vector multiplets has the following bosonic-field content:
\begin{itemize}
\setlength\itemsep{0em}
\item One gravitational field $g_{\mu\nu}$.
\item Three U(1) vector gauge fields $A^{I}_\mu$ and their field strengths $F^I= d_5 A^I$. One is coming from the graviton multiplet and is usually referred as the ``graviphoton" and the others come from the extra vector multiplets.
\item Three scalars $X^I$ in the symmetric space $SO(1,1) \times \left(SO(1,2)/SO(2) \right)$.
\end{itemize}
\noindent One can write down the five-dimensional action for the bosonic fields \cite{Gunaydin:1983bi,Gunaydin:1984ak}
\be 
(16\pi G_{5})\,S_{5} \= \int d^{5}x\, \sqrt{-g}\, R \- Q_{IJ}  \int \,\left(F^I\wedge \star_{5} F^J - d_5 X^I \wedge \star_{5} d_5X^J\right) \+ \frac{|\epsilon_{IJK}|}{6}\,\int A^I \wedge F^J\wedge F^K\,,
\label{eq:c25dAction}
\ee
where $\epsilon_{IJK}$ is the antisymmetric Levi-Civita tensor and the coupling $Q_{IJ}$ depends on the scalars via \cite{Gunaydin:1983bi,Gunaydin:1984ak}
\be 
Q_{IJ} \= \frac{9}{2}\, X_I X_J -\frac{1}{2} \, |\epsilon_{IJK}| X^K\, .
\ee
The dynamics of solutions of the action given in \eqref{eq:c25dAction} is governed by the following Einstein-Maxwell-scalar equations \cite{Gutowski:2004yv}
\be 
\begin{split}
R_{\mu \nu} \+ Q_{IJ} \left(\partial_\mu X^I \partial_\nu X^J \+ F^I_{\mu \rho}\, {F_{\nu}^J}^{ \rho} \- \frac{1}{6}\, g_{\mu \nu}\, F^I_{ \rho \sigma}\, {F^J}^{ \rho \sigma} \right) &\= 0\, ,\\
d_{5}\left(Q_{IJ}\,\, \star_{5} F^J \right)\+ \frac{1}{4}\,C_{IJK}\, F^J \wedge F^K &\= 0\,, \\
-d_5 \star_5 d_5 X_I \+ \left( C_{IJK} X_L X^K - \frac{1}{6} C_{ILJ}\right) \left( F^L\wedge \star_{5} F^J - dX^L \wedge \star_{5} dX^J \right) &\= 0\, .
\end{split}
\label{eq:c3EinsteinMaxScal5d}
\ee \\

In the floating-brane ansatz  \cite{Bena:2009fi}, we have
\be 
\begin{split}
X_I &\= \frac{Z_I}{(Z_1 Z_2 Z_3)^{1/3}}\= (X^I)^{-1} \,, \qquad F^I \= d_5 A^I  \= d_5 \left(-{dt+\varpi\over Z_I} + \left( K^I-\frac{\mu}{Z_I}\right) \left( d\psi-w^0\right)+w^I \right) \,,\\
ds_{5}^2 &\= -\left(Z_1 Z_2 Z_3\right)^{-\frac{2}{3}} \left(dt + \mu\,(d\psi -w^0) +\varpi \right)^2 \+ \left(Z_1 Z_2 Z_3\right)^{\frac{1}{3}} \,ds(\cB)^2\, ,
\end{split}
\label{eq:STUin5dmetric&GF}
\ee
where $ds(\cB)^2$ is the Gibbons-Hawking metric \eqref{eq:GHbase}. 

\subsection{Four-dimensional frame and the STU model}
\label{sec:4dframe}

The further reduction along the $\psi$ isometry direction leads to four-dimensional $\mathcal{N}=2$ supergravity coupled to three vector multiplets. Now, there are four gauge fields, $\cA^{\Lambda},$ for $\Lambda=\{0, I\}=\{0,1,2,3\},$ with one belonging to the supergravity multiplet ($\cA^0$ is induced by the metric fibration along $\psi$ in five dimensions). The four-dimensional metric for the floating-brane ansatz takes the form
\be
\begin{split}
d s_{4}^{2} &=- {\mathcal{I}_4}^{-\frac{1}{2}} \,\left( dt+\varpi \right)^2 \+ {\mathcal{I}_4}^{\frac{1}{2}} \,ds_3^2 \,, \\
\mathcal{I}_4 &=Z_{1} Z_{2} Z_{3} V -\mu^{2} V^2\,.
\label{eq:metricSTUapp}
\end{split}
\ee
We have three complex scalars
\be 
z^I \= K^I -  \frac{\mu}{Z_I} \-  i\, \frac{\sqrt{\mathcal{I}_4}}{V Z_I} \,,
\label{eq:STUscalars}
\ee
Then, the four gauge fields are
\be 
\begin{split}
\cA^0 \=  \frac{\mu V^2}{\mathcal{I}_4}\,(dt+\varpi)\+ w^0  \,,\qquad  \cA^I \=  - \frac{V}{\mathcal{I}_4\,Z_I}\left( Z_1 Z_2 Z_3 - \mu V K^I Z_I\right)\, (dt+\varpi) \+ w^I\,.
 \end{split}
\label{eq:STUgaugefields}
\ee

\noindent More generally, the reduction to four dimensions leads to the four-dimensional $S T U$ model, for a Lagrangian of the form
\be
\mathcal{L}_{4}=\frac{1}{2} R-g_{IJ} \partial_{\mu} z^I \partial^{\mu} \bar{z}^J+\frac{1}{8} \mathcal{I}_{\Lambda \Sigma} F_{\mu \nu}^{\Lambda} F^{\Sigma \mu \nu}+\frac{1}{8} \mathcal{R}_{\Lambda \Sigma} F_{\mu \nu}^{\Lambda}\left(*_{4} F\right)^{\Sigma \mu \nu}
\label{eq:LagrangianSTU}
\ee
with $\left(*_{4} F\right)_{\mu \nu}=\frac{1}{2} \sqrt{-g} \epsilon_{\mu \nu \rho \sigma} F^{\rho \sigma}$ and $F^\Lambda = d\cA^\Lambda$. Relabelling the scalar fields as $z^{I}=\{S=\sigma-i s, T=$
$\tau-i t, U=v-i u\},$ the metric of the scalar $\sigma$ -model $g_{I J}$ follows from the Kähler potential
\be
\cK=-\log (8 s t u)
\ee
the gauge kinetic couplings are
\be \label{eq:Iscalcoupling}
\mathcal{I}=-s t u\left(\begin{array}{cccc}
1+\frac{\sigma^{2}}{s^{2}}+\frac{\tau^{2}}{t^{2}}+\frac{v^{2}}{u^{2}} & -\frac{\sigma}{s^{2}} & -\frac{\tau}{l^{2}} & -\frac{v}{u^{2}} \\
-\frac{\sigma}{s^{2}} & \frac{1}{s^{2}} & 0 & 0 \\
-\frac{\tau}{l^{2}} & 0 & \frac{1}{t^{2}} & 0 \\
-\frac{v}{u^{2}} & 0 & 0 & \frac{1}{u^{2}}
\end{array}\right)
\ee
and the axionic couplings are
\be \label{eq:Rscalcoupling}
\mathcal{R}=\left(\begin{array}{cccc}
2 \sigma \tau v & -\tau v & -\sigma v & -\sigma \tau \\
-\tau v & 0 & v & \tau \\
-\sigma v & v & 0 & \sigma \\
-\sigma \tau & \tau & \sigma & 0
\end{array}\right)
\ee

We are interested in computing electric and magnetic charges in four dimensions, which requires us to compute the electromagnetic dual of $\cA^\Lambda$, $\cA_\Lambda$. From the Lagrangian, we see that the dual field strength is not simply given by the Hodge star of $F$. Instead, we have
\be 
G_\Lambda \= d\cA_\Lambda \=  \mathcal{R}_{\Lambda \Sigma} \,F^{\Sigma}\- \mathcal{I}_{\Lambda \Sigma}\, *_{4} F^{\Sigma}\,.
\label{eq:electromagDualGF}
\ee
One can then rewrite the STU Lagrangian with the dual field and obtain a more usual  Maxwell term, with a trivial electric coupling:
\be
\mathcal{L}_{4}=\frac{1}{2} R-g_{IJ} \partial_{\mu} z^I \partial^{\mu} \bar{z}^J+\frac{1}{8} F_{\mu \nu}^{\Lambda}\, (*_{4}G_\Lambda)^{\, \mu \nu}
\label{eq:LagrangianSTUbis}
\ee
This non-trivial electromagnetic duality gives a non-standard charge lattice obtained from 
\be 
\Gamma= -\frac{1}{4 \pi} \int_{S^{2}_\infty} \mathcal{F}=\left(\begin{array}{c}q_{0} \\ q_I \\ p^I \\ p^{0}\end{array}\right)\,,
\ee
where $S^{2}_\infty$ is the asymptotic two-sphere parametrized by $\theta$ and $\phi$ and where
\be
\mathcal{F}_{\mu \nu}=\left(\begin{array}{c}
G_{\Lambda \mu \nu} \\
F_{\mu \nu}^{\Lambda}
\end{array}\right)\,.
\ee

Thus,  we need to compute the value of $\cA_\Lambda$ as well. We have
\be 
\cA_\Lambda \= \zeta_\Lambda (dt+\varpi) \+ v_\Lambda\,,
\label{eq:dualSTUgaugefields}
\ee
where the important parts encoding the electric charges, $ v_\Lambda$, are present in the expressions of the tensors of the six-dimensional solution described above.

The magnetic charges $p^\Lambda$ and the electric charges $q_\lambda$ are obtained by integrating: 
\be 
p^\Lambda \= -\frac{1}{4\pi} \int_{S^{2}_\infty} d\cA^\Lambda \= - \frac{1}{4\pi} \int_{S^{2}_\infty} dw^\Lambda\,,\qquad q_\lambda \= -\frac{1}{4\pi } \int_{S^{2}_\infty} d\cA_\Lambda \= -\frac{1}{4\pi} \int_{S^{2}_\infty} dv_\Lambda\,.
\label{eq:4dchargesgeneric}
\ee

\subsection{Equations of motion}
The almost-BPS solutions are stationary solutions governed by the following reduced equations of motion
\begin{equation}
\begin{split}
& d\star_3 d Z_I ~=~  \frac{|\epsilon_{IJK}|}{2} \, V \,d\star_3d (K^J K^K )\,,\qquad d(\mu V) \- \star_3 d \varpi ~=~ V Z_I \,dK^I \,, \\
&  \star_3 dw^0 ~=~ d V \,,\hspace{4.6cm}  \star_3 d w^I ~=~ K^I \,dV- V \,dK^I , \\
& \star_3 dv_0 ~=~ Z_I dK_I-K_I dZ_I \+ V d(K^1K^2K^3)-K^1K^2K^3 \,dV\,, \\
& \star_3 dv_I ~=~ dZ_I \- \frac{|\epsilon_{IJK}|}{2}  \left(V d(K^JK^K) - K^JK^K \,dV \right)\,,
\label{eq:EOM}
\end{split}
\end{equation}
where $\epsilon_{IJK}$ is the three-dimensional Levi-Civita tensor and $\star_3$ is the Hodge star in the three-dimensional flat base. When considering more general $U(1)^N$ supergravities in five dimensions and the four-dimensional reductions thereof, $|\epsilon_{IJK}|$ is replaced by the corresponding symmetric tensor $C_{IJK}$.

By resolving these equations, one can extract the four-dimensional metric \eqref{eq:metricSTUapp}, the three scalars, $z^I$ \eqref{eq:STUscalars}, the four gauge fields, $\cA^\Lambda$ \eqref{eq:STUgaugefields}, and their duals $\cA_\Lambda$ \eqref{eq:dualSTUgaugefields}, using the fact that \cite{DallAgata:2010srl}
\begin{equation}
\begin{split}
\zeta_0& \= {\mathcal{I}_4}^{-1} \, \left[Z_1 Z_2 Z_3 - \mu V \left( V K^1 K^2 K^3 +\sum_I Z_I K^I \right) + V \sum_{J<K} K^J K^K Z_J Z_K \right]\,,\\
\zeta_I& \= {\mathcal{I}_4}^{-1} \, \left[ Z_I \left( \mu -\sum_{J\neq I} K^J Z_J \right) \+ \frac{|\epsilon_{IJK}|}{2} V\mu K^J K^K  \right]\,.
\end{split}
\end{equation}

\section{Almost-BPS black hole: details}
\label{sec:ABPSBHdetails}

To complement the presentation of the metric of the almost-BPS black hole in the four-dimensional frame on section \ref{sec:BHsol}, we give the relevant expressions for the gauge fields and scalars. We have considered a single-center solution given by 
\be 
V \= h +\frac{Q_0}{\rho}\,,\qquad Z_I \= \frac{1}{h}+\frac{Q_I}{\rho}\,,\qquad K^I \= 0 \,, \qquad \mu \,V \=  m_\infty  \+ \alpha \frac{\cos\theta}{\rho^2}\,, \qquad \varpi \= -\alpha {\sin^2\theta\over \rho} d\phi\,.
\label{eq:BHHarmonicFuncApp}
\ee
The equations of motion \eqref{eq:EOM} are solved by 
\be 
 \varpi \= -\alpha {\sin^2\theta\over \rho} d\phi\,,\quad w^0 \= Q_0 \cos\theta \,d\phi\,,\quad w^I \= 0 \,,\quad v_0 \= 0 \,,\quad v_I \= Q_I \cos \theta \,d\phi\,.
\ee

\subsection*{Scalars}

The three scalars are given by \eqref{eq:STUscalars}, which gives for the almost-BPS black hole:
\begin{equation}
z^I \=- \frac{m_\infty\,\rho^2 +\alpha \,\cos \theta}{\left( Q_0 +h \rho \right) \left( Q_I + \frac{\rho}{h} \right)} \-  i\, \frac{\Delta}{\left( Q_0 +h \rho \right) \left( Q_I + \frac{\rho}{h} \right) }\,,
\label{eq:scalarBHApp}
\end{equation}
where $\Delta$ has been defined in \eqref{eq:DeltaDef}. 

\subsection*{Gauge fields}

The gauge fields are given by the generic equation \eqref{eq:STUgaugefields}, which for the black hole gives\footnote{Note that we have gauged away the irrelevant asymptotics of the gauge fields by adding $-hm_\infty dt$ and $h^{-1}dt$ to $\cA^0$ and $\cA^I$ respectively.}
\be 
\begin{split}
 \mathcal{A}^0&\= Q_0 \,\cos\theta \, d\phi \+ \frac{(m_\infty \, \rho^2+\alpha\, \cos\theta)(Q_0 +h\,\rho)}{\Delta^2}\,\left(\rho\,dt - \alpha \sin^2 \theta \,d\phi \right) - hm_\infty \, dt\,,\\
 \mathcal{A}^I &\= - \frac{Q_0 + h \, \rho}{\Delta^2}\, \prod_{J\neq I} (Q_J+\frac{\rho}{h}) \, \left(\rho\,dt- \alpha \sin^2 \theta \,d\phi \right) + h^{-1} \, dt\,.
 \label{eq:metric&GFBlackhole}
\end{split}
\ee
Using \eqref{eq:dualSTUgaugefields}, we can also derive the dual gauge fields, $\cA_\Lambda$:
\be 
\begin{split}
 \mathcal{A}_0 &\= \,\frac{ \prod_{I} (Q_I+\frac{\rho}{h})}{\Delta^2}\,\left(\rho\,dt- \alpha \sin^2 \theta \,d\phi \right) -h^{-3}\,dt\,, \\
 \mathcal{A}_I &\= Q_I \,\cos\theta \, d\phi \+ \frac{(m_\infty \, \rho^2+\alpha\, \cos\theta)(Q_I + \,\frac{\rho}{h})}{\Delta^2}\,\left(\rho\,dt - \alpha \sin^2 \theta \,d\phi \right)-h^{-1}m_\infty \,dt\,.
\end{split}
\label{eq:dualguagefieldsBH}
\ee


\section{Axisymmetric almost-BPS multicenter solutions in Taub-NUT}
\label{sec:ABPSmulticenter}

In this section, we review the solutions derived in \cite{Bena:2009en} for axisymmetric multi-center configurations in Taub-NUT where the centers are at positions $a_j$ on the $z$ axis of the $\mathbb{R}^3$ base, $j=1...n$. We consider the Taub-NUT harmonic function sourced at the center of the $\mathbb{R}^3$ spherical coordinates $(\rho,\theta,\phi)$: $$V \= h + \frac{Q_0}{\rho}\,, \qquad w^0 \=  \,Q_0\,\cos\theta \, d\phi\,.$$ We assume that the centers that source the vector fields are all distinct from the Taub-NUT center, $a_j\neq0$. The shifted spherical coordinates around the $j^\text{th}$ center, $(\rho_j,\theta_j,\phi)$, are given by
\be
\rho_j \= \sqrt{\rho^2+a_j^2-2 \rho\, a_j\, \cos\theta} \, \qquad \cos \theta_j \= \frac{\rho \cos\theta- a_j}{\rho_j} \,.
\ee
We will sometimes use the index $j=0$ to denote the Taub-NUT center $(\rho_0, \theta_0)=(\rho,\theta)$ and $a_0=0$. We proceed step by step by solving first the magnetic field strengths before the warp factors and the angular momentum one-form. We end the discussion by deriving the regularity constraints.

\newpage
\begin{itemize}
\item \underline{The anti-self dual magnetic two-forms $\Theta^I$:}
\end{itemize}

The two-form field strengths, $\Theta^I$,  are given by \eqref{eq:oneform6dApp} with $\Theta^3 \= d\beta$. We will assume for simplicity that the $K^I$ are harmonic with no constant terms and no source at the Taub-NUT center
\be 
K^I \=  \sum_{j=1}^n \frac{k^{(j)}_I}{\rho_j}\,.
\ee
For axisymmetric center configurations, we have
\bea
*_3 d w^I \= K^I dV-V d K^I  \qquad \Rightarrow\qquad w^I \= -\sum_{j=1}^n k^{(j)}_I \Big(h \cos\theta_j \+ Q_0 \,{\rho-a_j \cos\theta\over \rho_j\, a_j}\Big) d\phi\,.
\eea

\begin{itemize}
\item \underline{The warp factors $Z_I$:}
\end{itemize}

The warp factors, $Z_I$, are determined by the harmonic equations with quadratic sources \eqref{eq:EOM}. For axisymmetric centers in Taub-NUT, the generic solutions are
\bea
Z_I ~=~ L_I + {|\epsilon_{IJK}| \over 2}\sum_{j,k=1}^n
\Bigl(h+{Q_0\, \rho\over a_j a_k}\Bigr) {k^{(j)}_J k^{(k)}_K \over \rho_j \rho_k}\,.
\label{eq2:Zgeneralform}
\eea
The functions $L_I$ are the electric harmonic functions one can freely add to the  $Z_I$:
$$L_I \= l^\infty_I \+\frac{Q^{(0)}_I}{\rho} \+\sum_{j=1}^n \frac{Q^{(j)}_I}{\rho_j} \, .
$$

\begin{itemize}
\item \underline{The angular momentum one-form $\omega$:}
\end{itemize}

The last equation of the first line in \eqref{eq:EOM} determines the two components, $\mu$ and $\varpi$, of the angular momentum one-form, $\omega$. The source terms are
\be
\begin{split}
 \hspace*{1cm} V Z_I  d_3 K^I \= & \sum_{j=1}^n l_I^\infty k^{(j)}_I \left(h\, s_j^{(1)} + Q_0\, s_j^{(2)}  \right) + \sum_{j=1}^n Q^{(j)}_I k^{(j)}_I \left(h\, s_{j}^{(3)} + Q_0\, s_{j}^{(4)}  \right)   \\ 
 &\hspace*{-3cm}\+ \sum_{i=0}^n\sum_{j=1,j\neq i}^n Q^{(i)}_I k^{(j)}_I \left(h\, s_{ij}^{(5)} + Q_0\, s_{ij}^{(6)}  \right) +  {|\epsilon_{IJK}|\over 2} \sum_{i,j,k=1}^n k^{(i)}_I k^{(j)}_J k^{(k)}_K \left(h^2\, s^{(7)}_{ijk} + Q_0^2 \,s^{(8)}_{ijk} + h Q_0 \,s^{(9)}_{ijk}\right)  \,.
\end{split}
\ee
where we have defined 9 generating functions, $s^{(\alpha)}$,
\be 
\begin{split}
s_j^{(1)} & \equi d_3 \left(\frac{1}{\rho_j} \right)\,,\qquad s_j^{(2)}  \equi \frac{1}{\rho}d_3 \left(\frac{1}{\rho_j} \right)\,, \qquad s_j^{(3)}  \equi \frac{1}{\rho_j} d_3 \left(\frac{1}{\rho_j} \right)\,,\qquad 
s_j^{(4)}  \equi \frac{1}{\rho\,\rho_j} d_3 \left(\frac{1}{\rho_j} \right)\,,\\
s_{ij}^{(5)} & \equi \frac{1}{\rho_i} d_3 \left(\frac{1}{\rho_j} \right)\,,\qquad s_{ij}^{(6)}  \equi \frac{1}{\rho\,\rho_i} d_3 \left(\frac{1}{\rho_j} \right)\,,\qquad s_{ijk}^{(7)}  \equi \frac{1}{\rho_i\,\rho_j} d_3 \left(\frac{1}{\rho_k} \right)+ \text{perm.}\,,\\
s_{ijk}^{(8)}  &\equi \frac{1}{a_i \,a_j\rho_i\,\rho_j} d_3 \left(\frac{1}{\rho_k} \right)+ \text{perm.}\,, \qquad s_{ijk}^{(9)}  \equi \left(\frac{1}{\rho}+  \frac{\rho}{a_i \,a_j} \right)\,\frac{1}{\rho_i\,\rho_j}d_3 \left(\frac{1}{\rho_k} \right)+ \text{perm.}\,.
\end{split}
\ee
We define the corresponding pairs of solutions $(f^{(\alpha)},\,t^{(\alpha)})$ that solve $$d_3 \,f^{(\alpha)} \- \star_3 d_3\, t^{(\alpha)} ~=~ s^{(\alpha)} \,.$$
One can also freely add the pair of solution $(f^{(10)},\,t^{(10)})$ of the homogeneous equation 
\be 
f^{(10)} \= M\,, \qquad \star_3 d_3 t^{(10)} \= d_3 M\,,
\ee
where $M$ is a harmonic function that generically takes the form
\be 
M \= m_\infty \+\frac{m^{(0)}}{\rho} \+ \alpha \frac{\cos\theta}{\rho^2}\+\sum_{j=1}^n \frac{m^{(j)}}{\rho_j} \+\alpha_j \frac{\cos\theta_j}{\rho_j^2}\, ,
\ee
which leads to
\be \label{eq:expression_t10}
t^{(10)} \= \varpi_0 \, d\phi \+ m^{(0)} \cos\theta\,d\phi \- \alpha \,\frac{\sin^2\theta}{\rho}\, d\phi \+ \sum_{j} \left(m^{(j)} \cos\theta_j - \alpha_j\, \frac{\rho^2\sin^2\theta_j}{\rho_j^3}\right)d\phi 
\ee

Solving for each generating functions $s^{(\alpha)}$ gives
\be \label{eq:solution_nine_generating_functions}
\begin{split}
f^{(1)}_j & \= \frac{1}{2 \,\rho_j}\,,\qquad t^{(1)}_j \= - \frac{1}{2}\cos\theta_j \,d\phi\,, \qquad \quad f^{(2)}_j  \= \frac{1}{2 \,\rho\,\rho_j}\,,\qquad t^{(2)}_j \= - \frac{1}{2}\,\frac{\rho-a_j \cos \theta}{a_j\,\rho_j}\,d\phi\,, \\
f^{(3)}_j & \= \frac{1}{2\, {\rho_j}^2}\,,\qquad t^{(3)}_j \= 0\,, \qquad \quad f^{(4)}_j  \= \frac{\cos\theta }{2\,a_j\,{\rho_j}^2}\,,\qquad t^{(4)}_j \= - \frac{\rho \,\sin^2 \theta}{2\,a_j\,{\rho_j}^2}\,d\phi\,, \\
f^{(5)}_{ij} & \= \frac{1}{2 \,\rho_i\,\rho_j}\,,\qquad t^{(5)}_{ij} \= \frac{\rho^2 + a_i a_j -(a_i+a_j) \rho \cos\theta}{2(a_i -a_j)\rho_i \rho_j}\,d\phi\,, \quad \qquad f^{(7)}_{ijk} \= \frac{1}{\rho_i \rho_j \rho_k} \,,\qquad t^{(7)}_{ijk} \= 0 \,,\\
f^{(6)}_{ij} & \= \frac{\rho^2 +a_i a_j - 2 a_j \rho \cos\theta}{2\,a_j\,(a_i -a_j)\rho \rho_i \rho_j}\,,\qquad t^{(6)}_{ij} \= \frac{\rho \left( a_i +a_j \cos 2 \theta \right) - \left(\rho^2 +a_i a_j \right) \cos\theta }{2a_j(a_j -a_i)\rho_i \rho_j}\,d\phi\,, \\
f^{(8)}_{ijk} & \= \frac{\rho \,\cos\theta}{a_i a_j a_k \rho_i \rho_j \rho_k} \,,\qquad t^{(8)}_{ijk} \= - \frac{\rho^2 \sin^2\theta}{a_i a_j a_k \rho_i \rho_j \rho_k} \,d\phi \,, \\
f^{(9)}_{ijk} & \= \frac{\rho^2 \left(a_i + a_j+ a_k \right) +a_i a_j a_k}{2 a_i a_j a_k \,\rho \,\rho_i \rho_j \rho_k} \,,\\
& t^{(9)}_{ijk} \= - \frac{\rho^3 \+ \rho \left( a_i a_j + a_i a_k + a_j a_k\right)- \left(\rho^2 (a_i +a_j +a_k)+a_i a_j a_k \right) \,\cos \theta}{2 a_i a_j a_k \, \rho_i \rho_j \rho_k} \,d\phi \,,
\end{split}
\ee
The complete expression for $\mu$ and $\varpi$ is then
\bea
&&\mu \= \sum_{j=1}^n {l^\infty_I k^{(j)}_I\over 2 \rho_j}+\sum_{j=1}^n {Q^{(j)}_I k^{(j)}_I\over 2 V \rho_j^2}\Bigl(h+{Q_0\cos\theta\over a_j}\Bigr)+ \sum_{i=0}^n\sum_{j=1,j\neq i}^n{Q^{(i)}_I k^{(j)}_I\over 2 V \rho_i \rho_{j}} \Bigl(h+Q_0 {\rho^2 + a_i a_{j} - 2 a_{j} \rho \cos\theta\over a_{j} (a_i - a_{j})\,\rho}\Bigr)\nonumber\\
&&\qquad \+\sum_{i,j,k=1}^n {k^{(i)}_1 k^{(j)}_2 k^{(k)}_3\over V \rho_i \rho_j \rho_k} \Bigl(h^2+{Q_0}^2\,{\rho\cos\theta\over a_i a_j a_k}+h Q_0 {\rho^2 (a_i + a_j + a_k)+ a_i a_j  a_k\over 2 a_i a_j a_k \rho }\Bigr)+{M\over V}\,,
\label{fullmu-exp}
\eea
\bea
&&\varpi \= - \sum_{j=1}^n { l^\infty_I  k^{(j)}_I\over 2}\Bigl(h\cos\theta_j+Q_0 {\rho-a_j \cos\theta\over a_j\rho_j }\Bigr)d\phi- \sum_{j=1}^n Q^{(j)}_I k^{(j)}_I {Q_0\, \rho \sin^2\theta\over 2 a_j \rho_j^2} d\phi\nonumber\\
&&\qquad\+\sum_{i=0}^n\sum_{j=1,j\neq i}^n {Q^{(i)}_I k^{(j)}_I\over 2 (a_i-a_j)\rho_i \rho_{j}}\Bigl(h (\rho^2 + a_i a_{j} -  (a_i+a_{j}) \rho \cos\theta)\label{fullomega-exp}\\
&&\hspace{7cm}\- Q_0{\rho(a_i+a_{j} \cos2\theta)-(\rho^2 + a_i a_{j})\cos\theta\over a_{j} }\Bigr)d\phi\nonumber\\
&&\qquad\-\sum_{i,j,k=1}^n {k^{(i)}_1 k^{(j)}_2 k^{(k)}_3\over a_i a_j a_k \rho_i \rho_j \rho_k} \Bigl({Q_0}^2 \rho^2 \sin^2\theta \nonumber\\
&&\hspace{4cm}\+h Q_0 {\rho^3 +\rho(a_i a_j+ a_i a_k + a_j a_k)-(\rho^2(a_i+a_j+a_k)+a_i a_j a_k) \cos\theta\over 2 }\Bigr)d\phi\nonumber\\
&&\qquad \+\varpi_0 \, d\phi \+ m^{(0)} \cos\theta\,d\phi \- \alpha \,\frac{\sin^2\theta}{\rho}\, d\phi \+ \sum_{j=1}^n \left(m^{(j)} \cos\theta_j - \alpha_j\, \frac{\rho^2\sin^2\theta_j}{\rho_j^3}\right)d\phi \,. \nonumber
\eea
\begin{itemize}
\item \underline{The electric one-forms $v^\Lambda$:}
\end{itemize}
Since we are interested in the profile of the solutions in four dimensions, we need to derive the electromagnetic dual gauge fields $A_\Lambda$ \eqref{eq:dualSTUgaugefields}. For this purpose we need to integrate the equations for the one-forms $v_\Lambda$ \eqref{eq:EOM}. We first decompose the source terms by defining some generating functions:
\be 
\begin{split}
\star_3 d_3 v_I &\= Q^{(0)}_I \,\star_3 d_3 T^{(0)} \+ \sum_{j=1}^n Q^{(j)}_I \,\star_3 d_3 T^{(1)}_j \+ \frac{|\epsilon_{IJK}|}{2}\, \sum_{j,k=1}^n \frac{Q_0 k^{(j)}_J k^{(k)}_K }{a_j a_k} \,\star_3 d_3 T^{(3)}_{jk}\,, \\
\star_3 d_3 v_0 &\= l_I^\infty  \sum_{j=1}^n k^{(j)}_I \,\star_3 d_3 T^{(1)}_j \+ Q^{(0)}_I  \sum_{j=1}^n k^{(j)}_I  \,\star_3 d_3 T^{(2)}_j \+  \sum_{j,k=1}^n Q^{(j)}_I k^{(k)}_I \,\star_3 d_3 T^{(4)}_{jk} \\
&\phantom{\=} \+ \frac{|\epsilon_{IJK}|}{6}\, \sum_{i,j,k=1}^n Q_0 k^{(i)}_I k^{(j)}_J k^{(k)}_K \,\star_3 d_3 T^{(5)}_{ijk}\,,
\end{split}
\ee 
where $T^{(0)}$, $T_j^{(1)}$, $T_j^{(2)}$, $T_{jk}^{(3)}$, $T_{jk}^{(4)}$ and $T_{ijk}^{(5)} $ satisfy
\begin{alignat}{2}
\star_3 d T^{(0)}  &~=~&& d\left(\frac{1}{\rho}\right)\,,\qquad \star_3 d T_j^{(1)}~=~ d\left(\frac{1}{\rho_j}\right) \,,\qquad \star_3 d T_j^{(2)}~=~ \frac{1}{\rho} d\left(\frac{1}{\rho_j}\right) - \frac{1}{\rho_j} d\left(\frac{1}{\rho}\right) \,,\nonumber \\
\star_3 d T_{jk}^{(3)} &~=~&&\left(1-\frac{a_j a_k}{\rho^2} \right) d\left(\frac{\rho}{\rho_j \rho_k} \right)  \,,\qquad \star_3 d T_{jk}^{(4)} ~=~  \frac{1}{\rho_j} d\left(\frac{1}{\rho_k}\right) - \frac{1}{\rho_k} d\left(\frac{1}{\rho_j}\right)\,,
\end{alignat}
\begin{alignat}{2}
\hspace{-5cm} \star_3 d T_{ijk}^{(5)} &~=~&& \left(\frac{1}{a_i a_j} +\frac{1}{\rho^2}- \frac{1}{a_i a_k}- \frac{1}{a_ja_k}\ \right) \,\frac{\rho}{\rho_i \rho_j} d\left(\frac{1}{\rho_k}\right) \,\nonumber\\
& && \,~ + \left(\frac{1}{a_i a_k} +\frac{1}{\rho^2}- \frac{1}{a_i a_j}- \frac{1}{a_j a_k}\ \right) \,\frac{\rho}{\rho_i \rho_k} d\left(\frac{1}{\rho_j}\right) \,\nonumber\\
& && \,~ + \left(\frac{1}{a_j a_k} +\frac{1}{\rho^2}- \frac{1}{a_i a_j}- \frac{1}{a_i a_k}\ \right) \,\frac{\rho}{\rho_j \rho_k} d\left(\frac{1}{\rho_i}\right) \,\nonumber\\
& && \,~ + \left(-\frac{1}{\rho^2} +\frac{1}{a_i a_j} + \frac{1}{a_i a_k}+ \frac{1}{a_j a_k}\ \right) \,\frac{\rho^2}{\rho_i \rho_j \rho_k} d\left(\frac{1}{\rho}\right)  \,.\hspace{2.5cm}\,\nonumber\\
\end{alignat}
We find
\begin{alignat}{1}
T^{(0)} &\= \cos \theta \, d\phi  \,, \qquad  T^{(1)}_j \= \cos \theta_j \, d\phi  \,, \qquad T^{(2)}_I \= \frac{\rho-a_j \cos\theta}{a_j\,\rho_j} \,d\phi\,, \nonumber\\
 T^{(3)}_{jk} &\=  \frac{(\rho^2 + a_j a_k) \cos\theta -(a_j+a_k)\rho }{\rho_j \rho_k}\,d\phi\,,\qquad T^{(4)}_{jk} \= \frac{\rho^2 + a_j a_k -(a_j+a_k)\rho\cos \theta }{(a_k-a_j) \,\rho_j \rho_k}\,d\phi\,, \\
T^{(5)}_{ijk} &\= \frac{\rho^3 + \rho (a_i a_j+a_i a_k+a_j a_k)- \left(\rho^2(a_i + a_j+ a_k) +a_i a_j a_k \right) \cos \theta}{a_i a_j a_k\,\rho_i \rho_j \rho_k}\,d\phi\,.\nonumber
\end{alignat}
Thus, $v_0$ and $v_I$ are given by
\be 
\begin{split}
 v_I &\= Q^{(0)}_I \,\cos \theta \,d\phi \+ \sum_{j=1}^n Q^{(j)}_I \,\cos \theta_j \,d\phi \+ \frac{|\epsilon_{IJK}|}{2}\, \sum_{j,k=1}^n Q_0 k^{(j)}_J k^{(k)}_K\,\frac{(\rho^2 + a_j a_k) \cos\theta -(a_j+a_k)\rho }{a_j a_k\,\rho_j \rho_k}\,d\phi\,, \\
v_0 &\= l_I^\infty  \sum_{j=1}^n k^{(j)}_I \,\cos \theta_j \,d\phi \+ Q^{(0)}_I  \sum_{j=1}^n k^{(j)}_I \,\frac{\rho-a_j \cos\theta}{a_j\,\rho_j} \,d\phi \+  \sum_{(j\neq k)=1}^n Q^{(j)}_I k^{(k)}_I \,\frac{\rho^2 + a_j a_k -(a_j+a_k)\rho\cos \theta }{(a_k-a_j) \,\rho_j \rho_k}\,d\phi \\
&\phantom{\=} \+ \frac{|\epsilon_{IJK}|}{6}\, \sum_{i,j,k=1}^n Q_0 k^{(i)}_I k^{(j)}_J k^{(k)}_K \,\frac{\rho^3 + \rho (a_i a_j+a_i a_k+a_j a_k)- \left(\rho^2(a_i + a_j+ a_k) +a_i a_j a_k \right) \cos \theta}{a_i a_j a_k\,\rho_i \rho_j \rho_k}\,d\phi\,.
\end{split}
\ee 

\begin{itemize}
\item \underline{The regularity constraints:}
\end{itemize}
The solutions constructed above are regular if:
\begin{itemize}
\setlength\itemsep{0em}
\item[-] The one-form $\varpi$ does not have Dirac-Misner string and must vanish on the $z$-axis.
\item[-] The absence of closed timelike curves requires the positivity of some metric components. It leads to
\be 
\label{eq:c3CTC}
Z_I \, V \geq 0\,,\qquad  \mathcal{I}_4 \equi  Z_1 Z_2 Z_3 \, V \- \mu^2 V^2 \,\geq\, |\varpi |^2\, .
\ee
\end{itemize} 
The first condition implies $n+1$ algebraic equations. One can make these constraints explicit, for example, by solving them with respect to the $n+1$ variables $\varpi_0$, $m^{(0)}$ and
$m^{(i)}$ for $i=1,\ldots,n$.  If one considers, for definiteness, a
configuration in which all the poles $a_i$ lie on one side of the
Taub-NUT center ($0<a_1<\ldots<a_n$), then the regularity constraints
are:
\bea
\varpi_0&\=& Q_0\sum_{j=1}^n{ l_I^\infty k^{(j)}_I\over 2 a_j}\+h \sum_{i=0}^n\sum_{j=1,j\neq i}^n {Q^{(i)}_I k^{(j)}_I\over 2 (a_{j}-a_i)}\+ h Q_0 \sum_{i,j,k=1}^n {k^{(i)}_1 k^{(j)}_2 k^{(k)}_3\over 2 a_i a_j a_k}\,,\nonumber\\
m^{(0)}&\=& -Q_0\sum_{j=1}^n {l_I^\infty  k^{(j)}_I\over 2 a_j}-h \sum_{j=1}^n {Q^{(0)}_I k^{(j)}_I\over 2 a_j}+Q_0 \sum_{i=0}^n\sum_{j=1,j\neq i}^n {Q^{(i)}_I k^{(j)}_I\over 2 a_{j} (a_{j}-a_i)}-h Q_0 \sum_{i,j,k=1}^n {k^{(i)}_1 k^{(j)}_2 k^{(k)}_3\over 2 a_i a_j a_k}\,,\nonumber\\
m^{(i)}&\=& {l_I^\infty  k^{(i)}_I\over 2} \Bigl(h+{Q_0\over a_i}\Bigr)+\sum_{j=1}^n {1\over 2 |a_i - a_j|} \Bigl[Q^{(j)}_I k^{(i)}_I \Bigl(h+{Q_0\over a_i}\Bigr)-Q^{(i)}_I k^{(j)}_I \Bigl(h+{Q_0\over a_j}\Bigr)\Bigr]\nonumber\\
&&+{h Q_0\over 2} \Bigl[{k^{(i)}_1 k^{(i)}_2 k^{(i)}_3 \over a_i^3} +{|\epsilon_{IJK}|\over 2} {k^{(i)}_I\over a_i} \sum_{j,k=1}^n  \mathrm{sign}(a_j-a_i)  \mathrm{sign}(a_k-a_i){k^{(j)}_J k^{(k)}_K\over a_j a_k} \Bigr]\quad (i\ge1)\,.\label{regularity1}
\eea
Those equations are equivalent to the Denef equations or bubble equations \cite{Denef:2000nb,Denef:2002ru,Bates:2003vx} but for almost-BPS solutions. The requirement that the quartic invariant be everywhere positive does not translate in a set of algebraic conditions and must be verified for all $(\rho,\theta)$.

\subsection{Three-center solution}
\label{sec:3centerdetails}

For the family of three-center solutions used in the text in section \ref{sec:concreteMGs}, the one-forms $(v_\Lambda,w^\Lambda)$ that encodes the charges of the gauge fields are given by:
\be
\begin{split}
w^0 &\= Q_0 \cos\theta \,d\phi,\qquad w^1 \= 0 \,, \\
w^2 &\= -k^{(2)} \left(h \cos\theta_2 + Q_0 \frac{\rho-a_2 \cos\theta}{\rho_2 a_2} \right)\,d\phi\,,\qquad w^3 \= - k^{(3)} \left(h \cos\theta_3 + Q_0 \frac{\rho-a_3 \cos\theta}{\rho_3 a_3} \right)\,d\phi\,,\\
v_0 &\= \frac{k^{(2)}}{h} \cos\theta_2 \,d\phi \+ \frac{k^{(3)}}{h}  \cos\theta_3 \,d\phi \- \left(k^{(2)} Q^{(3)}_2-k^{(3)} Q^{(2)}_3 \right)\frac{\rho^2 + a_2 a_3 -(a_2+a_3)\rho\cos \theta }{(a_3-a_2) \,\rho_2 \rho_3}\,d\phi \,,\\
v_1 & \=  Q^{(2)}_1 \cos\theta_2 \,d\phi \+ Q^{(3)}_1  \cos\theta_3 \,d\phi \+ Q_0 k^{(2)} k^{(3)} \frac{(\rho^2 + a_2 a_3) \cos\theta -(a_2+a_3)\rho }{a_2 a_3\,\rho_2 \rho_3}\,d\phi\,,\\
v_2 & \=  Q^{(3)}_2 \cos\theta_3 \,d\phi\,,\qquad v_3  \=  Q^{(2)}_3 \cos\theta_2 \,d\phi\,.
\end{split}
\ee

\section{Multipole moments of multicenter almost-BPS solutions in Taub-NUT}
\label{sec:MultipoleMulticenter}

In this section, we will compute the multipole moments of axisymmetric multicenter almost-BPS solutions.
Given the expression of the fields $Z_I$ (\ref{eq2:Zgeneralform}) and $\mu$ \ref{fullmu-exp}), the procedure constructed in \cite{Bena:2020uup} to compute multipole moments of BPS multicenter solutions does not apply here, so in the following we generalize this procedure to almost-BPS multicenter solutions.

\subsection{Algebra of multipole-decomposable functions}

We expand every pole $1/\rho_i$ on the $z$-axis with the multipole expansion in Legendre polynomials $P_l$:
\be \frac{1}{\rho_i} = \frac{1}{\sqrt{\rho^2 + a_i^2-2r a_i \cos\theta}} = \sum_{l=0}^{\infty} {a_i}^l \frac{P_l(\cos\theta)}{\rho^{l+1}}\,.\ee
Let $F$ be a function such as
\be
F=f_\infty+ \sum_{l=0}^{\infty} \frac{1}{\rho^{l+1}}\[P_l(\cos\theta)D_l(F)+(\textrm{lower harmonics than }P_l(\cos\theta))\] \,,
\ee
where the ``lower harmonics'' are comprised of products of Legendre polynomials $P_{l_1}\dots P_{l_m}$ with $\sum_j l_j<l$. In other terms, the polynomial degree of lower harmonic terms $P_{l_1}\dots P_{l_m}$ is at most $X^{l-1}$. We say that these functions ``decompose into a multipole expansion'', and will call the class of these functions to be ``multipole-decomposable''.
For instance, a harmonic function $G$ expands as 
\be \label{eq:harmfuncexpand} G = g^\infty + \sum_i \frac{g^{(i)}}{\rho_i} = g^\infty + \sum_i g^{(i)} \sum_{l=0}^\infty  {a_i}^l \frac{P_l(\cos\theta)}{\rho^{l+1}} \,,\ee
so $H$ is multipole-decomposable and its multipole decomposition at order $l$ is $D_l(H)= \sum_i g^{(i)} a_i^l$.

Applied on the vector space of multipole-decomposable functions, the multipole decomposition operator $D_l$ satisfies linearity.
When we multiply two multipole-decomposable functions together, at $O(\rho^{-(l+1)})$ we get:
\be \left(F_A F_B\right)_{\mathcal{O}(\rho^{-l-1})} = (f_A^\infty D_l(F_B)+f_B^\infty D_l(F_A)) \frac{P_l(\cos\theta)}{\rho^{l+1}} + \frac{\mathcal{LH}_{l}}{\rho^{l+1}},\ee
where $\mathcal{LH}_{l}$ denotes terms with lower harmonics than $P_l$ (which are polynomials in $\cos\theta$ of degree less or equal to $l-1$).
Thus, the vector space is an algebra and we read
\be
D_l(F_AF_B)=f_A^\infty D_l(F_B)+f_B^\infty D_l(F_A) \,.
\ee

To extract the $l$-th multipole from a generic functional $f(F_1,\dots,F_N)$ of multipole-decomposable functions $F_A$, the formula above generalizes to:
\be
\label{eq:D_l_general_formula}
D_l\[ f(F_1,\dots,F_N) \] = \sum_{B=1}^N \partial_{f_B^\infty}\left[f(F_1,\dots,F_N)_\infty\right] D_l(F_B) \,,
\ee
where we introduced the notation $f(F_1,\dots,F_N)_\infty:=\lim_{r\rightarrow\infty}f(F_1,\dots,F_N)$ to denote the functional evaluated when the radius $r$ is taken to infinity; this can be thought as a function of the moduli $f_A^\infty$.

\subsection{Mass multipoles}

We want to compute the multipole moments of a class of almost-BPS bubbling multi-center solutions whose four-dimensional metric is:
\be \label{eq:ds2multicenter} ds^2 = - (\mathcal{Q}(F_I))^{-1/2}(dt +  \omega)^2 + (\mathcal{Q}(F_I))^{1/2}\left( d\rho^2 + \rho^2d\theta^2 + \rho^2\sin^2\theta d\phi^2\right) \,,\ee
where the warp factor of the four-dimensional solution $\mathcal{Q}(H)$ is given by the expression:
\be \label{eq:quarticinvdef} \mathcal{Q}(F_1,\dots,F_N) = Z_1 Z_2 Z_3 V - (\mu V)^2.\ee

In this class of almost-BPS solutions, only five moduli $f^\infty$ are potentially turned on: $l^\infty_1$, $l^\infty_2$, $l^\infty_3$, $h\equiv v^\infty$ and $m_\infty$.\footnote{By comparison, the class of BPS solutions in the same STU model will admit in addition the magnetic moduli $k^\infty_1$, $k^\infty_2$, $k^\infty_3$.}
Therefore, we will use the following five multipole-decomposable functions $F_I=(Z_1,Z_2,Z_3,V,\mu V)$:
\be
\label{eq:five_decomposable_functions}
\begin{split}
Z_I \= &~ l^\infty_I  \+\frac{Q^{(0)}_I}{\rho} \+\sum_{j=1}^n \frac{Q^{(j)}_I}{\rho_j} + {|\epsilon_{IJK}| \over 2}\sum_{j,k=1}^n
\Bigl(h+{Q_0\, \rho\over a_j a_k}\Bigr) {k^{(j)}_J k^{(k)}_K \over \rho_j \rho_k}\,,\\
V \=  & ~h  \+ \frac{Q_0}{\rho}\\
\mu V \= &~m_\infty  \+\frac{m^{(0)}}{\rho} \+\sum_{j=1}^n \frac{m^{(j)}}{\rho_j} \+ \alpha \frac{\cos\theta}{\rho^2} \+ \sum_{j=1}^n \alpha_j \frac{\cos\theta_j}{\rho_j^2}\\
\+ &\sum_j l_I^\infty k^{(j)}_I \left(h\, f_j^{(1)} + Q_0\, f_j^{(2)}  \right) + \sum_{j} Q^{(j)}_I k^{(j)}_I \left(h\, f_{j}^{(3)} + Q_0\, f_{j}^{(4)}  \right) \\ 
\+& \sum_{i=0}^n\sum_{j=1,j\neq i}^n Q^{(i)}_I k^{(j)}_I \left(h\, f_{ij}^{(5)} + Q_0\, f_{ij}^{(6)}  \right) +  {C_{IJK}\over 6} \sum_{i,j,k} k^{(i)}_I k^{(j)}_J k^{(k)}_K \left({h}^2\, f^{(7)}_{ijk} + {Q_0}^2 \,f^{(8)}_{ijk} + h Q_0 \,f^{(9)}_{ijk}\right)  \\
\end{split}
\ee
where the functions $f^{(m)}_{i}$, $f^{(m)}_{ij}$ and $f^{(m)}_{ijk}$ are given in appendix \ref{sec:ABPSmulticenter} and are multipole-decomposable.

We wish to apply (\ref{eq:D_l_general_formula}) with $f(Z_1,Z_2,Z_3,V,\mu V)=-\mathcal{Q}^{-\frac{1}{2}}=-\( Z_1 Z_2 Z_3 V - (\mu V)^2 \)^{-\frac{1}{2}}$. As the quartic invariant at infinity equals  
\be \mathcal{Q}_\infty =  l^\infty_1 l^\infty_2 l^\infty_3 h - {m_\infty}^2  \,, \ee
we deduce
\be \label{eq:Dl_quartic_invariant}
D_l\[-\mathcal{Q}^{-\frac{1}{2}}\] =  
 \frac{1}{2}l^\infty_1 l^\infty_2 l^\infty_3 D_l[V] + \frac{1}{2}h 
\( l^\infty_1 l^\infty_2 D_l[Z_3] +l^\infty_2 l^\infty_3 D_l[Z_1] +l^\infty_3 l^\infty_1 D_l[Z_2] \)
- m_\infty D_l[\mu V]\,,
\ee
where we have used that $\mathcal{Q}_\infty^{-\frac{3}{2}}=1$ since we want to have an asymptotic flat space $\mathbb{R}^3$.
Thus, it remains to compute $D_l[F_I]$; by linearity, we only need to compute $D_l[f^{(m)}_{i}]$, $D_l[f^{(m)}_{ij}]$ and $D_l[f^{(m)}_{ijk}]$. 
\newline

By reading off the coefficient of the leading-degree monomial in $\cos\theta$, we deduce that $\frac{1}{\rho_i \rho_j}$ expands as
\be
\begin{split}
\frac{1}{\rho_i \rho_j} =& \sum_{l=1}^{\infty} \frac{1}{\rho^{l+1}} \sum_{p+q=l-1}^{p,q\geq 0} {a_i}^p {a_j}^q P_p P_q\\
=& \sum_{l=1}^{\infty} \frac{1}{\rho^{l+1}} \[ q^{(2)}_{l-1}(a_i,a_j) P_{l-1} + \mathcal{LH}_{l-1} \] \,,
\end{split}
\ee
where the bivariate polynomial $q^{(2)}_n$ is defined as
\be
q^{(2)}_{n}(a_i,a_j)\equiv\frac{1}{\binom{2n}{n}} \sum_{p+q=n} \binom{2p}{p} \binom{2q}{q} {a_i}^p {a_j}^q \,.
\ee
Similarly, we define the multivariate polynomial $q^{(3)}_n$:
\be
q^{(3)}_{n}(a_i,a_j,a_k)\equiv\frac{1}{\binom{2n}{n}} \sum_{p+q+s=n} \binom{2p}{p} \binom{2q}{q} \binom{2s}{s} {a_i}^p {a_j}^q {a_k}^s \,.
\ee
Note that
\be q^{(2)}_{n}(1,1)=\frac{4^n}{\binom{2n}{n}} \qquad , \qquad q^{(3)}_{n}(1,1,1)=2n+1 \,. \ee
The following functions then expand as
\bea
\frac{\rho}{\rho_i \rho_j} &=& \sum_{l=0}^{\infty} \frac{1}{\rho^{l+1}} \[ q^{(2)}_l(a_i,a_j) P_{l} + \mathcal{LH}_{l} \]\\
\frac{1}{\rho_i \rho_j \rho_k} &=& \sum_{l=2}^{\infty} \frac{1}{\rho^{l+1}} \[ q^{(3)}_{l-2}(a_i,a_j,a_k) P_{l-2} + \mathcal{LH}_{l-2} \]\\
\frac{\rho}{\rho_i \rho_j \rho_k} &=& \sum_{l=1}^{\infty} \frac{1}{\rho^{l+1}} \[ q^{(3)}_{l-1}(a_i,a_j,a_k) P_{l-1} + \mathcal{LH}_{l-1} \]\\
\frac{\rho^2}{\rho_i \rho_j \rho_k} &=& \sum_{l=0}^{\infty} \frac{1}{\rho^{l+1}} \[ q^{(3)}_{l}(a_i,a_j,a_k) P_{l} + \mathcal{LH}_{l} \] \,.
\eea
Following Bonnet's recursion formula, we have, for $l\geq 1$, $XP_{l-1}= \frac{l}{2l-1}P_l+ \mathcal{LH}_l$. We deduce
\bea
\frac{\cos\theta}{\rho_i \rho_j} &=& \sum_{l=1}^{\infty} \frac{1}{\rho^{l+1}} \[ q^{(2)}_{l-1}(a_i,a_j) \frac{l}{2l-1}P_l + \mathcal{LH}_{l} \] \label{eq:expansion_cos_rr}\\
\frac{\rho\cos\theta}{\rho_i \rho_j \rho_k} &=& \sum_{l=1}^{\infty} \frac{1}{\rho^{l+1}} \[ q^{(3)}_{l-1}(a_i,a_j,a_k) \frac{l}{2l-1}P_l + \mathcal{LH}_{l} \] \label{eq:expansion_rcos_rrr}\\
\frac{\cos\theta_i}{\rho_i^2}=\frac{r\cos\theta-a_i}{\rho_i^3} &=& \sum_{l=1}^{\infty} \frac{1}{\rho^{l+1}} \[ q^{(3)}_{l-1}(1,1,1) \frac{l}{2l-1} a_i^{l-1} P_l + \mathcal{LH}_{l} \] \,.
\eea

Thus, we deduce the expansions $D_l[f^{(m)}_{i}]$, $D_l[f^{(m)}_{ij}]$ and $D_l[f^{(m)}_{ijk}]$, which in turn, provide the mass multipoles from (\ref{eq:Dl_quartic_invariant}):
\be \label{eq:almost-BPS_microstates_mass_multipoles}
\begin{split}
4 \tilde{M}_l \= 
&l^\infty_1 l^\infty_2 l^\infty_3 Q_0 {a_0}^l 
+ h \frac{|\varepsilon_{IJK}|}{2}l^\infty_I l^\infty_J \sum_{j=0}^{n} Q^{(j)}_K {a_j}^l
+ h Q_0 \sum_{J\neq K}\frac{l^\infty_J l^\infty_K}{2} \sum_{j,k=1}^n k^{(j)}_J k^{(k)}_K \frac{q^{(2)}_l(a_j,a_k)}{a_ja_k}\\
&-2 m_\infty \sum_{j=0}^n m^{(j)} {a_j}^l 
\- 2m_\infty\sum_{j=0}^n \alpha_j \,l {a_j}^{l-1}
\- m_\infty h  l_I^\infty\sum_{j=1}^n k^{(j)}_I {a_j}^l \\
&- m_\infty Q_0\sum_{j=1}^n Q^{(j)}_I k^{(j)}_I \frac{4^{l-1}}{\binom{2(l-1)}{l-1}}\frac{l}{2l-1}{a_j}^{l-2} \\ 
&\- 2m_\infty Q_0 \sum_{i=0}^n\sum_{j=1,j\neq i}^n Q^{(i)}_I k^{(j)}_I \(\frac{q^{(2)}_l(a_i,a_j)}{2a_j\,(a_i-a_j)} -  \frac{l}{2l-1} \frac{q^{(2)}_{l-1}(a_i,a_j)}{a_i-a_j} \)\\
&\- 2m_\infty{Q_0}^2  \sum_{1\leq i,j,k\leq n} k^{(i)}_1 k^{(j)}_2 k^{(k)}_3\frac{l}{2l-1} \frac{q^{(3)}_{l-1}(a_i,a_j,a_k)}{a_ia_ja_k} \,. 
\end{split}
\ee
The position of the Taub-NUT center is fixed at $a_0=0$, and ${a_0}^l=\delta_{l,0}$.
The binomial coefficient in the third line of (\ref{eq:almost-BPS_microstates_mass_multipoles}) is not defined for $l=0$ but, as it is multiplied by $l$, one can set this term to be zero for $l=0$ and extend the formula (\ref{eq:almost-BPS_microstates_mass_multipoles}) for all $l\in \mathbb{N}$.
In particular,
\be \label{eq:almost-BPS_microstates_M0}
\begin{split}
4 \tilde{M}_0 \= 
&l^\infty_1 l^\infty_2 l^\infty_3 Q_0
+ h \frac{|\varepsilon_{IJK}|}{2}l^\infty_I l^\infty_J \sum_{j=0}^{n} Q^{(j)}_K 
+ h Q_0 \sum_{J\neq K} \frac{l^\infty_J l^\infty_K}{2} \sum_{j,k=1}^n k^{(j)}_J k^{(k)}_K \frac{1}{a_ja_k}\\
&-2 m_\infty \sum_{j=0}^n m^{(j)}
\- m_\infty h  l_I^\infty\sum_{j=1}^n k^{(j)}_I 
- m_\infty Q_0 \sum_{i=0}^n\sum_{j=1,j\neq i}^n Q^{(i)}_I k^{(j)}_I \frac{1}{a_j(a_i-a_j)} \,. 
\end{split}
\ee
and
\be \label{eq:almost-BPS_microstates_tilde_M1}
\begin{split}
4 \tilde{M}_1 \= 
& h \frac{|\varepsilon_{IJK}|}{2}l^\infty_I l^\infty_J \sum_{j=1}^{n} Q^{(j)}_K a_j
+ h Q_0 \sum_{J\neq K} \frac{l^\infty_J l^\infty_K}{2} \sum_{j,k=1}^n k^{(j)}_J k^{(k)}_K \frac{a_j+a_k}{a_ja_k}\\
&-2 m_\infty \sum_{j=1}^n m^{(j)} a_j 
\- 2m_\infty\sum_{j=0}^n \alpha_j 
\- m_\infty h  l_I^\infty\sum_{j=1}^n k^{(j)}_I a_j 
- m_\infty Q_0\sum_{j=1}^n Q^{(j)}_I k^{(j)}_I \frac{1}{a_j} \\ 
&\- 2m_\infty Q_0 \sum_{i=0}^n\sum_{j=1,j\neq i}^n Q^{(i)}_I k^{(j)}_I \( \frac{a_i+a_j}{2a_j(a_i-a_j)} - \frac{1}{a_i-a_j}\)
\- 2m_\infty{Q_0}^2  \sum_{1\leq i,j,k\leq n} \frac{k^{(i)}_1 k^{(j)}_2 k^{(k)}_3}{a_ia_ja_k} \,. 
\end{split}
\ee

In the scaling limit, we scale the intercenter distances $a_i\rightarrow\lambda d_i$, with $d_i\sim \mathcal{O}(1)$, and we keep the charges $\hat{\kappa}^{(j)}_I\equiv k^{(j)}_I/a_j$ fixed, according to (\ref{eq:effectivedipolecharges}). The last term of (\ref{eq:almost-BPS_microstates_M0}) naively scales like $- \lambda^{-1}\, m_\infty Q_0 \sum_{i=0}^n\sum_{j=1,j\neq i}^n Q^{(i)}_I \hat{\kappa}^{(j)}_I (d_i-d_j)^{-1}$. However, using the regularity constraint (\ref{regularity1}), the scaling limit of the mass multipole $M_0=\tilde{M}_0$ has no $\lambda^{-1}$ term and is in fact of the form
\be
\tilde{M}_0=\tilde{M}_0^{(0)}+\lambda\tilde{M}_0^{(1)} \,.
\ee
We see from (\ref{eq:almost-BPS_microstates_mass_multipoles}) that in the scaling limit, the dominant term in $\tilde{M}_l$ is of order $\lambda^{l-1}$ for $l\geq1$.

\subsection{Current multipoles}

To get the expression of the current multipoles $\tilde{S}_l$ defined in (\ref{eq:gtphimultipoles}), we need to determine the $\sin^2\theta P'_l(\cos\theta)$ expansions in $1/\rho^l$ ($l\leq 0$) of $\varpi\equiv\omega_\phi \:d\phi$ in (\ref{fullomega-exp})
\be
\begin{split}
\omega_\phi \= & \sum_j l_I^\infty k^{(j)}_I \left(h\, \tau_j^{(1)} + Q_0\, \tau_j^{(2)}  \right) + \sum_{j} Q^{(j)}_I k^{(j)}_I \left(h\, \tau_{j}^{(3)} + Q_0\, \tau_{j}^{(5)}  \right) \\ 
 &\+ \sum_{i=0}^n\sum_{j=1,j\neq i}^n Q^{(i)}_I k^{(j)}_I \left(h\, \tau_{ij}^{(4)} + Q_0\, \tau_{ij}^{(6)}  \right) +  {C_{IJK}\over 6} \sum_{i,j,k} k^{(i)}_I k^{(j)}_J k^{(k)}_K \left(h^2\, \tau^{(7)}_{ijk} + {Q_0}^2 \,\tau^{(8)}_{ijk} + h Q_0 \,\tau^{(9)}_{ijk}\right)\\
 &\+ \tau^{(10)}  \,, 
\end{split}
\ee
where the one-forms $\tau^{(m)}d\phi\equiv t^{(m)}$ are given in equations (\ref{eq:solution_nine_generating_functions}) and (\ref{eq:expression_t10}).

The additional difficulty in the computation of the current multipoles with respect to the mass multipoles is that in order to check that the the metric is AC-$\infty$ in its $g_{t\phi}$ part, one needs to be able to factorize by $\sin^2\theta$ in the multipole development of $g_{t\phi}$. 
To do this, we consider $\omega_\phi$ as a polynomial in $X \equiv \cos\theta$, before doing any expansion in $\frac{1}{\rho^l}$. The charges and the poles $a_i$ satisfy the regularity constraints in the like of (\ref{regularity1}) to avoid Dirac-Misner strings on the $z$-axis, that is to say $\omega_\phi(\cos\theta=1)=0$ and $\omega_\phi(\cos\theta=-1)=0$. Therefore $\omega_\phi$ is divisible by $(1-X)(1+X)=1-X^2= \sin^2\theta$.

We can now read off all the current multipoles simply by looking at the coefficient in front of the dominant term, $X^{l+1}$.
Any polynomial of degree $l$ or less is then counted among the the lower harmonics. For instance,
\be
\frac{\rho^2}{\rho_i \rho_j} = \sum_{l=0}^{\infty} \frac{1}{\rho^{l}} \[ q^{(2)}_l(a_i,a_j) P_{l} + \mathcal{LH}_{l} \]
\ee
is counted among the lower harmonics $\mathcal{LH}_{l+1}$ and does not contribute to the current multipole. It is easy to see that a generic function of the form 
\be \label{generic_function_expansion}
\frac{\rho^n \cos^m\theta}{\rho_{i_1}\cdots \rho_{i_p}} = \sum_{l=p-n}^\infty \frac{1}{\rho^{l}} \[ q^{(p)}_{l+n-p}(a_{i_1},\dots,a_{i_p}) X^m P_{l+n-p} + \mathcal{LH}_{l+n+m-p} \]
\ee
contributes to a polynomial of degree $l+(n+m-p)$ in the $\frac{1}{\rho^l}$-expansion. (In practice, $m=1$ or $m=2$.)
The dominant multipole is a polynomial of degree $l+1$, which is achieved when $n+m-p=1$.

The last paragraph shows that in the expression $g_{t\phi}=-2\mathcal{Q}^{-1/2} \omega_\phi$, $\mathcal{Q}^{-1/2}$ does not contribute to the current multipoles. Indeed, because $\mathcal{Q}_\infty=1$, we can write $\mathcal{Q}=1+\mathcal{R}$, where $\mathcal{R}$ comprises only functions of the form (\ref{generic_function_expansion}) with $n+m-p\leq -1$. Thus, the same applies to $\mathcal{Q}^{-1/2}$, and the functions of $\mathcal{R}$ are absorbed in the lower harmonics of $\omega_\phi$, whose highest harmonic generating functions verifiy $n+m-p=1$.

The relation between $P_n$ and $P'_n$
\be \frac{X^2-1}{n} P_n' = X P_n - P_{n-1} \ee
ensures that
\be
X^mP_{l+n-p}= (X^2-1) \frac{c_{l+n-p}}{c_{l+n-p+m-1}} \frac{1}{l+n-p+m-1} P'_{l+n-p+m-1} + \mathcal{LH}_{l+n+m-p}\,,
\ee
where $c_n=\frac{1}{2^n}\binom{2n}{n}$ is the coefficient of the leading order $X^n$ in $P_n$.
In practice, the functions that contribute to the dominant current multipole can be developed as
\be \label{eq:dominant_functions_for_current}
\begin{split}
\frac{\rho^p \cos\theta}{\rho_{i_1}\cdots \rho_{i_p}} &= \cos\theta +\sum_{l=1}^\infty \frac{1}{\rho^{l}} \[ (X^2-1) \frac{1}{l}q^{(p)}_{l}(a_{i_1},\dots,a_{i_p})  P'_{l} + \mathcal{LH}_{l+1} \] \,,\\
\frac{\rho^{p-1} \cos^2\theta}{\rho_{i_1}\cdots \rho_{i_p}} &= \sum_{l=1}^\infty \frac{1}{\rho^{l}} \[ (X^2-1) \frac{1}{2l-1}q^{(p)}_{l-1}(a_{i_1},\dots,a_{i_p}) P'_{l} + \mathcal{LH}_{l+1} \] \,.
\end{split}
\ee
Notice that $XP_0=X=\cos\theta$ is not divisible by $X^2-1$, so it cannot contribute to any current multipole.

Using (\ref{eq:dominant_functions_for_current}), one can derive the expansion of the functions appearing in $\varpi$ involving a degree $l+1$ polynomial, and by linearity, the expansion of $\omega_\phi$ itself.
Finally, factorizing $g_{t\phi}=-2\mathcal{Q}^{-1/2} \omega_\phi$ by $\sin^2\theta$, one deduces the current multipoles from (\ref{eq:gtphimultipoles})
\be \label{eq:almost-BPS_microstates_current_multipoles} 
\begin{split}
\tilde{S}_l \= & \frac{h}{4} \sum_{j=1}^n l_I^\infty k^{(j)}_I \,{a_j}^l 
- \frac{Q_0}{4}\sum_{j=1}^n Q^{(j)}_I k^{(j)}_I  \frac{l}{2l-1} \frac{4^{l-1}}{\binom{2(l-1)}{l-1}} {a_j}^{l-2} \\  
&\- \frac{Q_0}{2} \sum_{i=0}^n\sum_{j=1,j\neq i}^n Q^{(i)}_I k^{(j)}_I  \( \frac{l}{2l-1}\frac{q^{(2)}_{l-1}(a_i,a_j)}{a_j-a_i}  
-\frac{q^{(2)}_{l}(a_{i},a_{j})}{2a_j(a_j-a_i)}   \) \\
&\-  \frac{{Q_0}^2}{2} \sum_{1\leq i,j,k\leq n} k^{(i)}_1 k^{(j)}_2 k^{(k)}_3   \, \frac{l}{2l-1} \frac{q^{(3)}_{l-1}(a_i,a_j,a_k)}{a_ia_ja_k} 
\- \frac12 \sum_{j=0}^n m^{(j)} {a_j}^l
\- \frac12 \sum_{j=0}^n \alpha_j \, l \,{a_j}^{l-1} \,. 
\end{split}
\ee

It is easy to check that in the scaling limit, one gets the equality
\be \tilde{M}_1=m_\infty \tilde{S}_1 \,, \ee
which is consistent with equations (\ref{eq:BHACmassmult}) and (\ref{eq:BHACcurrentmult}) for the almost-BPS black hole (\ref{eq:almostBPSBHmetric}) with a parameter $\alpha$.

\subsection{Multipole ratios}\label{sec:multipoleratios}
As it was shown in \cite{Bena:2020see,Bena:2020uup}, if one wants to calculate \emph{ratios of vanishing multipoles} for certain black holes, there exist two methods: the \indirect and the \direct one. In this section we review these methods and apply them briefly to almost-BPS black holes, as a proof of concept that the analysis of \cite{Bena:2020see,Bena:2020uup} for supersymmetric black holes and their multicentered microstates can be straightforwardly generalized to the almost-BPS case (albeit only for $h=1$).

\subsubsection{Brief overview of \indirect and \direct methods}\label{sec:ratiosindirect}

In the \indirect method, we consider a generic STU black hole in maximal ungauged supergravity \cite{Chow:2014cca}; these black holes are characterized by 10 parameters: a non-extremality parameter, $m$, a rotation parameter, $a$, and four electric and four magnetic charge parameters, $\delta_I,\gamma_I$. For such a generic STU black hole, \emph{all} of the multipoles $M_\ell,S_\ell$ are non-zero, and so any multipole ratio, which we can generically denote as $\mathcal{M}$, is well-defined. Then, one can take the limit of $\mathcal{M}$ as we go from the generic STU black hole to the black hole in question. Defined in this way as a limit, any ratio of vanishing multipoles, $\mathcal{M}$, becomes well-defined for any black hole. For example, in \cite{Bena:2020see,Bena:2020uup} it was found that:
\be  \frac{M_2 S_2}{M_3S_1} = 1,\ee
for \emph{any} black hole; this includes the Kerr black hole for which this ratio is strictly speaking undefined (since $S_2=M_3=0$ for Kerr). The multipoles of a generic STU black hole can be seen as functions of four variables: the mass, $M$, the angular momentum, $J$, the rotation parameter, $a$, and the dipole moment, $M_1$.\footnote{Note that we are using a slightly different parameterization than in \cite{Bena:2020uup}, where $D=M_1/a$ was used instead; the one we use here is more natural for almost-BPS black holes and microstate geometries.} The \indirect method of calculating a ratio of vanishing multipoles, $\mathcal{M}$, for a black hole with parameters $(M_0,J_0,(M_1)_0,a_0)$ can then be summarized as: 
\be\label{eq:indirectmethoddef} \mathcal{M}^\text{ind}\left(M_0,J_0,(M_1)_0,a_0\right) = \lim_{\left(M,J,M_1,a\right)\rightarrow \left(M_0,J_0,(M_1)_0,a_0\right)} \mathcal{M}\left(M,J,M_1,a\right).\ee
For further discussion, see \cite{Bena:2020uup} (and especially appendix B therein for the subtleties of this \indirect method).

The \direct method instead uses scaling microstate geometries which tend to the black hole geometry in the scaling limit $\lambda\rightarrow 0$. This method calculates the ratios of vanishing multipoles, $\mathcal{M}$, by considering the multipole ratios of microstate geometries as one takes the scaling limit:
\be \mathcal{M}^\text{dir} = \lim_{\lambda\rightarrow 0} \mathcal{M}(\lambda).\ee
It was shown in \cite{Bena:2020see,Bena:2020uup} that ratios of vanishing multipoles can be computed in this way for supersymmetric black holes, in which all all multipole moments (except $M_0$) vanish.

\subsubsection{Almost-BPS multipole ratios at \texorpdfstring{$h=1$}{h=1}}
The Chow-Compere STU black holes \cite{Chow:2014cca} and thus the \indirect method \cite{Bena:2020see,Bena:2020uup} are only defined for the family of our almost-BPS black holes and microstates when we take the 4D modulus $h=1$.\footnote{This can be seen for example by comparing the asymptotic values of the matrices of scalar couplings (\ref{eq:Iscalcoupling})-(\ref{eq:Rscalcoupling}) for our ansatz to that of the STU black hole in \cite{Chow:2014cca}.} We therefore restrict ourself to these solutions; as mentioned above in section \ref{sec:MSmult},  this implies $M_\ell\sim\mathcal{O}(\lambda^\ell)$ and $S_\ell\sim\mathcal{O}(\lambda^{\ell-1})$ for microstate geometries in the scaling limit $\lambda\rightarrow 0$.

The generic underrotating limit (to which the almost-BPS black hole belongs) of the generic STU black hole family with parameters $(M,J,M_1,a)$ corresponds to the $a\rightarrow 0$ limit keeping $M,M_1,J$ finite. However, as we saw in (\ref{eq:almostBPSmultipoles}), our almost-BPS black hole with $h=1$ has \emph{all} multipoles vanishing except $M_0$ and $S_1$. Thus, the \indirect limit for our almost-BPS black hole family involves also taking the limit $M_1\rightarrow 0$, which we will always consider taking \emph{after} the underrotating limit.

Note that, precisely because all multipoles (except $M_0,S_1$) vanish for our almost-BPS black holes with $h=1$, any multipole ratio that involves two monomials in $M_\ell,S_\ell$ is a non-trivial quantity that we can compute and compare using both the \direct and \indirect method.
This is similar to the analysis of the difference between the \indirect and \direct calculations of multipole ratios for supersymmetric black holes, which was discussed in detail in \cite{Bena:2020see,Bena:2020uup}. There, it was found that the two methods only gave (closely) matching results for certain classes of (SUSY) black holes.

\begin{figure}[ht]\centering
\begin{subfigure}{0.32\textwidth}\centering
 \includegraphics[width=\textwidth]{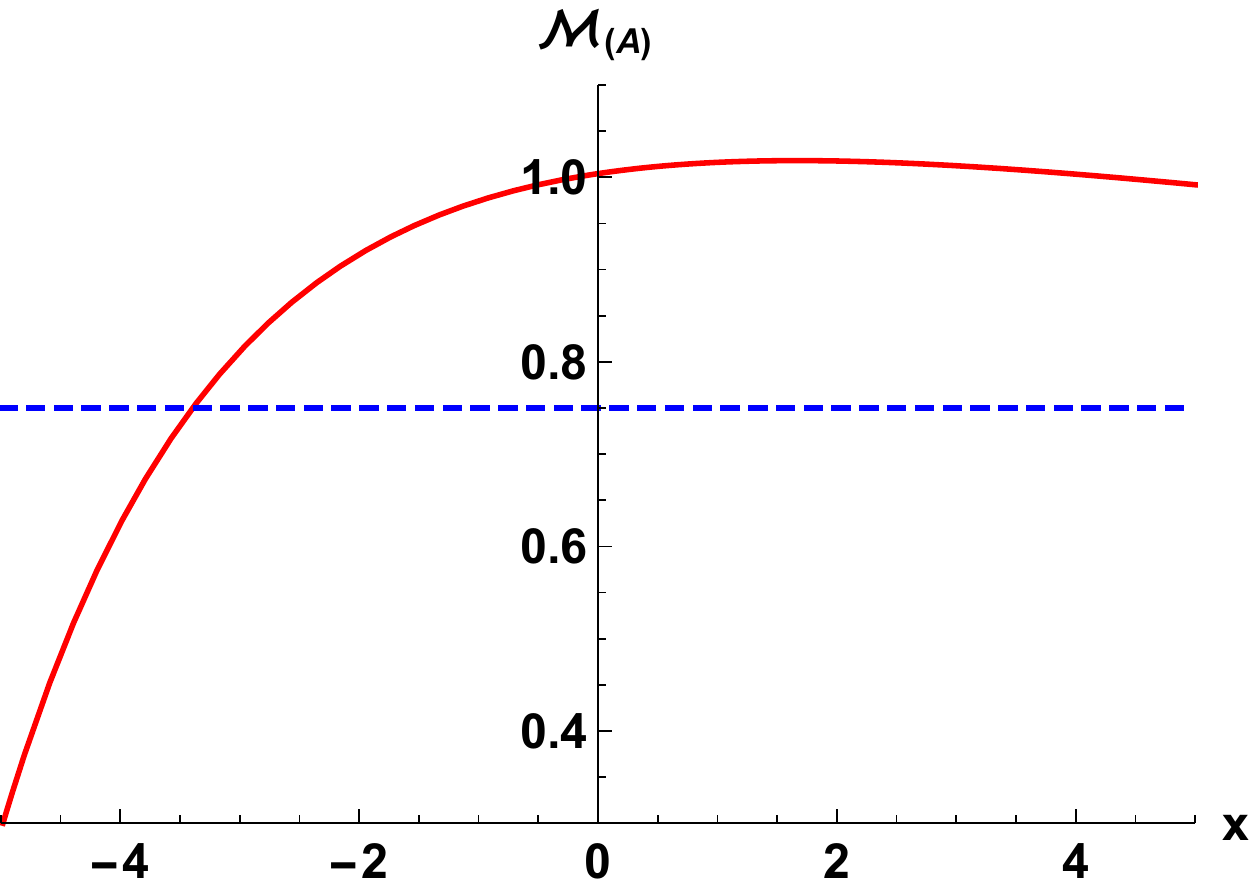}
 \caption{$\mathcal{M}_{(A)}$}
 \end{subfigure}
 \begin{subfigure}{0.32\textwidth}\centering
   \includegraphics[width=\textwidth]{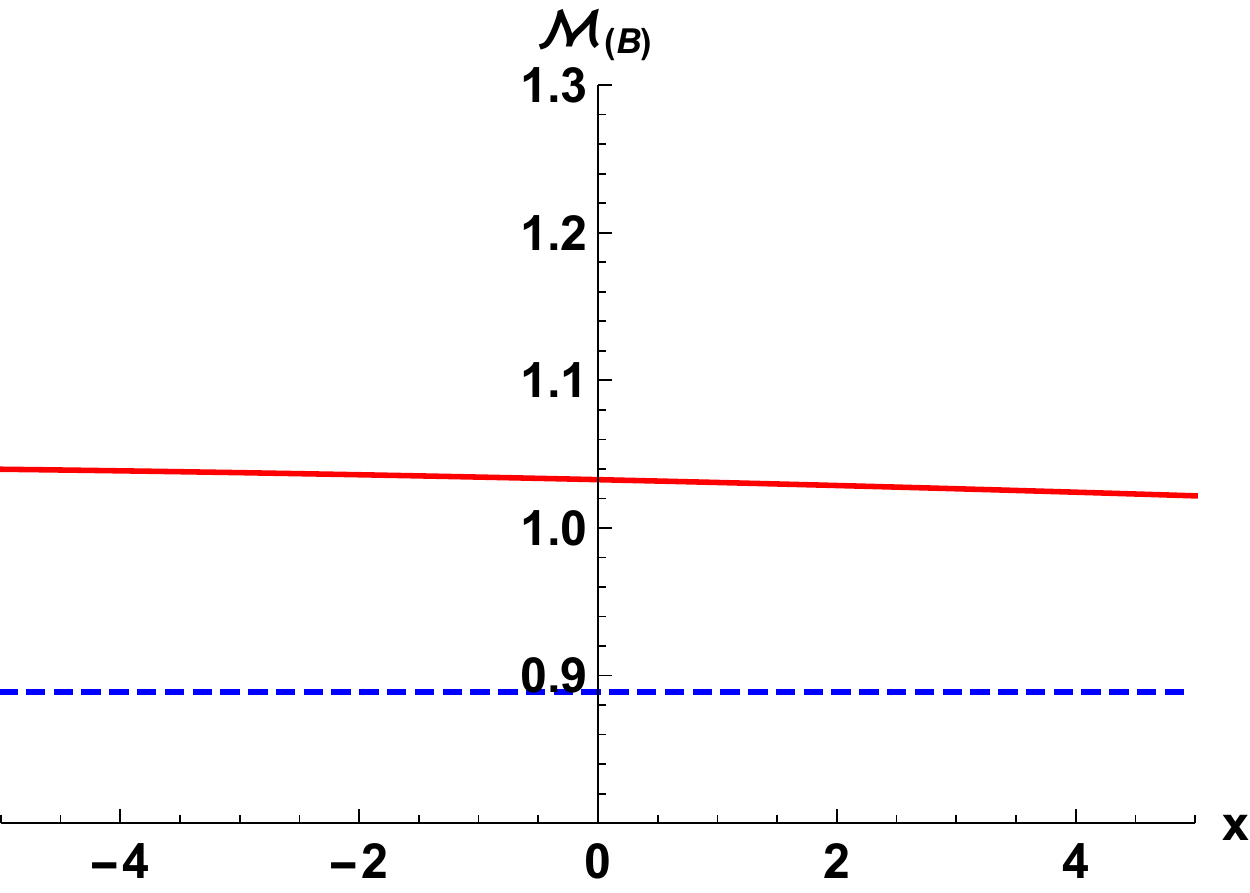}
    \caption{$\mathcal{M}_{(B)}$}
 \end{subfigure}
  \begin{subfigure}{0.32\textwidth}\centering
   \includegraphics[width=\textwidth]{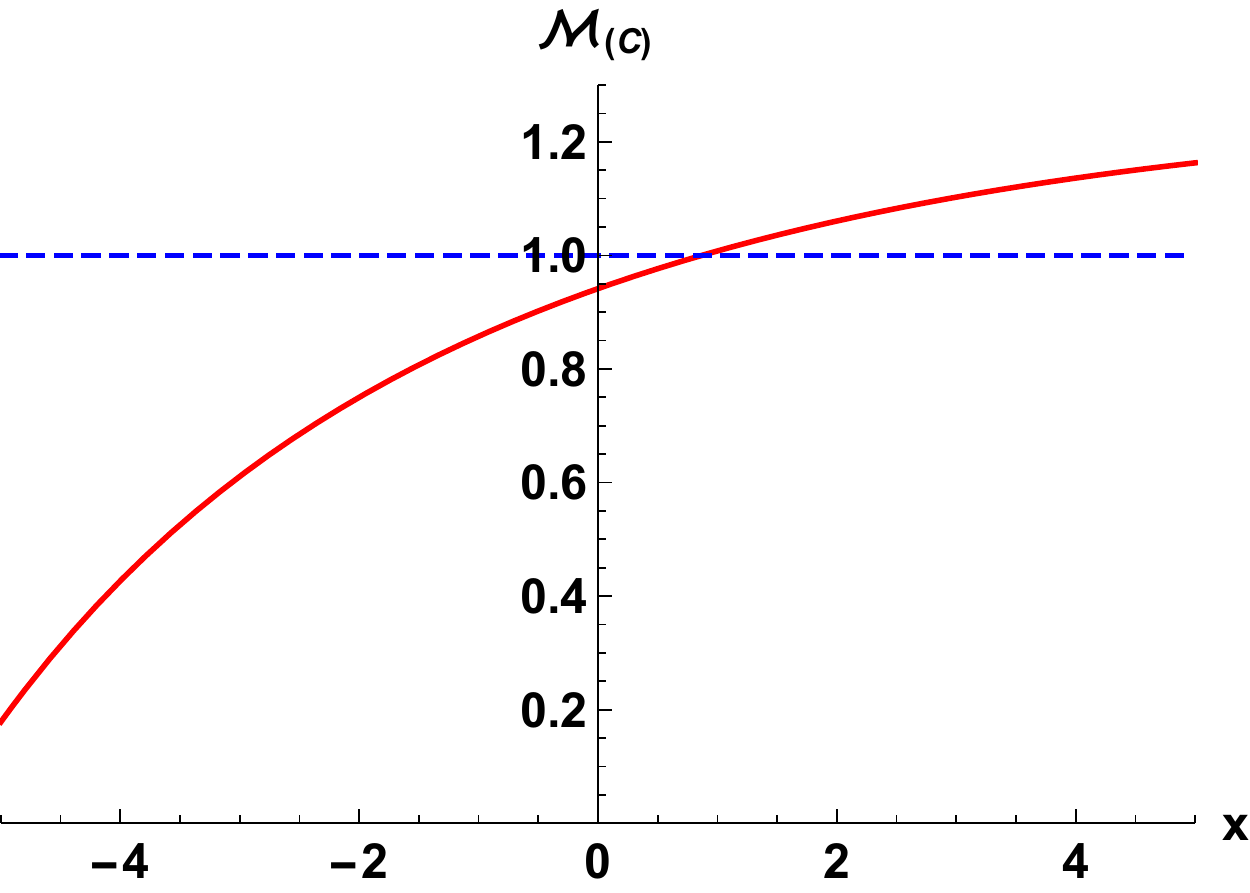}
    \caption{$\mathcal{M}_{(C)}$}
 \end{subfigure}
 \caption{Plots of the \direct method ratios $\mathcal{M}_{(A,B,C)}^\text{dir}$ (red lines) for the microstate family (\ref{eq:fammicrostates}) with $h=1$ as a function of $x$. The blue dashed lines indicate the \indirect method values given in (\ref{eq:indmethodratios}).}
 \label{fig:multipolesratios}
\end{figure}

As an example, let us consider the following three multipole ratios:
\be  \mathcal{M}_{(A)} \equiv \frac{M_2 M_4}{M_3 M_3}, \qquad  \mathcal{M}_{(B)} \equiv \frac{S_2 S_4}{S_3 S_3}, \qquad \mathcal{M}_{(C)} \equiv \frac{M_2 S_2}{M_3 S_1}. \ee
These have simple values when calculated with the \indirect method:
\be \label{eq:indmethodratios} \mathcal{M}_{(A)}^\text{ind} \equiv \lim_{M_1\rightarrow 0}\lim_{a\rightarrow 0} \frac{M_2 M_4}{M_3 M_3} = \frac34, \qquad \mathcal{M}_{(B)}^\text{ind} = \frac89, \qquad \mathcal{M}_{(C)}^\text{ind} = 1. \ee
Note that these \indirect ratios are pure numbers and do not depend on the charges of the black hole; this is a general feature of \indirect ratios for the almost-BPS black hole we consider here. However, the same ratios calculated in the \direct method \emph{do} depend on the specifics of the microstate. As an illustration, in fig. \ref{fig:multipolesratios} we consider the three ratios $\mathcal{M}_{(A,B,C)}$ for the family of microstates defined by (\ref{eq:fammicrostates}) with varying $x$ (and $h=1$).

Clearly, the \indirect and \direct methods do not necessarily agree with their predictions for the values of multipole ratios of the almost-BPS black hole. It would be interesting to conduct an analysis similar to \cite{Bena:2020see,Bena:2020uup} to understand if there is a certain condition that the black hole must satisfy in order for the two methods to give similar answers; we leave this for future work.

\begin{adjustwidth}{-1mm}{-1mm} 
\bibliographystyle{utphys}      
\bibliography{microstates}       

\providecommand{\href}[2]{#2}\begingroup\raggedright\begin{thebibliography}{10}

\bibitem{Abbott:2016blz}
{\bfseries LIGO Scientific, Virgo} , B.~P. Abbott {\em et~al.}, ``{Observation
  of Gravitational Waves from a Binary Black Hole Merger},''
  \href{http://dx.doi.org/10.1103/PhysRevLett.116.061102}{{\em Phys. Rev.
  Lett.} {\bfseries 116} no.~6, (2016) 061102},
\href{http://arxiv.org/abs/1602.03837}{{\ttfamily arXiv:1602.03837 [gr-qc]}}.

\bibitem{Abbott:2007kv}
{\bfseries LIGO Scientific} , B.~P. Abbott {\em et~al.}, ``{LIGO: The Laser
  interferometer gravitational-wave observatory},''
  \href{http://dx.doi.org/10.1088/0034-4885/72/7/076901}{{\em Rept. Prog.
  Phys.} {\bfseries 72} (2009) 076901},
  \href{http://arxiv.org/abs/0711.3041}{{\ttfamily arXiv:0711.3041 [gr-qc]}}.

\bibitem{EHT2019a}
{EHT~Collaboration}, ``{First M87 Event Horizon Telescope Results. I. The
  Shadow of the Supermassive Black Hole},''
  \href{http://dx.doi.org/10.3847/2041-8213/ab0ec7}{{\em ApJL} {\bfseries 875}
  no.~1, (Apr., 2019) L1}.

\bibitem{Audley:2017drz}
{\bfseries LISA} , P.~Amaro-Seoane {\em et~al.}, ``{Laser Interferometer Space
  Antenna},'' \href{http://arxiv.org/abs/1702.00786}{{\ttfamily
  arXiv:1702.00786 [astro-ph.IM]}}.

\bibitem{Punturo:2010zz}
M.~Punturo {\em et~al.}, ``{The Einstein Telescope: A third-generation
  gravitational wave observatory},''
  \href{http://dx.doi.org/10.1088/0264-9381/27/19/194002}{{\em Class. Quant.
  Grav.} {\bfseries 27} (2010) 194002}.

\bibitem{Mathur:2009hf}
S.~D. Mathur, ``{The information paradox: A pedagogical introduction},''
  \href{http://dx.doi.org/10.1088/0264-9381/26/22/224001}{{\em Class. Quant.
  Grav.} {\bfseries 26} (2009) 224001},
\href{http://arxiv.org/abs/0909.1038}{{\ttfamily arXiv:0909.1038 [hep-th]}}.

\bibitem{Almheiri:2012rt}
A.~Almheiri, D.~Marolf, J.~Polchinski, and J.~Sully, ``{Black Holes:
  Complementarity or Firewalls?},''
  \href{http://dx.doi.org/10.1007/JHEP02(2013)062}{{\em JHEP} {\bfseries 1302}
  (2013) 062},
\href{http://arxiv.org/abs/1207.3123}{{\ttfamily arXiv:1207.3123 [hep-th]}}.

\bibitem{Strominger:1996sh}
A.~Strominger and C.~Vafa, ``{Microscopic Origin of the Bekenstein-Hawking
  Entropy},'' \href{http://dx.doi.org/10.1016/0370-2693(96)00345-0}{{\em Phys.
  Lett.} {\bfseries B379} (1996) 99--104},
\href{http://arxiv.org/abs/hep-th/9601029}{{\ttfamily arXiv:hep-th/9601029}}.

\bibitem{Bena:2007kg}
I.~Bena and N.~P. Warner, ``{Black holes, black rings and their microstates},''
  \href{http://dx.doi.org/10.1007/978-3-540-79523-0}{{\em Lect. Notes Phys.}
  {\bfseries 755} (2008) 1--92},
\href{http://arxiv.org/abs/hep-th/0701216}{{\ttfamily arXiv:hep-th/0701216}}.

\bibitem{Bena:2016ypk}
I.~Bena, S.~Giusto, E.~J. Martinec, R.~Russo, M.~Shigemori, D.~Turton, and
  N.~P. Warner, ``{Smooth horizonless geometries deep inside the black-hole
  regime},'' \href{http://dx.doi.org/10.1103/PhysRevLett.117.201601}{{\em Phys.
  Rev. Lett.} {\bfseries 117} no.~20, (2016) 201601},
\href{http://arxiv.org/abs/1607.03908}{{\ttfamily arXiv:1607.03908 [hep-th]}}.

\bibitem{Goldstein:2008fq}
K.~Goldstein and S.~Katmadas, ``{Almost BPS black holes},''
  \href{http://dx.doi.org/10.1088/1126-6708/2009/05/058}{{\em JHEP} {\bfseries
  05} (2009) 058},
\href{http://arxiv.org/abs/0812.4183}{{\ttfamily arXiv:0812.4183 [hep-th]}}.

\bibitem{Bena:2009en}
I.~Bena, S.~Giusto, C.~Ruef, and N.~P. Warner, ``{Multi-Center non-BPS Black
  Holes - the Solution},''
  \href{http://dx.doi.org/10.1088/1126-6708/2009/11/032}{{\em JHEP} {\bfseries
  11} (2009) 032},
\href{http://arxiv.org/abs/0908.2121}{{\ttfamily arXiv:0908.2121 [hep-th]}}.

\bibitem{Bena:2009ev}
I.~Bena, G.~Dall'Agata, S.~Giusto, C.~Ruef, and N.~P. Warner, ``{Non-BPS Black
  Rings and Black Holes in Taub-NUT},''
  \href{http://dx.doi.org/10.1088/1126-6708/2009/06/015}{{\em JHEP} {\bfseries
  06} (2009) 015},
\href{http://arxiv.org/abs/0902.4526}{{\ttfamily arXiv:0902.4526 [hep-th]}}.

\bibitem{DallAgata:2010srl}
G.~Dall'Agata, S.~Giusto, and C.~Ruef, ``{U-duality and non-BPS solutions},''
  \href{http://dx.doi.org/10.1007/JHEP02(2011)074}{{\em JHEP} {\bfseries 02}
  (2011) 074},
\href{http://arxiv.org/abs/1012.4803}{{\ttfamily arXiv:1012.4803 [hep-th]}}.

\bibitem{Vasilakis:2011ki}
O.~Vasilakis and N.~P. Warner, ``{Mind the Gap: Supersymmetry Breaking in
  Scaling, Microstate Geometries},''
  \href{http://dx.doi.org/10.1007/JHEP10(2011)006}{{\em JHEP} {\bfseries 1110}
  (2011) 006},
\href{http://arxiv.org/abs/1104.2641}{{\ttfamily arXiv:1104.2641 [hep-th]}}.

\bibitem{Bena:2020see}
I.~Bena and D.~R. Mayerson, ``{Multipole Ratios: A New Window into Black
  Holes},'' \href{http://dx.doi.org/10.1103/PhysRevLett.125.221602}{{\em Phys.
  Rev. Lett.} {\bfseries 125} no.~22, (2020) 221602},
  \href{http://arxiv.org/abs/2006.10750}{{\ttfamily arXiv:2006.10750
  [hep-th]}}.

\bibitem{Bianchi:2020bxa}
M.~Bianchi, D.~Consoli, A.~Grillo, J.~F. Morales, P.~Pani, and G.~Raposo,
  ``{Distinguishing fuzzballs from black holes through their multipolar
  structure},'' \href{http://dx.doi.org/10.1103/PhysRevLett.125.221601}{{\em
  Phys. Rev. Lett.} {\bfseries 125} no.~22, (2020) 221601},
  \href{http://arxiv.org/abs/2007.01743}{{\ttfamily arXiv:2007.01743
  [hep-th]}}.

\bibitem{Bena:2020uup}
I.~Bena and D.~R. Mayerson, ``{Black Holes Lessons from Multipole Ratios},''
  \href{http://dx.doi.org/10.1007/JHEP03(2021)114}{{\em JHEP} {\bfseries 03}
  (2021) 114}, \href{http://arxiv.org/abs/2007.09152}{{\ttfamily
  arXiv:2007.09152 [hep-th]}}.

\bibitem{Bianchi:2020miz}
M.~Bianchi, D.~Consoli, A.~Grillo, J.~F. Morales, P.~Pani, and G.~Raposo,
  ``{The multipolar structure of fuzzballs},''
  \href{http://dx.doi.org/10.1007/JHEP01(2021)003}{{\em JHEP} {\bfseries 01}
  (2021) 003}, \href{http://arxiv.org/abs/2008.01445}{{\ttfamily
  arXiv:2008.01445 [hep-th]}}.

\bibitem{Mayerson:2020tpn}
D.~R. Mayerson, ``{Fuzzballs and Observations},''
  \href{http://dx.doi.org/10.1007/s10714-020-02769-w}{{\em Gen. Rel. Grav.}
  {\bfseries 52} no.~12, (2020) 115},
  \href{http://arxiv.org/abs/2010.09736}{{\ttfamily arXiv:2010.09736
  [hep-th]}}.

\bibitem{Cardoso:2017njb}
V.~Cardoso and P.~Pani, ``{The observational evidence for horizons: from echoes
  to precision gravitational-wave physics},''
  \href{http://arxiv.org/abs/1707.03021}{{\ttfamily arXiv:1707.03021 [gr-qc]}}.

\bibitem{Cardoso:2019rvt}
V.~Cardoso and P.~Pani, ``{Testing the nature of dark compact objects: a status
  report},'' \href{http://dx.doi.org/10.1007/s41114-019-0020-4}{{\em Living
  Rev. Rel.} {\bfseries 22} no.~1, (2019) 4},
  \href{http://arxiv.org/abs/1904.05363}{{\ttfamily arXiv:1904.05363 [gr-qc]}}.

\bibitem{Buchdahl:1959zz}
H.~A. Buchdahl, ``{General Relativistic Fluid Spheres},''
  \href{http://dx.doi.org/10.1103/PhysRev.116.1027}{{\em Phys. Rev.} {\bfseries
  116} (1959) 1027}.

\bibitem{Dimitrov:2020txx}
V.~Dimitrov, T.~Lemmens, D.~R. Mayerson, V.~S. Min, and B.~Vercnocke,
  ``{Gravitational Waves, Holography, and Black Hole Microstates},''
  \href{http://arxiv.org/abs/2007.01879}{{\ttfamily arXiv:2007.01879
  [hep-th]}}.

\bibitem{deBoer:2008zn}
J.~de~Boer, S.~El-Showk, I.~Messamah, and D.~Van~den Bleeken, ``{Quantizing N=2
  Multicenter Solutions},''
  \href{http://dx.doi.org/10.1088/1126-6708/2009/05/002}{{\em JHEP} {\bfseries
  05} (2009) 002},
\href{http://arxiv.org/abs/0807.4556}{{\ttfamily arXiv:0807.4556 [hep-th]}}.

\bibitem{Li:2021gbg}
Y.~Li, ``{Black Holes and the Swampland: the Deep Throat revelations},''
  \href{http://arxiv.org/abs/2102.04480}{{\ttfamily arXiv:2102.04480
  [hep-th]}}.

\bibitem{Bossard:2012ge}
G.~Bossard, ``{Octonionic black holes},''
  \href{http://dx.doi.org/10.1007/JHEP05(2012)113}{{\em JHEP} {\bfseries 05}
  (2012) 113}, \href{http://arxiv.org/abs/1203.0530}{{\ttfamily arXiv:1203.0530
  [hep-th]}}.

\bibitem{Bossard:2011kz}
G.~Bossard and C.~Ruef, ``{Interacting non-BPS black holes},''
  \href{http://dx.doi.org/10.1007/s10714-011-1256-9}{{\em Gen. Rel. Grav.}
  {\bfseries 44} (2012) 21--66},
  \href{http://arxiv.org/abs/1106.5806}{{\ttfamily arXiv:1106.5806 [hep-th]}}.

\bibitem{Bossard:2012xsa}
G.~Bossard and S.~Katmadas, ``{Duality covariant non-BPS first order
  systems},'' \href{http://dx.doi.org/10.1007/JHEP09(2012)100}{{\em JHEP}
  {\bfseries 09} (2012) 100}, \href{http://arxiv.org/abs/1205.5461}{{\ttfamily
  arXiv:1205.5461 [hep-th]}}.

\bibitem{Cremmer:1984hj}
E.~Cremmer, C.~Kounnas, A.~Van~Proeyen, J.~P. Derendinger, S.~Ferrara,
  B.~de~Wit, and L.~Girardello, ``{Vector Multiplets Coupled to N=2
  Supergravity: SuperHiggs Effect, Flat Potentials and Geometric Structure},''
  \href{http://dx.doi.org/10.1016/0550-3213(85)90488-2}{{\em Nucl. Phys. B}
  {\bfseries 250} (1985) 385--426}.

\bibitem{Duff:1995sm}
M.~J. Duff, J.~T. Liu, and J.~Rahmfeld, ``{Four-dimensional
  string-string-string triality},''
  \href{http://dx.doi.org/10.1016/0550-3213(95)00555-2}{{\em Nucl. Phys. B}
  {\bfseries 459} (1996) 125--159},
  \href{http://arxiv.org/abs/hep-th/9508094}{{\ttfamily arXiv:hep-th/9508094}}.

\bibitem{Chow:2014cca}
D.~D.~K. Chow and G.~Comp\`ere, ``{Black holes in N=8 supergravity from SO(4,4)
  hidden symmetries},''
  \href{http://dx.doi.org/10.1103/PhysRevD.90.025029}{{\em Phys. Rev. D}
  {\bfseries 90} no.~2, (2014) 025029},
  \href{http://arxiv.org/abs/1404.2602}{{\ttfamily arXiv:1404.2602 [hep-th]}}.

\bibitem{Thorne:1980ru}
K.~S. Thorne, ``{Multipole Expansions of Gravitational Radiation},''
  \href{http://dx.doi.org/10.1103/RevModPhys.52.299}{{\em Rev. Mod. Phys.}
  {\bfseries 52} (1980) 299--339}.

\bibitem{Cardoso:2016olt}
V.~Cardoso, C.~F.~B. Macedo, P.~Pani, and V.~Ferrari, ``{Black holes and
  gravitational waves in models of minicharged dark matter},''
  \href{http://dx.doi.org/10.1088/1475-7516/2016/05/054}{{\em JCAP} {\bfseries
  05} (2016) 054}, \href{http://arxiv.org/abs/1604.07845}{{\ttfamily
  arXiv:1604.07845 [hep-ph]}}. [Erratum: JCAP 04, E01 (2020)].

\bibitem{KNGRMHDpaper}
B.~Ripperda, J.~Davelaar, H.~Olivares, D.~R. Mayerson, F.~Bacchini,
  B.~Vercnocke, and T.~Hertog, ``Can we detect signatures of dark matter and
  hidden dimensions in black-hole shadows?,'' 2021.
\newblock in preparation.

\bibitem{bozzola2020general}
G.~Bozzola and V.~Paschalidis, ``General relativistic simulations of the
  quasi-circular inspiral and merger of charged black holes: Gw150914 and
  fundamental physics implications,'' 2020.

\bibitem{Sotiriou:2004ud}
T.~P. Sotiriou and T.~A. Apostolatos, ``{Corrected multipole moments of
  axisymmetric electrovacuum spacetimes},''
  \href{http://dx.doi.org/10.1088/0264-9381/21/24/003}{{\em Class. Quant.
  Grav.} {\bfseries 21} (2004) 5727--5733},
  \href{http://arxiv.org/abs/gr-qc/0407064}{{\ttfamily arXiv:gr-qc/0407064}}.

\bibitem{Babak:2017tow}
S.~Babak, J.~Gair, A.~Sesana, E.~Barausse, C.~F. Sopuerta, C.~P.~L. Berry,
  E.~Berti, P.~Amaro-Seoane, A.~Petiteau, and A.~Klein, ``{Science with the
  space-based interferometer LISA. V: Extreme mass-ratio inspirals},''
  \href{http://dx.doi.org/10.1103/PhysRevD.95.103012}{{\em Phys. Rev. D}
  {\bfseries 95} no.~10, (2017) 103012},
  \href{http://arxiv.org/abs/1703.09722}{{\ttfamily arXiv:1703.09722 [gr-qc]}}.

\bibitem{Barack:2006pq}
L.~Barack and C.~Cutler, ``{Using LISA EMRI sources to test off-Kerr deviations
  in the geometry of massive black holes},''
  \href{http://dx.doi.org/10.1103/PhysRevD.75.042003}{{\em Phys. Rev. D}
  {\bfseries 75} (2007) 042003},
  \href{http://arxiv.org/abs/gr-qc/0612029}{{\ttfamily arXiv:gr-qc/0612029}}.

\bibitem{Gair:2012nm}
J.~R. Gair, M.~Vallisneri, S.~L. Larson, and J.~G. Baker, ``{Testing General
  Relativity with Low-Frequency, Space-Based Gravitational-Wave Detectors},''
  \href{http://dx.doi.org/10.12942/lrr-2013-7}{{\em Living Rev. Rel.}
  {\bfseries 16} (2013) 7}, \href{http://arxiv.org/abs/1212.5575}{{\ttfamily
  arXiv:1212.5575 [gr-qc]}}.

\bibitem{Barausse:2020rsu}
E.~Barausse {\em et~al.}, ``{Prospects for Fundamental Physics with LISA},''
  \href{http://dx.doi.org/10.1007/s10714-020-02691-1}{{\em Gen. Rel. Grav.}
  {\bfseries 52} no.~8, (2020) 81},
  \href{http://arxiv.org/abs/2001.09793}{{\ttfamily arXiv:2001.09793 [gr-qc]}}.

\bibitem{Cunha:2018uzc}
P.~V.~P. Cunha, C.~A.~R. Herdeiro, and E.~Radu, ``{Isolated black holes without
  $\mathbb Z_2$ isometry},''
  \href{http://dx.doi.org/10.1103/PhysRevD.98.104060}{{\em Phys. Rev. D}
  {\bfseries 98} no.~10, (2018) 104060},
  \href{http://arxiv.org/abs/1808.06692}{{\ttfamily arXiv:1808.06692 [gr-qc]}}.

\bibitem{Raposo:2018xkf}
G.~Raposo, P.~Pani, and R.~Emparan, ``{Exotic compact objects with soft
  hair},'' \href{http://dx.doi.org/10.1103/PhysRevD.99.104050}{{\em Phys. Rev.
  D} {\bfseries 99} no.~10, (2019) 104050},
  \href{http://arxiv.org/abs/1812.07615}{{\ttfamily arXiv:1812.07615 [gr-qc]}}.

\bibitem{Junior:2021atr}
H.~C. D.~L. Junior, L.~C.~B. Crispino, P.~V.~P. Cunha, and C.~A.~R. Herdeiro,
  ``{Mistaken identity: can different black holes cast the same shadow?},''
  \href{http://arxiv.org/abs/2102.07034}{{\ttfamily arXiv:2102.07034 [gr-qc]}}.

\bibitem{Chen:2020aix}
C.-Y. Chen, ``{Rotating black holes without $\mathbb{Z}_2$ symmetry and their
  shadow images},'' \href{http://dx.doi.org/10.1088/1475-7516/2020/05/040}{{\em
  JCAP} {\bfseries 05} (2020) 040},
  \href{http://arxiv.org/abs/2004.01440}{{\ttfamily arXiv:2004.01440 [gr-qc]}}.

\bibitem{Bates:2003vx}
B.~Bates and F.~Denef, ``{Exact solutions for supersymmetric stationary black
  hole composites},'' \href{http://dx.doi.org/10.1007/JHEP11(2011)127}{{\em
  JHEP} {\bfseries 1111} (2011) 127},
\href{http://arxiv.org/abs/hep-th/0304094}{{\ttfamily arXiv:hep-th/0304094
  [hep-th]}}.

\bibitem{Bena:2006kb}
I.~Bena, C.-W. Wang, and N.~P. Warner, ``{Mergers and Typical Black Hole
  Microstates},'' \href{http://dx.doi.org/10.1088/1126-6708/2006/11/042}{{\em
  JHEP} {\bfseries 11} (2006) 042},
\href{http://arxiv.org/abs/hep-th/0608217}{{\ttfamily arXiv:hep-th/0608217}}.

\bibitem{Bena:2007qc}
I.~Bena, C.-W. Wang, and N.~P. Warner, ``{Plumbing the Abyss: Black Ring
  Microstates},'' \href{http://dx.doi.org/10.1088/1126-6708/2008/07/019}{{\em
  JHEP} {\bfseries 07} (2008) 019},
\href{http://arxiv.org/abs/0706.3786}{{\ttfamily arXiv:0706.3786 [hep-th]}}.

\bibitem{Bena:2008dw}
I.~Bena, N.~Bobev, C.~Ruef, and N.~P. Warner, ``{Supertubes in Bubbling
  Backgrounds: Born-Infeld Meets Supergravity},''
  \href{http://dx.doi.org/10.1088/1126-6708/2009/07/106}{{\em JHEP} {\bfseries
  07} (2009) 106},
\href{http://arxiv.org/abs/0812.2942}{{\ttfamily arXiv:0812.2942 [hep-th]}}.

\bibitem{Mateos:2001qs}
D.~Mateos and P.~K. Townsend, ``{Supertubes},''
  \href{http://dx.doi.org/10.1103/PhysRevLett.87.011602}{{\em Phys. Rev. Lett.}
  {\bfseries 87} (2001) 011602},
\href{http://arxiv.org/abs/hep-th/0103030}{{\ttfamily arXiv:hep-th/0103030}}.

\bibitem{Emparan:2001ux}
R.~Emparan, D.~Mateos, and P.~K. Townsend, ``{Supergravity supertubes},'' {\em
  JHEP} {\bfseries 07} (2001) 011,
\href{http://arxiv.org/abs/hep-th/0106012}{{\ttfamily arXiv:hep-th/0106012}}.

\bibitem{Bena:2008wt}
I.~Bena, N.~Bobev, and N.~P. Warner, ``{Spectral Flow, and the Spectrum of
  Multi-Center Solutions},''
  \href{http://dx.doi.org/10.1103/PhysRevD.77.125025}{{\em Phys. Rev.}
  {\bfseries D77} (2008) 125025},
\href{http://arxiv.org/abs/0803.1203}{{\ttfamily arXiv:0803.1203 [hep-th]}}.

\bibitem{Bena:2009fi}
I.~Bena, S.~Giusto, C.~Ruef, and N.~P. Warner, ``{Supergravity Solutions from
  Floating Branes},'' \href{http://dx.doi.org/10.1007/JHEP03(2010)047}{{\em
  JHEP} {\bfseries 03} (2010) 047},
\href{http://arxiv.org/abs/0910.1860}{{\ttfamily arXiv:0910.1860 [hep-th]}}.

\bibitem{Bena:2005va}
I.~Bena and N.~P. Warner, ``{Bubbling supertubes and foaming black holes},''
  \href{http://dx.doi.org/10.1103/PhysRevD.74.066001}{{\em Phys. Rev.}
  {\bfseries D74} (2006) 066001},
\href{http://arxiv.org/abs/hep-th/0505166}{{\ttfamily arXiv:hep-th/0505166}}.

\bibitem{Berglund:2005vb}
P.~Berglund, E.~G. Gimon, and T.~S. Levi, ``{Supergravity microstates for BPS
  black holes and black rings},''
  \href{http://dx.doi.org/10.1088/1126-6708/2006/06/007}{{\em JHEP} {\bfseries
  0606} (2006) 007},
\href{http://arxiv.org/abs/hep-th/0505167}{{\ttfamily arXiv:hep-th/0505167
  [hep-th]}}.

\bibitem{danieloddparity}
K.~Fransen, L.~K\"uchler, and D.~R. Mayerson.
\newblock in preparation.

\bibitem{Bacchini:2021fig}
F.~Bacchini, D.~R. Mayerson, B.~Ripperda, J.~Davelaar, H.~Olivares, T.~Hertog,
  and B.~Vercnocke, ``{Fuzzball Shadows: Emergent Horizons from
  Microstructure},'' \href{http://arxiv.org/abs/2103.12075}{{\ttfamily
  arXiv:2103.12075 [hep-th]}}.

\bibitem{Ikeda:2021uvc}
T.~Ikeda, M.~Bianchi, D.~Consoli, A.~Grillo, J.~F. Morales, P.~Pani, and
  G.~Raposo, ``{Black-hole microstate spectroscopy: ringdown, quasinormal
  modes, and echoes},'' \href{http://arxiv.org/abs/2103.10960}{{\ttfamily
  arXiv:2103.10960 [gr-qc]}}.

\bibitem{Nishino:1984gk}
H.~Nishino and E.~Sezgin, ``{Matter and Gauge Couplings of N=2 Supergravity in
  Six-Dimensions},''
\href{http://dx.doi.org/10.1016/0370-2693(84)91800-8}{{\em Phys. Lett.}
  {\bfseries 144B} (1984) 187--192}.

\bibitem{Ferrara:1997gh}
S.~Ferrara, F.~Riccioni, and A.~Sagnotti, ``{Tensor and vector multiplets in
  six-dimensional supergravity},''
  \href{http://dx.doi.org/10.1016/S0550-3213(97)00837-7}{{\em Nucl. Phys.}
  {\bfseries B519} (1998) 115--140},
\href{http://arxiv.org/abs/hep-th/9711059}{{\ttfamily arXiv:hep-th/9711059
  [hep-th]}}.

\bibitem{Gutowski:2003rg}
J.~B. Gutowski, D.~Martelli, and H.~S. Reall, ``{All supersymmetric solutions
  of minimal supergravity in six dimensions},''
  \href{http://dx.doi.org/10.1088/0264-9381/20/23/008}{{\em Class. Quant.
  Grav.} {\bfseries 20} (2003) 5049--5078},
\href{http://arxiv.org/abs/hep-th/0306235}{{\ttfamily arXiv:hep-th/0306235}}.

\bibitem{Cariglia:2004kk}
M.~Cariglia and O.~A.~P. Mac~Conamhna, ``{The General form of supersymmetric
  solutions of N=(1,0) U(1) and SU(2) gauged supergravities in
  six-dimensions},'' \href{http://dx.doi.org/10.1088/0264-9381/21/13/006}{{\em
  Class. Quant. Grav.} {\bfseries 21} (2004) 3171--3196},
\href{http://arxiv.org/abs/hep-th/0402055}{{\ttfamily arXiv:hep-th/0402055
  [hep-th]}}.

\bibitem{deLange:2015gca}
P.~de~Lange, D.~R. Mayerson, and B.~Vercnocke, ``{Structure of Six-Dimensional
  Microstate Geometries},''
  \href{http://dx.doi.org/10.1007/JHEP09(2015)075}{{\em JHEP} {\bfseries 09}
  (2015) 075},
\href{http://arxiv.org/abs/1504.07987}{{\ttfamily arXiv:1504.07987 [hep-th]}}.

\bibitem{Antoniadis:1995vz}
I.~Antoniadis, S.~Ferrara, and T.~R. Taylor, ``{N=2 heterotic superstring and
  its dual theory in five-dimensions},''
  \href{http://dx.doi.org/10.1016/0550-3213(95)00659-1}{{\em Nucl. Phys.}
  {\bfseries B460} (1996) 489--505},
\href{http://arxiv.org/abs/hep-th/9511108}{{\ttfamily arXiv:hep-th/9511108
  [hep-th]}}.

\bibitem{Gunaydin:1983bi}
M.~Gunaydin, G.~Sierra, and P.~K. Townsend, ``{The Geometry of N=2
  Maxwell-Einstein Supergravity and Jordan Algebras},''
\href{http://dx.doi.org/10.1016/0550-3213(84)90142-1}{{\em Nucl. Phys.}
  {\bfseries B242} (1984) 244--268}.

\bibitem{Gunaydin:1984ak}
M.~Gunaydin, G.~Sierra, and P.~K. Townsend, ``{Gauging the d = 5
  Maxwell-Einstein Supergravity Theories: More on Jordan Algebras},''
  \href{http://dx.doi.org/10.1016/0550-3213(85)90547-4}{{\em Nucl. Phys.}
  {\bfseries B253} (1985) 573}.
[,573(1984)].

\bibitem{Gutowski:2004yv}
J.~B. Gutowski and H.~S. Reall, ``{General supersymmetric AdS(5) black
  holes},'' \href{http://dx.doi.org/10.1088/1126-6708/2004/04/048}{{\em JHEP}
  {\bfseries 04} (2004) 048},
\href{http://arxiv.org/abs/hep-th/0401129}{{\ttfamily arXiv:hep-th/0401129}}.

\bibitem{Denef:2000nb}
F.~Denef, ``{Supergravity flows and D-brane stability},''
  \href{http://dx.doi.org/10.1088/1126-6708/2000/08/050}{{\em JHEP} {\bfseries
  0008} (2000) 050},
\href{http://arxiv.org/abs/hep-th/0005049}{{\ttfamily arXiv:hep-th/0005049
  [hep-th]}}.

\bibitem{Denef:2002ru}
F.~Denef, ``{Quantum quivers and Hall / hole halos},''
  \href{http://dx.doi.org/10.1088/1126-6708/2002/10/023}{{\em JHEP} {\bfseries
  0210} (2002) 023},
\href{http://arxiv.org/abs/hep-th/0206072}{{\ttfamily arXiv:hep-th/0206072
  [hep-th]}}.

\end{thebibliography}\endgroup

\end{adjustwidth}

\end{document}